\documentclass[]{ufdissertation}

\usepackage{tikz}
\usepackage{pgfplots}
\usepackage{algpseudocode}
\usepackage[all]{nowidow}

\usepackage{scrextend}
\deffootnote{1.5em}{0em}{\thefootnotemark\quad}

\usepackage[compatibility=false]{caption}
\usepackage{subcaption}
\usepackage{natbib}
\usepackage{hyperref}

\haveTablestrue
\haveFigurestrue

\title{QUANTUM GRAVITATIONAL CORRECTIONS TO ELECTROMAGNETISM AND BACKREACTION}

\degreeType{Doctor of Philosophy}
\major{Physics}
\author{Sanjib Katuwal}
\thesisType{Dissertation}
\degreeYear{2023}
\degreeMonth{August}
\chair{Richard P. Woodard}

\setDedicationFile{dedicationFile}
\setAcknowledgementsFile{acknowledgementsFile}
\setAbstractFile{abstractFile}
\setReferenceFile{refFile}{ieeetr}

\setBiographicalFile{biographyFile}

\setAppendixFile{appendix}
\multipleAppendixtrue

\begin{document}



\chapter{INTRODUCTION and opening remarks} \label{intro}
\counterwithin{algorithm}{chapter}
\renewcommand*{\thealgorithm}{\thechapter-\arabic{algorithm}}

In the catalogue of fundamental interactions, the two for which mankind has the greatest familiarity are gravity
and electromagnetism. The theme of this thesis is to explore the interactions between these two fundamental forces on 
the quantum level. This thesis has six chapters, with the Chapter \ref{intro} being this introduction and Chapter \ref{conclusion} being the conclusion.

\section{Effects of Quantum Gravity on Electromagnetism}
In Chapter \ref{gaugePaper} and Chapter \ref{perturbative-scalar-coupling}, we will look at the effects of gravity on electromagnetism. In classical electrodynamics, we tackle almost all problems by solving Maxwell's equations. However, as soon as quantum corrections are taken into consideration, received wisdon is that we must abandon the classical approach and infer all physics from scattering amplitudes. This approach has been very successful, but it cannot be easily generalized to Cosmology. In Chapter \ref{gaugePaper}, we will quantum correct Maxwell's equation using quantum gravitational corrections to the vacuum polarization in flat space background. These equations can be used just as one would use Maxwell's equations to solve problems in classical electrodynamics. The power of this method lies in its ability to generalize to cosmology.

The next effect of quantum gravity on electromagnetism is discussed in Chapter \ref{perturbative-scalar-coupling}. It has recently been suggested that quantum gravitationally induced scalar couplings to electromagnetism might be detected by atom interferometers. Widespread belief is that such couplings can only be generated using nonperturbative effects. In Chapter \ref{perturbative-scalar-coupling}, we will show a completely perturbative mechanism to calculate the coefficient through which quantum gravity induces a dimension six coupling to a massive scalar.

\section{Effects of Electromagnetism on Quantum Gravity}
In Chapter \ref{coleman-weinberg} and Chapter \ref{reheatingChapter}, we will consider the effects of electromagnetism on gravity. As the inflaton is part of the gravitational sector, in Chapter \ref{coleman-weinberg}, we calculate the one-loop photon contribution to the inflaton effective potential. On a general cosmological background with arbitrary first slow-roll parameter $\epsilon$, this leads to a calculation of generalized Coleman-Weinberg potential.

During primordial inflation, the temperature of the universe decreases by factor of about 100,000. The universe regains its temperature through the process of reheating during which the inflaton oscillates and its kinetic energy is transferred to ordinary matter. In Chapter \ref{reheatingChapter}, we will generalize our result from Chapter \ref{coleman-weinberg} for constant photon mass to variable photon mass and use it to consider reheating for a charged inflaton minimally coupled to electromagnetism. 

Quantum Gravity is a non-renormalizable theory and we will treat general relativity as an effective field theory in all of the calculations that follow. We will perform them perturbatively, without worrying about its UV completion, the necessity for which does not arise for physics at distances larger than about a Planck length, $10^{-31}$ m.
\chapter{Gauge Independent Quantum Gravitational Corrections to Maxwell's Equation} \label{gaugePaper}

\section{Introduction}

The greatest story ever told in physics is how a century of brilliant 
experimental extemporization culminated in the development of Maxwell's
equations.\footnote{This chapter has been adapted from a published article in JHEP\cite{Katuwal:2020rkv}.} This was humanity's first relativistic, unified field theory
and it set the stage for the discoveries of general relativity and 
non-Abelian gauge theories. Electrodynamics is still one of the core 
subjects in the study of physics. Most western physicists recall the 
ingenuity and perseverance required of them as graduate students to 
solve Maxwell's equations in the wide variety of settings treated in 
the classic text by the late J. D. Jackson \cite{Jackson:1998nia}.

Quantum loop corrections to electrodynamics are small at low frequencies,
and those from quantum gravity are unobservable. One might therefore
expect that including these effects causes only a small change in 
electrodynamics. The math is simple enough: one first computes the 1PI 
(one-particle-irreducible) 2-photon function, $i[\mbox{}^{\mu} 
\Pi^{\nu}](x;x')$, known as the ``vacuum polarization''. Then Maxwell's 
equations are supplemented by the integral of the vacuum polarization 
contracted into the vector potential $A_{\nu}(x')$,
\begin{equation}
\partial_{\nu} F^{\nu\mu}(x) + \int \!\! d^4x' \Bigl[\mbox{}^{\mu} 
\Pi^{\nu}\Bigr](x;x') A_{\nu}(x') = J^{\mu}(x) \; , \label{QMax}
\end{equation}
where $F_{\mu\nu} \equiv \partial_{\mu} A_{\nu} - \partial_{\nu} A_{\mu}$
is the field strength tensor and $J^{\mu}(x)$ is the current density.
However, students of quantum field theory are strongly enjoined that
they cannot think of solving the quantum-corrected equation the same
as its classical analog; they must instead abandon the concept of local 
fields and infer physics entirely from scattering amplitudes. Although 
basing physics on scattering amplitudes is valid for most situations on 
flat space background, it does seem to be an over-reaction, and it is 
not even possible in cosmology. The purpose of this chapter is to provide 
a version of the quantum-corrected field equation (\ref{QMax}) which can 
be solved as in classical electrodynamics.

Part of the reason for the curious dichotomy between classical and 
quantum is the prevalence of the ``in-out'' formalism so elegantly
summarized by the Feynman rules. The in-out vacuum polarization is 
neither real, nor is it causal in the sense of vanishing for points
${x'}^{\mu}$ outside the past light-cone of $x^{\mu}$. Those two 
properties are not errors; in-out amplitudes are precisely the right 
objects of study for computing scattering amplitudes. However, the 
absence of reality and causality is certainly problematic if one 
wishes to regard solutions to the quantum-corrected field equation
(\ref{QMax}) as electric and magnetic fields. 

Julian Schwinger long ago devised a method for computing true 
expectation values which is almost as simple to use as the Feynman 
rules \cite{Schwinger:1960qe}. When the vacuum polarization of the 
Schwinger-Keldysh formalism is employed in equation (\ref{QMax}) the 
effective field equations become manifestly real and causal 
\cite{Mahanthappa:1962ex,Bakshi:1962dv,Bakshi:1963bn,Keldysh:1964ud,
Chou:1984es,Jordan:1986ug,Calzetta:1986ey,Ford:2004wc}. However, 
there is still an obstacle: the propagators of vector and tensor 
fields require gauge fixing, and loop corrections involving these 
propagators cause the vacuum polarization to depend on the choice 
of gauge. For example, single graviton loop corrections to the vacuum 
polarization on a $D$-dimensional flat space background ($g_{\mu\nu} 
\equiv \eta_{\mu\nu} + \kappa h_{\mu\nu}$ with $\kappa^2 = 16 \pi G$) 
with the most general, Poincar\'e invariant gauge fixing functional,
\begin{equation}
\mathcal{L}_{GF} = -\frac1{2 a} \eta^{\mu\nu} F_{\mu} F_{\nu} \qquad ,
\qquad F_{\mu} = \eta^{\rho\sigma} \Bigl(h_{\mu \rho , \sigma} -
\frac{b}{2} h_{\rho\sigma , \mu}\Bigr) \; , \label{gauge}
\end{equation}
result in a primitive vacuum polarization of the form \cite{Leonard:2012fs},
\begin{equation}
i\Bigl[\mbox{}^{\mu} \Pi^{\nu}\Bigr](x;x') = -\frac{\kappa^2 
\mathcal{C}_0(D,a,b) (D\!-\!2) \Gamma^2(\frac{D}2 \!-\! 1)}{32 (D\!-\!1) 
\pi^D} \Bigl( \eta^{\mu\nu} \partial^2 \!-\! \partial^{\mu} \partial^{\nu}
\Bigr) \frac{1}{(x \!-\! x')^{2D-2}} \; , \label{vacpol}
\end{equation}
where the gauge dependent multiplicative factor is,
\begin{eqnarray}
\lefteqn{\mathcal{C}_0(D,a,b) = D (D \!-\! 2) (D\!-\!3) + \frac{(D\!-\!1) 
(D\!-\!2)^2 [(D\!-\!2) (a \!-\! 1) \!-\! D (b \!-\! 1)^2]}{2 (b \!-\! 2)^2} }
\nonumber \\
& & \hspace{5.5cm} + (D\!-\!1) (D\!-\!2)^2 (D\!-\! 4) \Bigl[ -
\frac{(a \!-\!1)}{2} \!+\! \frac{2}{D \!-\! 2} \Bigl( \frac{b \!-\! 1}{
b \!-\! 2}\Bigr)\Bigr] \; . \label{C0def} \qquad 
\end{eqnarray}
Although the tensor structure and spacetime dependence of (\ref{vacpol})
is universal, the multiplicative factor $\mathcal{C}(D,a,b)$ can be made
to range from $-\infty$ to $+\infty$ by adjusting the gauge parameters 
$a$ and $b$ \cite{Leonard:2012fs}.

John Donoghue has shown how to use general relativity as a low energy
effective field theory to reliably compute quantum gravitational 
corrections to the long-range potentials induced by the exchange of 
massless particles such as photons and gravitons \cite{Donoghue:1993eb,
Donoghue:1994dn}. His technique is to compute the scattering amplitude 
between two massive particles which interact with the massless field,
and then use inverse scattering theory to infer the exchange potential.
In this way one can derive gauge independent, single graviton loop
corrections to the Newtonian potential \cite{Bjerrum-Bohr:2002fji,
Bjerrum-Bohr:2002gqz} and to the Coulomb potential 
\cite{Bjerrum-Bohr:2002aqa}. 

It has recently been noted that Donoghue's S-matrix technique can be 
short-circuited to produce gauge independent effective field equations 
directly, without passing through the intermediate stages of computing 
scattering amplitudes and solving the inverse scattering problem 
\cite{Miao:2017feh}. The key is applying position space versions of a 
series of identities derived by Donoghue and collaborators for the 
purpose of isolating the nonlocal and nonanalytic parts of scattering 
amplitudes which correct long-range potentials \cite{Donoghue:1994dn,
Donoghue:1996mt}. These identities degenerate the massive propagators 
of the particles being scattered to delta functions, thus casting the
important parts of higher-point contributions to 2-particle scattering 
in a form that can be regarded as corrections to the 1PI 2-point 
function of the massless field. In this picture the gauge dependence 
of the original effective field equation derives from having omitted 
to include quantum gravitational interactions with the source which 
disturbs the effective field and from the observer who measures it; 
and the corrections to the 1PI 2-point function repair this omission. 
The new technique has already been implemented at one loop order for 
quantum gravitational corrections to a massless scalar on flat space
background, and its independence of the gauge parameters $a$ and $b$ 
explicitly demonstrated \cite{Miao:2017feh}. In this paper we do the
same for quantum gravitational corrections to electrodynamics, which
is a realistic system and one involving vector fields.

This chapter closely follows the analysis of Bjerrum-Bohr 
\cite{Bjerrum-Bohr:2002aqa} who applied Donoghue's technique to 
include one graviton loop corrections to electrodynamics on flat space 
background. Section 2 goes through the position-space version of each 
of the same diagrams he considered, including first order perturbations
of the gauge parameters (\ref{gauge}),
\begin{equation}
\left\{\begin{matrix}
a \equiv 1 + \delta a \cr b \equiv 1 + \delta b
\end{matrix} \right\} \qquad \Longrightarrow \qquad \mathcal{C}_0(4,a,b) 
= 8 + 12 \!\cdot\! \delta a + 0 \!\cdot\! \delta b + O(\delta^2) \; . 
\label{deltagauge}
\end{equation}
In each case we show how the Donoghue identities allow one to regard 
the diagram as a correction to the vacuum polarization. Of course the 
gauge dependence cancels when everything is summed up, and the result 
has the same form (\ref{vacpol}), but with the constant 
$\mathcal{C}_0(4,a,b)$ replaced by the gauge independent number $+40$.
Our conclusions comprise section 3. Three appendices give, respectively, 
the vertices, the propagators and the Donoghue identities, including
the new one we required for certain of the $\delta b$ contributions.

\section{Including the Source and the Observer}

In this section, we use the scattering of a pair of massive, charged
scalars to provide the source which disturbs the effective field and 
the observer who measures this disturbance. The Lagrangian that describes
the scattering is,
\begin{equation}\label{lagrangian1}
\mathcal{L} = \left[\frac{R}{16\pi G}-\frac{1}{4} g^{\alpha\mu} g^{\beta\nu} 
F_{\alpha\nu} F_{\mu\beta} - (D_\mu\phi)^* g^{\mu\nu} (D_\nu\phi) + m^2 \phi^* 
\phi \right] \sqrt{-g} \; ,
\end{equation}
where, $D_{\mu} = \nabla_{\mu} - i e A_{\mu}$ and $\nabla_\mu$ denotes the 
metric-compatible covariant derivative. (When acting on a scalar the covariant
derivative degenerates to the partial derivative, $D_{\mu} \phi = 
\partial_{\mu} \phi - i e A_{\mu} \phi$.) Unless otherwise stated, we work with 
the usual $c = \hbar =1$ convention of particle physics, however, we employ a
spacelike metric. General relativity plus SQED (Scalar Quantum Electrodynamics) is treated
as a low energy effective field theory in the sense of Donoghue 
\cite{Donoghue:1993eb,Donoghue:1994dn}. The perturbation is around flat space 
with the following definitions of the graviton field $h_{\mu\nu}$ and the loop 
counting parameter $\kappa^2$,
\begin{equation}
g_{\mu\nu}(x) \equiv \eta_{\mu\nu} + \kappa h_{\mu\nu} , \quad 
\kappa^2 \equiv 16 \pi G \; .
\end{equation}
The vertices we require are listed in Appendix \ref{appendix:vertices}, and the various propagators 
are given in Appendix \ref{appendix:propagators}.

\begin{figure}[H]
\centering
\includegraphics[width=0.95\textwidth]{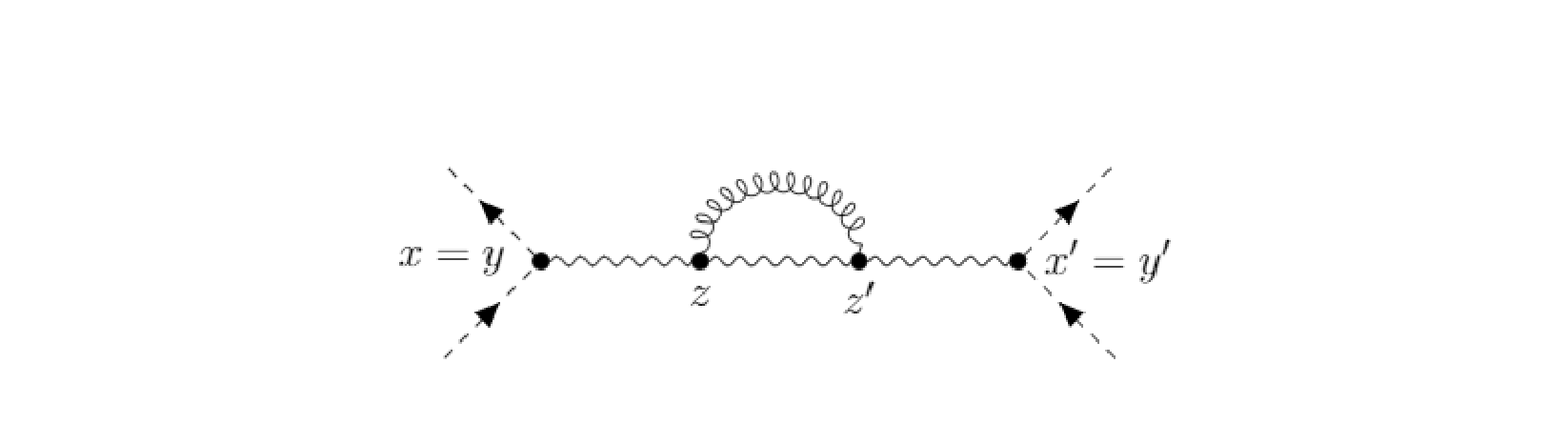}
\caption[Vacuum polarization contribution to amputated 4-scalar function]{\footnotesize This diagram shows how the vacuum polarization contributes to 
the amputated 4-scalar vertex function. Dashed lines represent massive scalars, wavy 
lines represent photons and curly lines represent gravitons. These graphs have the 
same topology as Bjerrum-Bohr's Diagram 8 \cite{Bjerrum-Bohr:2002aqa}.}
\label{fig:feyn_so}
\end{figure}

Our procedure for purging gauge dependence from the one loop vacuum polarization
is to write down position space representations for each of the order $e^2 \kappa^2$ 
contributions to the amputated 4-scalar function. Any external derivatives are assumed 
to act on the external scalar wave functions appropriate to 2-particle scattering. By 
exploiting the various Donoghue Identities of Appendix \ref{appendix:donoghue} to degenerate the (internal)
massive scalar propagators to Dirac delta functions, we reduce each contribution to
a form that can be interpreted as a correction to $i[\mbox{}^{\mu} \Pi^{\nu}](x;x')$.

We begin by considering the contribution of the original, gauge dependent vacuum 
polarization to the amputated 4-scalar function as shown in Figure \ref{fig:feyn_so}.
The expression for this diagram is
\begin{equation}\label{eq:so}
    \begin{split}
        iV_0(x;x') = & e (\partial_{x}\downarrow - \partial_{x} \uparrow)^{\alpha}
        \times e(\partial_{x'} \downarrow - \partial_{x'} \uparrow)^{\beta} \\
        & \times \int d^Dz\ i[_\alpha\Delta_\mu](x;z) \int d^Dz'\ i[_\beta\Delta_\nu](x';z')
        \times i[^\mu\Pi^\nu](z;z')
    \end{split}
\end{equation}
where the vacuum polarization was given in (\ref{vacpol}) and external derivatives with 
an up (down) arrow act on upper (lower) scalar wave functions at that vertex. First
note that Poincar\'e invariance and partial integration allows us to act all longitudinal
parts on the external legs, where (by current conservation) they vanish due to the 
on-shell condition,
\begin{equation}
(\partial \downarrow - \partial \uparrow)_{\mu} (\partial \downarrow + \partial 
\uparrow)^{\mu} = (\partial^2 \downarrow - m^2) - (\partial^2 \uparrow - m^2) \; .
\end{equation}
We can also use the relation,
\begin{equation}
\frac1{\Delta x^{2D-2}} = \frac{\partial^2}{2 (D\!-\! 2)^2} \frac1{\Delta x^{2D-4}}
= \frac{\partial^2}{2 (D\!-\!2)^2} \Biggl[ \frac{4 \pi^{\frac{D}2}}{\Gamma(
\frac{D}2 \!-\!1)} i\Delta(x;x')\Biggr]^2 \; ,
\end{equation}
to attain the form,
\begin{equation}\label{eq:vp_explicit}
\begin{split}
    iV_0(x;x') = & - \frac{e^2 \kappa^2 \mathcal{C}_0(D,a,b)}{4 (D\!-\! 1)(D\!-\!2)}
    \times (\partial_{x} \downarrow - \partial_{x} \uparrow)_{\mu}
    (\partial_{x'} \downarrow - \partial_{x'} \uparrow)^{\mu} \\
    & \times \int d^Dz\ i\Delta(x;z) \int d^Dz'\ i\Delta(x';z') \, 
    \partial_{z}^2 \partial_{z'}^2 \Bigl[ i\Delta(z;z')\Bigr]^2 \; .
\end{split}
\end{equation}
The final step is to partially integrate the factors of $\partial_{z}^2$ and 
$\partial_{z'}^2$ to act on the massless propagators, and use the delta functions
that result from the propagator equation (\ref{eq:equation_massless}) to eliminate
the integrations over $z^{\mu}$ and ${z'}^{\mu}$,
\begin{equation}\label{eq:vp_final}
iV_0(x;x') = \frac{\mathcal{C}_0(D,a,b)}{(D\!-\!1) (D\!-\!2)} \times 
\frac{e^2 \kappa^2}{4} (\partial_{x} \downarrow - \partial_{x} \uparrow) 
\!\cdot\! (\partial_{x'} \downarrow - \partial_{x'} \uparrow) \times 
[i\Delta(x;x')]^2 \; .
\end{equation}
After applying the appropriate Donoghue Identity from Appendix C it turns out 
that all contributions to the amputated 4-scalar function take this same form, with 
different gauge dependent multiplicative factors. To simplify the notation, we 
define a new gauge dependent constant which includes the factor of $1/(D\!-\!1)
(D\!-\!2)$, and we take $D=4$ because dimensional regularization plays no role, 
while also dropping higher order perturbations in the gauge parameters $a = 1 + 
\delta a$ and $b = 1 + \delta b$,
\begin{equation}
\frac{\mathcal{C}_0(D,a,b)}{(D\!-\!1) (D\!-\!2)} \equiv C_0(\delta a,\delta b) +
O\Bigl(D\!-\!4,\delta^2\Bigr) \; . \label{newC0def}
\end{equation}
In other words, $C_0(\delta a,\delta b) = \frac43 + 2 \delta a$.

\subsection{Correlation between Vertices}
\begin{figure}[H]
\centering
\includegraphics[width=0.9\textwidth]{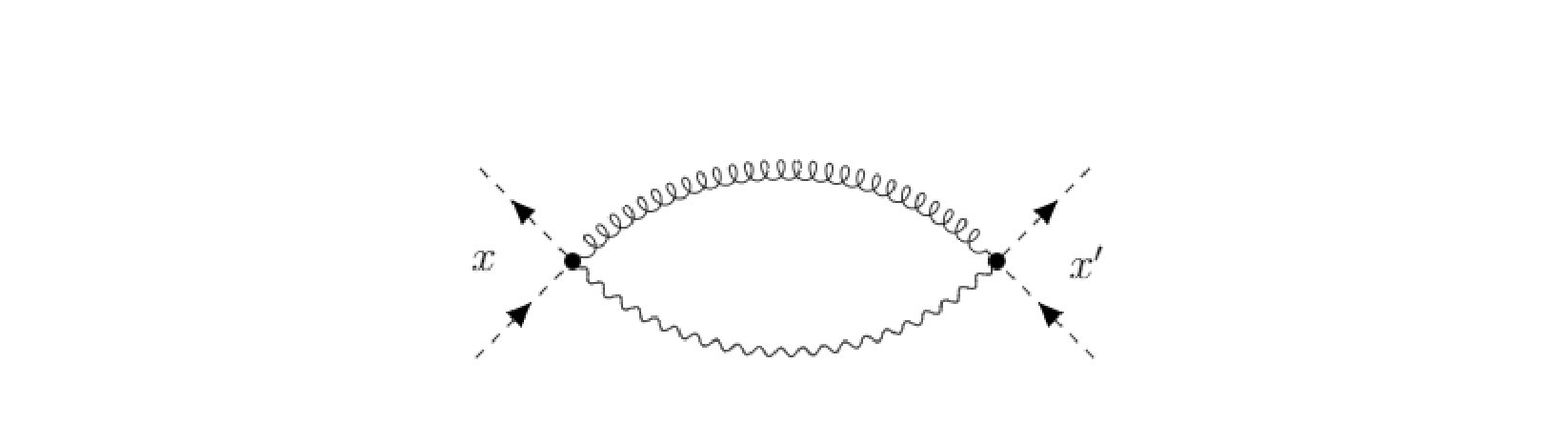}
\caption[Contribution of graviton correlation between two vertices]{\footnotesize This diagram shows the contribution of graviton correlation 
between two vertices. Dashed lines represent massive scalars, wavy lines represent 
photons and curly lines represent gravitons. These graphs have the same topology 
as Bjerrum-Bohr's Diagram 4 \cite{Bjerrum-Bohr:2002aqa}.}
\label{fig:feyn_betn_vert}
\end{figure}

The correlation between source (at ${x'}^{\mu}$) and observer (at $x^{\mu}$) 
vertices is the first extra contribution to the amputated 4-scalar function,
as shown in Figure \ref{fig:feyn_betn_vert}. This diagram corresponds to the
analytic expression,
\begin{equation}\label{eq:feyn_betn_vert}
\begin{split}
    iV_1(x;x')=&\frac{1}{2} e \kappa \left[\eta^{\delta\mu} \eta^{\nu\alpha}
    \!+\! \eta^{\delta\nu} \eta^{\mu\alpha} \!-\! \eta^{\mu\nu} \eta^{\delta\alpha}\right]
    (\partial_{x} \!\uparrow \!- \partial_{x}\!\downarrow)_{\alpha} \\
    & \times \frac{1}{2} e \kappa \left[\eta^{\gamma\rho} \eta^{\sigma\beta}
    \!+\! \eta^{\sigma\gamma} \eta^{\rho\beta} \!-\! \eta^{\rho\sigma} 
    \eta^{\gamma\beta}\right] (\partial_{x'} \!\uparrow \!- \partial_{x'}\!\downarrow)_{\beta} \\
    & \times i[_{\mu\nu}\Delta_{\rho\sigma}](x;x') \times i[_\gamma\Delta_\delta](x;x') \; .
\end{split}
\end{equation}
Substituting the appropriate propagators from Appendix B, contracting 
all the indices, simplifying and making use of the relation,
\begin{equation}\label{eq:prop_relation_1}
    i\Delta(x;x') \frac{\partial_\mu\partial_\nu}{\partial^2} i\Delta(x;x') = 
    \frac{1}{4} \!\times\! \frac{\Gamma^2(\frac{D}{2}-1)}{16\pi^D} 
    \left[\frac{\eta_{\mu\nu}}{\Delta x^{2D-4}} \!+\! \frac{\partial_{\mu}
    \partial_{\nu}}{(2D \!-\! 6)} \frac{1}{\Delta x^{2D-6}}\right] \; ,
\end{equation}
gives,
\begin{small}
\begin{equation}\label{eq:vert_corr_explicit}
    i V_1(x;x') = \frac{e^2 \kappa^2 \Gamma^2(\frac{D}{2} \!-\!1)}{16\pi^D}
    \left[D \!+\! \frac{(3D\!-\!2) \delta a}{4} \!-\! \frac{(D-2)^2 \delta b}{4}\right]
    (\partial_{x} \!\downarrow \!- \partial_{x} \!\uparrow)_{\mu} 
    (\partial_{x'} \!\downarrow \!- \partial_{x'} \!\uparrow)_{\nu} 
    \frac{\eta_{\mu\nu}}{\Delta x^{2D-4}} .
\end{equation}
\end{small}

\noindent Note that the second term in the square bracket of expression 
(\ref{eq:prop_relation_1}) drops out by current conservation.

Recognizing the massless scalar propagator (\ref{massless_scalar_prop}) provides 
a simpler form for (\ref{eq:vert_corr_explicit}),
\begin{equation}
i V_1(x;x') = \left[D \!+\! \frac{(3D\!-\!2) \delta a}{4} \!-\! \frac{(D-2)^2 
\delta b}{4}\right] \!\times\! e^2 \kappa^2 (\partial_{x} \!\downarrow \!- 
\partial_{x} \!\uparrow) \!\cdot\! (\partial_{x'} \!\downarrow \!- \partial_{x'}
\! \uparrow) \!\times\! [i\Delta(x;x')]^2 \; . \label{V1penult}
\end{equation}
As promised, expression (\ref{eq:vert_corr_explicit}) takes the same form as
the vacuum polarization contribution (\ref{eq:vp_final}), but with a different
gauge dependent, multiplicative constant. By comparison with (\ref{eq:vp_final})
we can recognize,
\begin{equation}
    \mathcal{C}_1(D,a,b) = 4 (D\!-\!1) (D\!-\!2)
    \left[D \!+\! \frac{(3D-2) \delta a}{4} \!-\! \frac{(D-2)^2 \delta b}{4}
    \!+\! O(\delta^2) \right] \; .
\end{equation}
Henceforth we will not bother with dimensional regularization, and we will make 
the same notational simplification as (\ref{newC0def}). This means that the
vertex-vertex correction is,
\begin{equation}\label{eq:vert_corr_final}
    i V_1(x;x') = C_1(\delta a,\delta b) \!\times\! \frac{e^2 \kappa^2}{4}
    (\partial_{x} \!\downarrow \!- \partial_{x}\! \uparrow) \!\cdot\! 
    (\partial_{x'} \!\downarrow \!- \partial_{x'}\! \uparrow)
    \!\times\! [i\Delta(x;x')]^2 \; ,
\end{equation}
where $C_1(\delta a,\delta b) = 16 + 10 \delta a - 4\delta b$.

\subsection{Vertex-Force Carrier Correlations}\label{sub_sec:vertex_force}

\begin{figure}[H]
\centering
\includegraphics[width=0.95\textwidth]{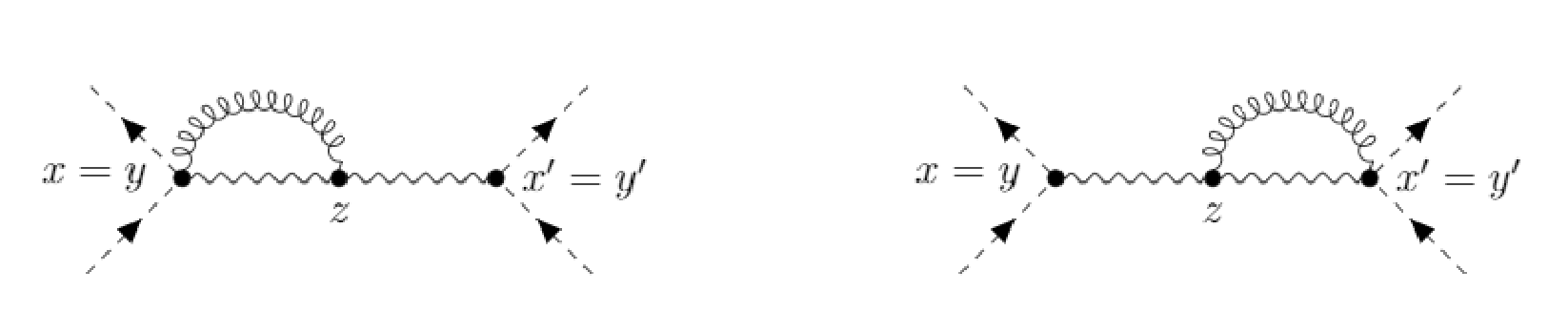}
\caption[Contributions from correlations between force carrier and vertices]{\footnotesize These diagrams show the contributions from correlations between 
the force carrier and one of the vertices. Dashed lines represent massive scalars, wavy 
lines represent photons and curly lines represent gravitons. These graphs have the same 
topology as Bjerrum-Bohr's Diagram 7 \cite{Bjerrum-Bohr:2002aqa}.}
\label{fig:photon_vertices}
\end{figure}

The next contribution comes from the correlations between a single vertex and
the exchange photon, as shown in Figure \ref{fig:photon_vertices}. The 
analytic form is,
\begin{equation}
    \begin{split}
        iV_2(x;x')=&\frac{1}{2}e\kappa\left[\eta^{\epsilon\mu}\eta^{\nu\beta}+\eta^{\epsilon\nu}\eta^{\mu\beta}-\eta^{\mu\nu}\eta^{\epsilon\beta}\right](\partial_x\uparrow-\partial_x\downarrow)_\beta\times e(\partial_{x'}\downarrow-\partial_{x'}\uparrow)^\theta\\
        &\times\int d^Dz\ (-i\kappa V^{\gamma\delta\alpha\tau\rho\sigma})\ \partial_{z\tau}i[_\epsilon\Delta_\delta](x;z)\ \partial_{z\alpha}i[_\gamma\Delta_\theta](z;x')\times i[_{\mu\nu}\Delta_{\rho\sigma}](x;z)\\
        &+(\text{Permutation}) \; .
    \end{split}
\end{equation}
For reducing this diagram it is useful to note how the product of a
massless propagator times one of the gauge variations can be expressed
as a differential operator acting on a single function of the Poincar\'e
interval,
\begin{align}
    i\Delta(x;x')\frac{\partial_\mu\partial_\nu}{\partial^2}i\Delta(x;x')=&\frac{1}{4}\eta_{\mu\nu}[i\Delta(x;x')]^2-\frac{1}{8}\partial_\mu\partial_\nu I\{[i\Delta(x;x')]^2\} \; , \label{eq:id1} \\ 
    \begin{split}\label{eq:id2}
    \partial_\kappa i\Delta(x;x')\frac{\partial_\alpha\partial_\beta}{\partial^2}i\Delta(x;x')&=\frac{D-2}{16(D-1)}\partial_\kappa\partial_\alpha\partial_\beta I\{[i\Delta(x;x')]^2\}
        +\frac{D}{8(D-1)}\eta_{\alpha\beta}\partial_\kappa[i\Delta(x;x')]^2 \\
        &-\frac{D-2}{8(D-1)}(\eta_{\kappa\alpha}\partial_\beta
        + \eta_{\kappa\beta}\partial_\alpha)[i\Delta(x;x')]^2 \; ,
            \end{split}
\end{align}
where the symbol $I\{\}$ represents indefinite integration of the argument with 
respect to $\Delta x^2$. The final result for these diagrams is,
\begin{equation}
    i V_2(x;x') = C_2(\delta a,\delta b) \!\times\! \frac{e^2 \kappa^2}{4}
    (\partial_{x} \!\downarrow \!- \partial_{x}\! \uparrow) \!\cdot\! 
    (\partial_{x'} \!\downarrow \!- \partial_{x'}\! \uparrow)
    \!\times\! [i\Delta(x;x')]^2 \; ,
\end{equation}
where $C_2(\delta a,\delta b) = -12 -16 \delta a + 4 \delta b$.

\subsection{Vertex-Source and Vertex-Observer Correlations}\label{sec:triangular}

\begin{figure}[H]
\centering
\includegraphics[width=0.95\textwidth]{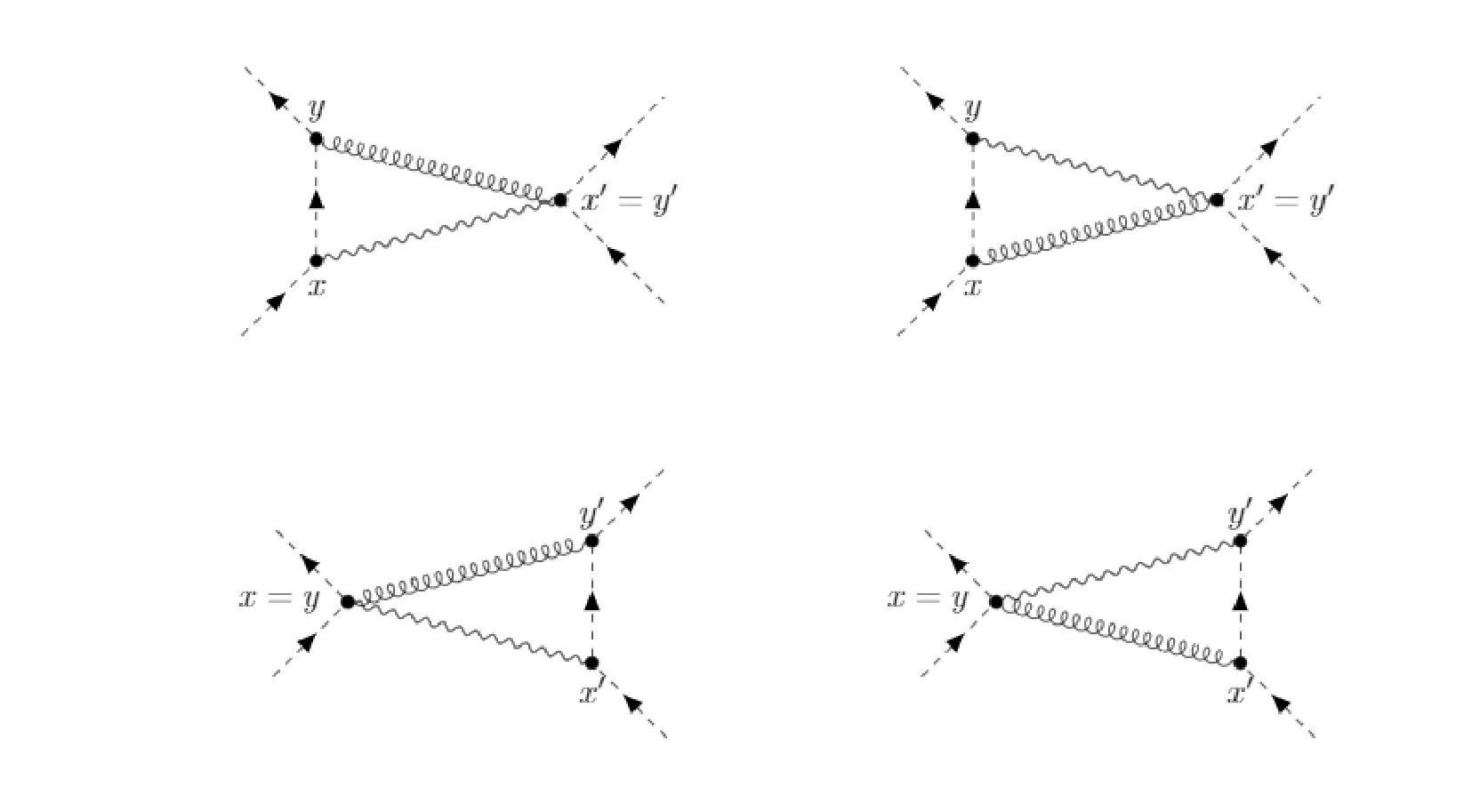}
\caption[Contributions from correlations between source or observer and vertices]{\footnotesize These diagrams show the contributions from correlations 
between the source (primed) or observer (unprimed) and the opposite vertex. 
Dashed lines represent massive scalars, wavy lines represent photons and curly 
lines represent gravitons. These graphs have the same topology as Bjerrum-Bohr's 
Diagram 3 \cite{Bjerrum-Bohr:2002aqa}.}
\label{fig:triangular}
\end{figure}

We next consider contribution from correlations between the source, or observer,
and the opposite vertex, as shown in Figure \ref{fig:triangular}. (Correlations 
with nearer vertices do not contribute because they are cancelled by field 
strength renormalization.) We use $x^\mu$ ($y^\mu$) for incoming (outgoing) 
observer, and ${x'}^\mu$ (${y'}^\mu$) for incoming (outgoing) source. We also
adopt the notation that a bar over a vertex with only a single external leg
denotes differentiation of the on-shell external wave function. With these
conventions we can write the analytic form of the diagrams in 
Figure~\ref{fig:triangular} as,
\begin{equation} \label{triangular}
    \begin{split}
        {\rm Figure\ \ref{fig:triangular}} =& \frac{i\kappa}{2} \left[
        \bar{\partial}^{\mu}_y \partial^{\nu}_y
        \!+\! \bar{\partial}^{\nu}_y \partial^{\mu}_y \!-\! \eta^{\mu\nu} \left( 
        \partial_{y} \!\cdot\! \bar{\partial}_{y} \!+\! m^2\right)\right] \!\times\!
        e(\bar{\partial}_{x} \!-\! \partial_{x})^{\delta} i\Delta_m(x;y) \\
        & \times \frac{e\kappa}{2} \left[\eta^{\gamma\rho} \eta^{\alpha\sigma} \!+\!
        \eta^{\gamma\sigma} \eta^{\alpha\rho} \!-\! \eta^{\rho\sigma} \eta^{\alpha\gamma}
        \right](\partial_{x'} \!\uparrow \!- \partial_{x'}\!\downarrow)_{\alpha} \!\times\! 
        i[_{\delta} \Delta_{\gamma}](x;x') \!\times\! i[_{\mu\nu} \Delta_{\rho\sigma}](y;x') \\
        & + (3\ \text{permutations}) \; .
    \end{split}
\end{equation}

As can be seen from Figure~\ref{fig:triangular}, these contributions involve an 
internal massive scalar propagator in the loop. This poses an obstacle to regarding 
expression (\ref{triangular}) as a correction to the vacuum polarization. This is 
overcome through the ``Donoghue Identities'' of Appendix \ref{appendix:donoghue},
which degenerate the massive scalar propagator to a Dirac delta function, and 
reduce expression (\ref{triangular}) to the same 2-point form (\ref{eq:vp_final})
as the contribution from the original vacuum polarization. The part of expression
(\ref{triangular}) which is independent of the gauge parameters $\delta a$ and
$\delta b$ reaches the desired form through the Donoghue Identities (\ref{eq:3pt})
and (\ref{eq:3pt_derivative}). 

As an example of the part of (\ref{triangular}) proportional to $\delta a$, we 
consider the term,
\begin{equation}\label{eq:trick1}
    \begin{split}
        \delta a & \times \frac{i\kappa}{2} \left[\bar{\partial}^{\mu}_y
        \partial^{\nu}_y \!+\! \bar{\partial}^{\nu}_y \partial^{\mu}_y \!-\!
        \eta^{\mu\nu} \left(\partial_y \!\cdot\! \bar{\partial}_y \!+\! m^2\right)
        \right] \!\times\! e(\bar{\partial}_x \!-\! \partial_x)^{\delta} i\Delta_m(x;y) \\
        & \times \frac{e\kappa}{2} \left[\eta^{\gamma\rho} \eta^{\alpha\sigma}
        \!+\! \eta^{\gamma\sigma} \eta^{\alpha\rho} \!-\! \eta^{\rho\sigma}
        \eta^{\alpha\gamma}\right](\partial_{x'} \!\uparrow \!- \partial_{x'}\! 
        \downarrow)_{\alpha} \!\times\! i[_{\delta} \Delta_{\gamma}](x;x') \!\times\!
        \eta_{\nu\rho} \frac{\partial_{\mu} \partial_{\sigma}} {\partial^2} 
        i \Delta(y;x') \; .
    \end{split}
\end{equation}
The derivative $\partial_\mu$ acting on the massless propagator $i\Delta(y;x')$ can 
be partially integrated to act on the massive propagator $i\Delta_m(x;y)$,
\begin{equation}\label{eq:conservation_trick}
    -(\bar{\partial}_y \!+\! \partial_y)_{\mu} \left[ \bar{\partial}^{\mu}_y
    \partial^{\nu}_y \!+\! \bar{\partial}^{\nu}_y \partial^{\mu}_y \!-\!
    \eta^{\mu\nu}\left(\partial_y \!\cdot\! \bar{\partial}_y \!+\! m^2\right)
    \right] = -\bar{\partial}_y^{\nu} i\delta^D(x \!-\! y) \; .
\end{equation}
The remaining factor of $\partial^{\sigma}/\partial^2$ can be reduced using,
\begin{align}
    & i\Delta(x;x') \!\times\! \frac{\partial_\alpha}{\partial^2} i\Delta(x;x')
    = \frac{1}{4} \partial_{\alpha} I\{[i\Delta(x;x')]^2\} \; , \\
    & \partial^2 I\{[i\Delta(x;x')]^2\} \longrightarrow -2 (D\!-\!4) 
    [i\Delta(x;x')]^2 \; .
\end{align}

From expression (\ref{eq:graviton_prop}) for the graviton propagator we see that
there are two parts proportional to $\delta b$. The second term with $\partial_{\mu}
\partial_{\nu}/\partial^2$ can be reduced just like (\ref{eq:trick1}). The other 
term requires additional effort,
\begin{equation}\label{eq:trick2}
    \begin{split}
        -2 \delta b & \times \frac{i\kappa}{2} \left[\bar{\partial}^{\mu}_y 
        \partial^{\nu}_y \!+\! \bar{\partial}^{\nu}_y \partial^{\mu}_y \!-\!
        \eta^{\mu\nu} \left(\partial_y \!\cdot\! \bar{\partial}_y \!+\! m^2\right)
        \right] \!\times\! e(\bar{\partial}_x \!-\! \partial_x)^{\delta} 
        i\Delta_m(x;y) \\
        & \times \frac{e\kappa}{2} \left[\eta^{\gamma\rho} \eta^{\alpha\sigma} \!+\!
        \eta^{\gamma\sigma} \eta^{\alpha\rho} \!-\! \eta^{\rho\sigma} \eta^{\alpha\gamma}
        \right](\partial_{x'}\! \uparrow \!- \partial_{x'} \!\downarrow)_{\alpha}
        \!\times\! i[_{\delta}\Delta_{\gamma}](x;x') \!\times\! \eta_{\mu\nu}
        \frac{\partial_{\rho} \partial_{\sigma}}{\partial^2} i \Delta(y;x') \; .
    \end{split}
\end{equation}
To reduce this term we distinguish between $y^{\mu}$ derivatives acting on the
external leg ($\bar{\partial}_{y}^{\mu}$), the massive propagator ($\partial_{y}^{\mu}$)
and the massless propagator of the graviton ($\tilde{\partial}_{y}^{\mu}$),
\begin{equation}
\bar{\partial}_{y}^{\mu} + \partial_{y}^{\mu} + \tilde{\partial}_{y}^{\mu} = 0 \; .
\end{equation}
Now note that,
\begin{equation}\label{eq:trick3}
    \eta_{\mu\nu} \left[\bar{\partial}^{\mu}_y \partial^{\nu}_y \!+\! 
    \bar{\partial}^{\nu}_y \partial^{\mu}_y \!-\! \eta^{\mu\nu} \left(\partial_y
    \!\cdot\! \bar{\partial}_y \!+\! m^2\right) \right] = (\bar{\partial}_y^2 \!-\! m^2)
    + (\partial_y^2 \!-\! m^2) - \tilde{\partial}_y^2 - 2 m^2 \; .
\end{equation}
The factor of $(\bar{\partial}_y^2 - m^2)$ vanishes due to the external leg being on shell.
The next term in (\ref{eq:trick3}) degenerates the massive scalar propagator, 
\begin{equation}
    (\partial_y^2 \!-\! m^2) i\Delta_m(x;y) = i \delta^D(x \!-\! y) \; .
\end{equation} 
Of course the factor of $\tilde{\partial}_y^2$ eliminates the troublesome inverse 
D`Alembertian, whereupon the Donoghue Identity (\ref{eq:3pt_2derivative}) completes the
reduction. The final term in (\ref{eq:trick3}) requires the newly Donoghue Identities 
(\ref{eq:new_dono_1}) and (\ref{eq:new_dono_2}). 

Putting everything together gives the final result for Figure~\ref{fig:triangular},
\begin{equation}
    i V_3(x;x') = C_3(\delta a,\delta b) \!\times\! \frac{e^2 \kappa^2}{4}
    (\partial_{x} \!\downarrow \!- \partial_{x}\! \uparrow) \!\cdot\! 
    (\partial_{x'} \!\downarrow \!- \partial_{x'}\! \uparrow)
    \!\times\! [i\Delta(x;x')]^2 \; ,
\end{equation}
where $C_3(\delta a,\delta b) = -32 -8 \delta a - 2 \delta b$.

\subsection{Source-Observer Correlations}
\begin{figure}[H]
\centering
\includegraphics[width=1\textwidth]{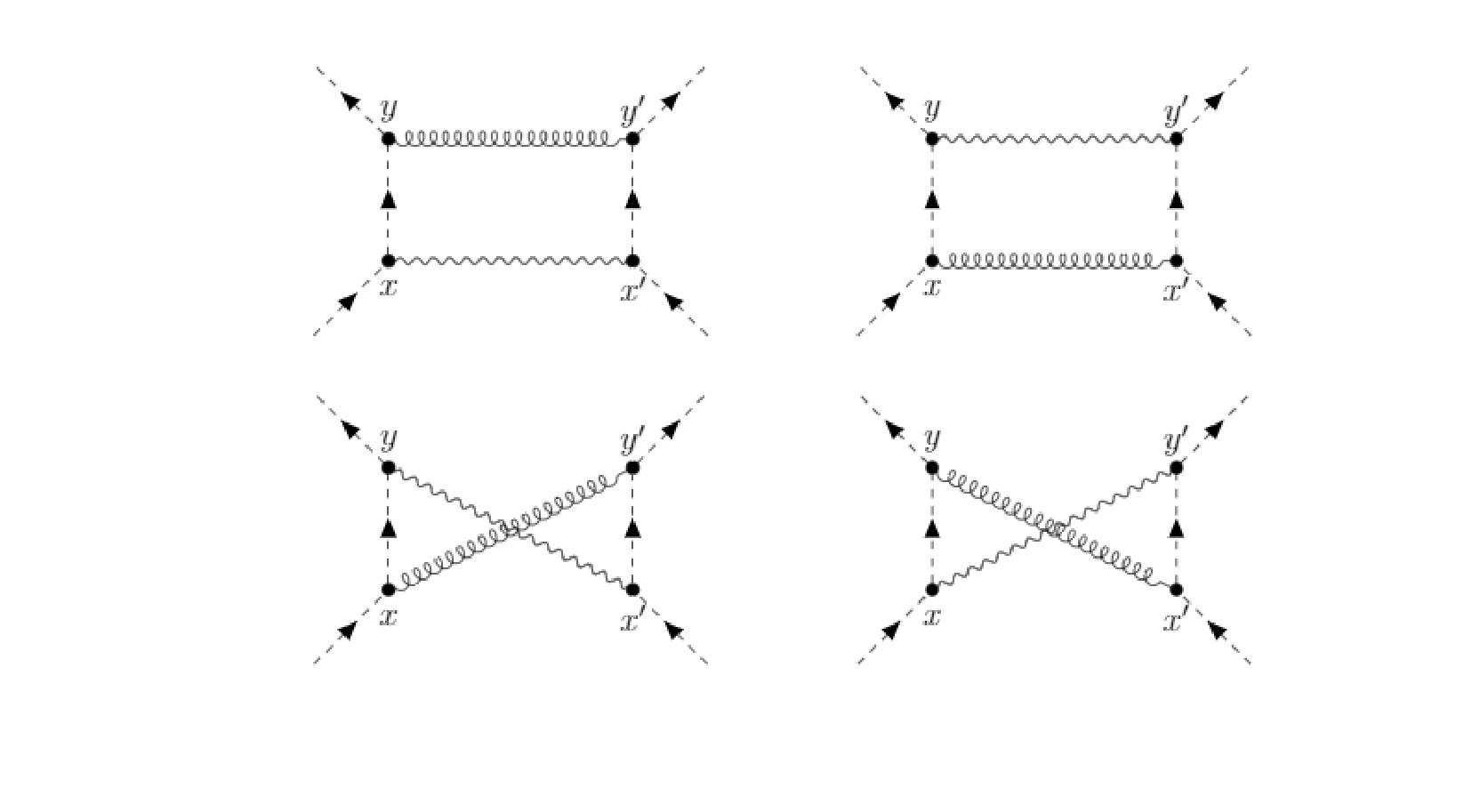}
\caption[Contributions from correlations between the source and observer]{\footnotesize These diagrams show the contributions from correlations 
between the source (primed) and observer (unprimed). Dashed lines represent 
massive scalars, wavy lines represent photons and curly lines represent 
gravitons. These graphs have the same topology as Bjerrum-Bohr's Diagram 2
\cite{Bjerrum-Bohr:2002aqa}.}
\label{fig:box}
\end{figure}

Figure~\ref{fig:box} shows contributions from correlations between source and 
observer. (Correlations between the source and itself, or the observer and itself,
do not correct the exchange photon.) The analytic form for these diagrams is,
\begin{equation} \label{box}
    \begin{split}
        (\rm Figure\ \ref{fig:box}) = & \frac{i\kappa}{2} \left[\bar{\partial}_y^{\mu}
        \partial_y^{\nu} \!+\! \bar{\partial}_y^{\nu} \partial_y^{\mu} \!-\!
        \eta^{\mu\nu} \left(\bar{\partial}_y \!\cdot\! \partial_y \!+\! m^2\right) 
        \right]\left(\bar{\partial}_x \!-\! \partial_x\right)^{\delta} i\Delta_m(x;y) \\
        & \times \frac{i\kappa}{2} \left[\bar{\partial}_{y'}^{\rho} \partial_{y'}^{\sigma}
        \!+\! \bar{\partial}_{y'}^{\sigma} \partial_{y'}^{\rho} \!-\! \eta^{\rho\sigma}
        \left(\bar{\partial}_{y'} \!\cdot\! \partial_{y'} \!+\! m^2\right) \right]
        \left(\bar{\partial}_{x'} \!-\! \partial_{x'}\right)^{\gamma} i\Delta_m(x';y') \\
        & \times i[_{\gamma} \Delta_{\delta}](x;x') \!\times\! i[_{\mu\nu} 
        \Delta_{\rho\sigma}](y;y') + (3 \text{ Permutations}) \; .
    \end{split}
\end{equation}
Note that the two permutations on the bottom line of Figure~\ref{fig:box} contain an
extra minus sign due to 2-scalar-1-photon vertex. 

The part of (\ref{box}) independent of $\delta a$ and $\delta b$ is accomplished by the
Donoghue Identity (\ref{eq:box_dono}). The reduction of the gauge dependent parts is
similar to what we have seen before with one difference: after using relation 
(\ref{eq:conservation_trick}), one must combine parts from the various diagrams to 
eliminate some troublesome terms. The final result for Figure~\ref{fig:box} is,
\begin{equation}
    i V_4(x;x') = C_4(\delta a,\delta b) \!\times\! \frac{e^2 \kappa^2}{4}
    (\partial_{x} \!\downarrow \!- \partial_{x}\! \uparrow) \!\cdot\! 
    (\partial_{x'} \!\downarrow \!- \partial_{x'}\! \uparrow)
    \!\times\! [i\Delta(x;x')]^2 \; ,
\end{equation}
where $C_4(\delta a,\delta b) = \frac{80}{3} + 0 \cdot \delta a +2 \delta b$.

\subsection{Force Carrier Correlations with Source and Observer}
\begin{figure}[H]
\centering
\includegraphics[width=0.95\textwidth]{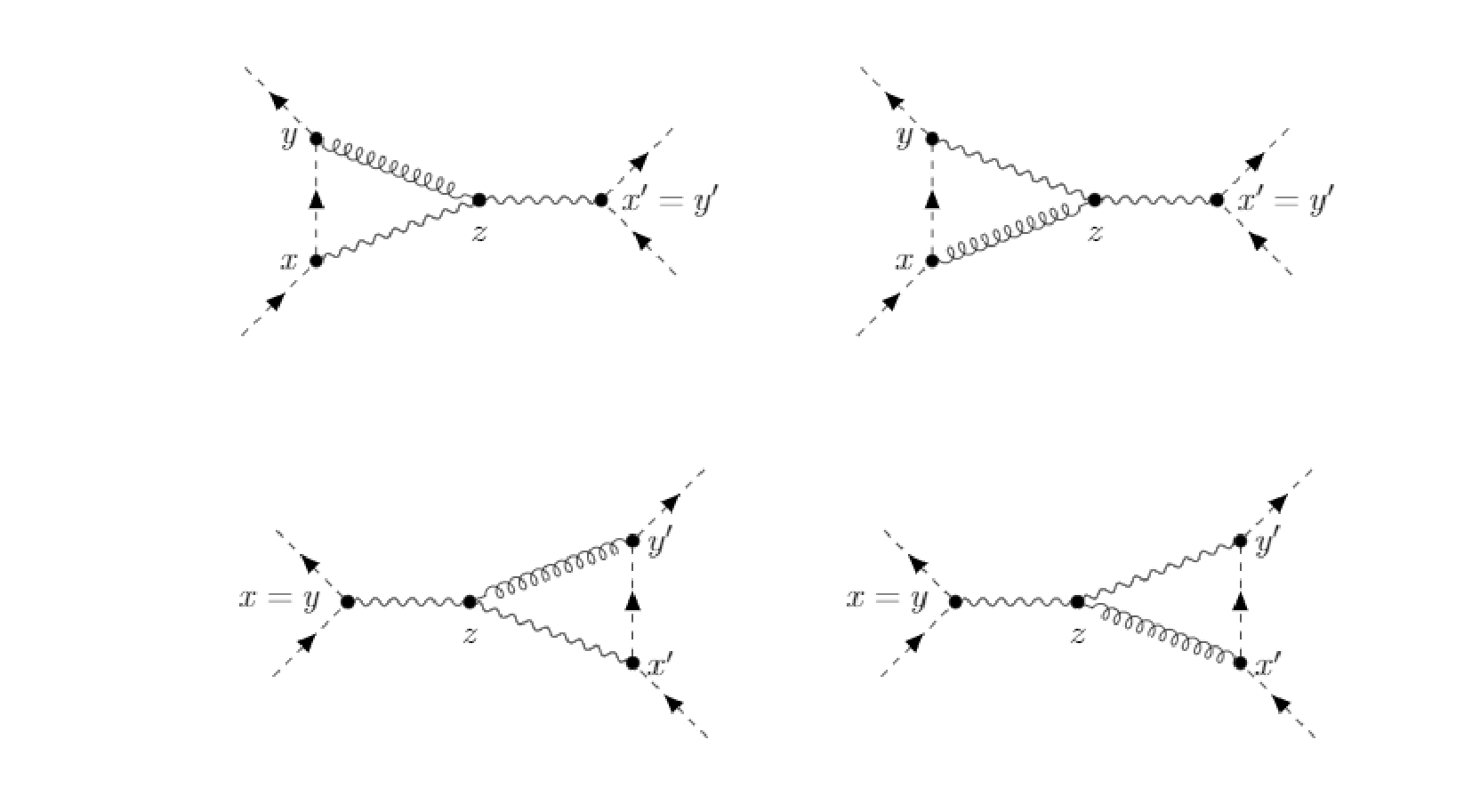}
\caption[Contributions from correlations between the source or observer and the force carrier]{\footnotesize These diagrams show the contributions from correlations 
between the source (primed) or observer (unprimed) and the force carrier. Dashed 
lines represent massive scalars, wavy lines represent photons and curly lines 
represent gravitons. These graphs have the same topology as Bjerrum-Bohr's 
Diagram 6 \cite{Bjerrum-Bohr:2002aqa}.}
\label{fig:force_source_obs}
\end{figure}

The next contribution comes from correlations between the source or observer and
the photon. The Feynman diagrams are given in Figure~\ref{fig:force_source_obs},
and the analytic expression is,
\begin{equation}
    \begin{split}
        {\rm (Figure\ \ref{fig:force_source_obs})} = & \frac{i\kappa}{2} \left[
        \bar{\partial}_y^{\mu} \partial_y^{\nu} \!+\! \bar{\partial}_y^{\nu}
        \partial_y^{\mu} \!-\! \eta^{\mu\nu} \left(\bar{\partial}_y \!\cdot\!
        \partial_y \!+\! m^2\right) \right] \left(\bar{\partial}_x \!-\! 
        \partial_x\right)^{\epsilon} i\Delta_m(x;y) \\
        & \times \int \!\! d^Dz (-i\kappa V^{\gamma\delta\alpha\tau\rho\sigma}) 
        \frac{\partial}{\partial z^{\tau}} i[_{\epsilon} \Delta_{\delta}](x;z) 
        \frac{\partial}{\partial z^{\alpha}} i[_{\gamma} \Delta_{\theta}](z;x') \!\times\!
        e\left(\partial_{x'} \!\downarrow \!- \partial_{x'} \!\uparrow\right)^\theta \\
        & \times i[_{\mu\nu} \Delta_{\rho\sigma}] + (3\text{ Permutations}) \; .
    \end{split}
\end{equation}
The reduction process is almost same as in Section \ref{sec:triangular}, the chief
difference being the extra photon propagator. We extract a D'Alembertian and then use 
(\ref{eq:equation_massless}) to eliminate this and the integration over $z^{\mu}$. 
The final contribution from these diagrams is,
\begin{equation}
    i V_5(x;x') = C_5(\delta a,\delta b) \!\times\! \frac{e^2 \kappa^2}{4}
    (\partial_{x} \!\downarrow \!- \partial_{x}\! \uparrow) \!\cdot\! 
    (\partial_{x'} \!\downarrow \!- \partial_{x'}\! \uparrow)
    \!\times\! [i\Delta(x;x')]^2 \; ,
\end{equation}
where $C_5(\delta a,\delta b) = 12 + 12 \delta a + 4 \delta b$.

\subsection{Gravitational 1-PR Vertex Corrections}

\begin{figure}[H]
\centering
\includegraphics[width=0.95\textwidth]{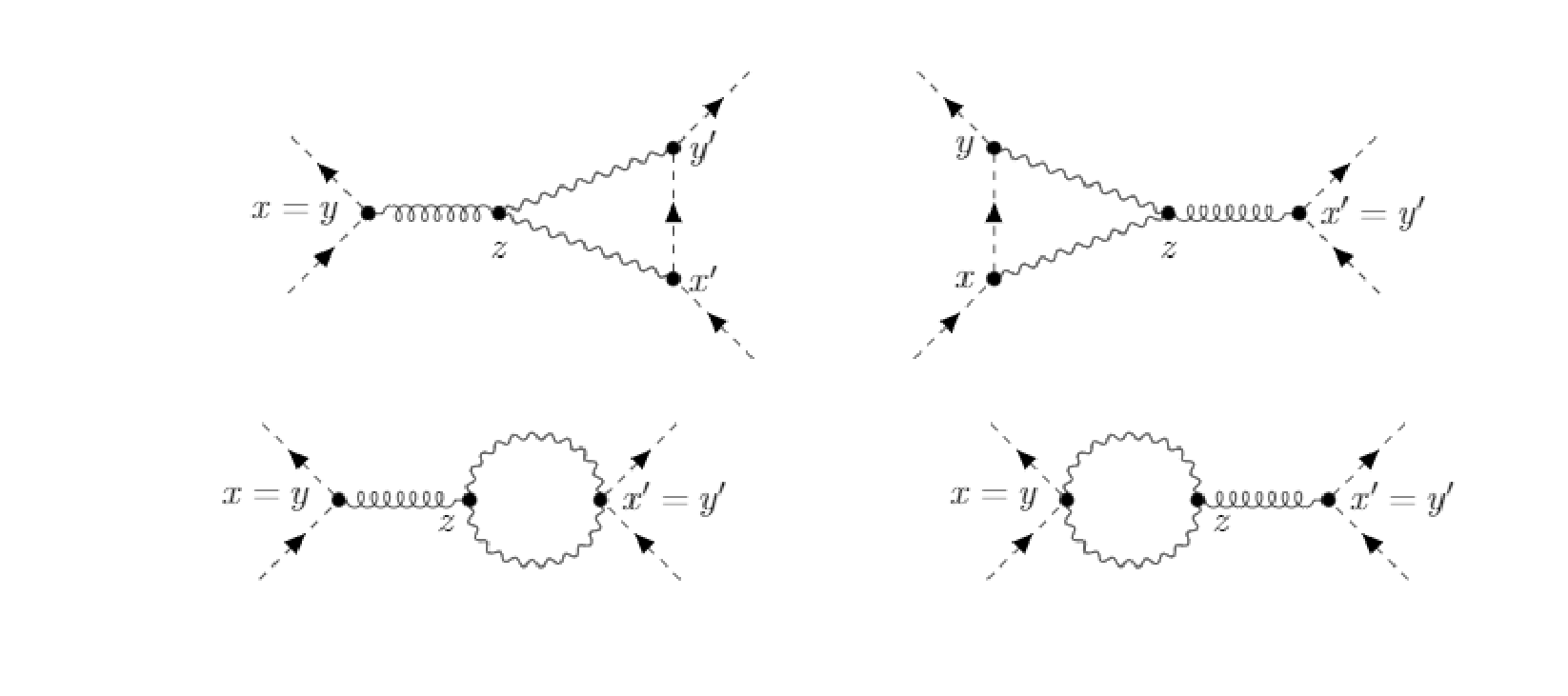}
\caption[Contributions from the 1PR diagrams corresponding to gravitational vertex corrections]{\footnotesize These diagrams show the contributions from the 1PR
(one particle reducible) diagrams corresponding to gravitational vertex 
corrections. Dashed lines represent massive scalars, wavy lines represent 
photons and curly lines represent gravitons. These graphs have the same 
topology as Bjerrum-Bohr's Diagram 5 \cite{Bjerrum-Bohr:2002aqa}.}
\label{fig:graviton_vertex}
\end{figure}

The final contribution to the amputated 4-scalar function comes from 
diagrams in which a loop of photons corrects one of the vertices and
the graviton carries the exchange force. The relevant diagrams are shown 
in Figure~\ref{fig:graviton_vertex}.
The analytic form for first two diagrams is,
\begin{equation}
    \begin{split}
        \text{Figure 6}_\text{up} = & \frac{i\kappa}{2} \left[\partial_x^{\mu} \!\uparrow
        \partial_x^{\nu} \! \downarrow \!+\! \partial_x^{\nu} \!\uparrow
        \partial_x^{\mu} \!\downarrow \!-\! \eta^{\mu\nu} \left(\partial_x
        \!\uparrow \!\cdot\! \partial_x \!\downarrow \!+\! m^2\right)\right]
        \!\times\! e\left(\bar{\partial}_{x'} \!-\! \partial_{x'}\right)^{\theta} \\
        & \times e\left(\partial_{y'} \!-\! \bar{\partial}_{y'}\right)^{\epsilon} 
        i\Delta_m(x';y') \int\!\! d^Dz\ (-i\kappa V^{\gamma\delta\alpha\tau\rho\sigma})
        \frac{\partial}{\partial z^{\tau}} i[_{\epsilon}\Delta_{\delta}](z;x') 
        \frac{\partial}{\partial z^{\alpha}} i[_\gamma\Delta_\theta](z;x') \\
        & \times i[_{\mu\nu} \Delta_{\rho\sigma}](x;z) + (\text{Permutation}) \; .
    \end{split}
\end{equation}
The second two diagrams are,
\begin{equation}
    \begin{split}
        \text{Figure 6}_\text{down} = & \frac{1}{2} \!\times\! \frac{i\kappa}{2} \left[\partial_x^{\mu}
        \!\uparrow \partial_x^{\nu} \!\downarrow \!+\! \partial_x^{\nu} \!\uparrow
        \partial_x^{\mu} \! \downarrow \!-\! \eta^{\mu\nu} \left(\partial_x \!\uparrow
        \!\cdot\! \partial_x \!\downarrow \!+\! m^2\right)\right] \!\times\! (-2 i e^2
        \eta^{\epsilon\theta}) \\
        &\int \!\! d^Dz (-i\kappa V^{\gamma\delta\alpha\tau\rho\sigma}) 
        \frac{\partial}{\partial z^{\tau}} i[_{\epsilon}\Delta_{\delta}](z;y')
        \frac{\partial}{\partial z^{\alpha}} i[_{\gamma} \Delta_{\theta}](z;x') \\
        & \times i[_{\mu\nu}\Delta_{\rho\sigma}](x;z) + (\text{Permutation}) \; .
    \end{split}
\end{equation}
Whenever possible, it is prudent to partially integrate and reflect a derivative 
through the graviton propagator to act on the 2-scalar-1-graviton vertex. One then makes
use of the relation,
\begin{equation}
    (\partial_x \!\downarrow \!+\! \partial_x\! \uparrow)_{\mu} \left[\partial_x^{\mu}
    \! \uparrow \partial_x^{\nu} \!\downarrow \!+\! \partial_x^{\nu} \!\uparrow
    \partial_x^{\mu} \!\downarrow \!-\! \eta^{\mu\nu} \left(\partial_x \!\uparrow \!\cdot\!
    \partial_x \!\downarrow \!+\! m^2\right)\right] = 0 \; .
\end{equation}
The rest of the reduction is same as in earlier sections. The final result is,
\begin{equation}
    i V_6(x;x') = C_6(\delta a,\delta b) \!\times\! \frac{e^2 \kappa^2}{4}
    (\partial_{x} \!\downarrow \!- \partial_{x}\! \uparrow) \!\cdot\! 
    (\partial_{x'} \!\downarrow \!- \partial_{x'}\! \uparrow)
    \!\times\! [i\Delta(x;x')]^2 \; ,
\end{equation}
where $C_6(\delta a,\delta b) = -\frac{16}{3} + 0 \cdot \delta a - 4 \delta b$.

\subsection{Sum Total}

As we have explained, the Donoghue Identities of Appendix \ref{appendix:donoghue} allow us to cast each 
contribution to the amputated 4-scalar function in the form,
\begin{equation}
    i V_i(x;x') = C_i(\delta a,\delta b) \!\times\! \frac{e^2 \kappa^2}{4} 
    (\partial_{x} \!\downarrow \!- \partial_{x}\! \uparrow) \!\cdot\! 
    (\partial_{x'} \!\downarrow \!- \partial_{x'}\! \uparrow)
    \!\times\! [i\Delta(x;x')]^2 \; ,
\end{equation}
where the gauge dependent constant is $C_i(\delta a,\delta b) =O_i + A_i \delta a
+ B_i \delta b$. Table~\ref{tab:sum_total} summarizes our results.
\begin{table}[H]
\caption[Cancellation of the gauge dependent contributions to vacuum polarization]{The entry on the $i^{\text{th}}$ row represent the gauge dependent factors for 
the contribution coming from the diagram in Figure $i+1$.}
\begin{tabular}[l]{rlrrr}
\hline
 $i$ & Description & $\quad O_i\quad $& $\quad A_i\quad$&$\quad B_i\quad$\\
\hline
     0 & Vacuum Polarization & $+\frac{4}{3}$ & $+2$ & $0$\\
  
    1 & Circular Diagram & $+16$ & $+10$ & $-4$\\
     
    2 & Vertex-Force Carrier & $-12$ & $-16$ & $+4$\\
     
     3 & Triangular Diagrams & $-32$ & $-8$ & $-2$\\
     
     4 & Box Diagrams & $+\frac{80}{3}$ & $0$ & $+2$\\
     
     5 & Source, Obs.- Force Carrier & $+12$ & $+12$ & $+4$\\
     
     6 & Graviton 1PR Vertex & $-\frac{16}{3}$ & $0$ & $-4$\\
     
      & Total & $+\frac{20}{3}$ & $0$ & $0$\\
      \hline
\end{tabular}
    \label{tab:sum_total}
\end{table}
Of course the point of the exercise is to total the $O_i$'s, and to show that the
terms proportional to $\delta a$ and $\delta b$ sum to zero,
\begin{equation}
    \sum_{i=0}^6 i V_{i}(x;x') = +\frac{20}{3} \times \frac{e^2 \kappa^2}{4}
    (\partial_{x} \!\downarrow \!- \partial_{x}\! \uparrow) \!\cdot\! 
    (\partial_{x'} \!\downarrow \!- \partial_{x'}\! \uparrow)
    \!\times\! [i\Delta(x;x')]^2 \; .
\end{equation}
We then reverse the steps that led from expression (\ref{eq:so}) to (\ref{eq:vp_final})
in order to conclude that the gauge independent vacuum polarization from a single
graviton loop is,
\begin{equation}
    i[^{\mu} \Pi^{\nu}](x;x') = -\frac{40}{3} \!\times\!
    \frac{\kappa^2}{16\pi^4} \left[\eta^{\mu\nu} \partial^2 \!-\! \partial^{\mu}
    \partial^{\nu}\right] \frac{1}{\Delta x^{2D-2}} \; .
\end{equation}

\section{Conclusions}

The main result of this chapter is that including quantum gravitational
corrections from the source which disturbs the effective field, and from
the observer who measures the disturbance, eliminates the massive gauge 
dependence of the quantum-corrected Maxwell equation (\ref{QMax}) that
was evident in the multiplicative constant $\mathcal{C}_0(D,a,b)$ of 
expression (\ref{C0def}). After renormalization, and application of the 
Schwinger-Keldysh formalism \cite{Ford:2004wc}, our final result for the 
one loop effective field equation is,
\begin{equation}
\partial_{\nu} F^{\nu\mu}(x) + \frac{5 \hbar G \partial^6}{48 \pi^2 c^3} 
\! \int \!\! d^4x' \, \theta(\Delta t \!-\! \Delta r) \Bigl\{ \ln[\mu^2 
(\Delta t^2 \!-\! r^2)] \!-\! 1\Bigr\} \partial_{\nu}' F^{\nu\mu}(x') = 
J^{\mu}(x) \; , \label{QMaxfinal}
\end{equation}
where $\Delta t \equiv t - t'$, $r \equiv \Vert \vec{x} - \vec{x}'\Vert$
and we have restored the factors $\hbar$ and $c$. Although equation 
(\ref{QMaxfinal}) is not local, it is real and causal.

For a static point charge $J^{\mu}(t,\vec{x}) = q \delta^{\mu}_{~0} 
\delta^3(\vec{x} \!-\! \vec{x}')$ the quantum-corrected Coulomb potential is,
\begin{equation}
\Phi(r) = \frac{q}{4\pi r} \left\{ 1 + \frac{10 \hbar G}{3\pi r^2 c^3}
+ \mathcal{O}(G^2) \right\} .
\end{equation}
This result agreees with Bjerrum-Bohr \cite{Bjerrum-Bohr:2002aqa}, but we now
have the ability to solve for quantum gravitational corrections to the full
range of problems one encounters in classical electrodynamics. These 
corrections are bound to be quite small under ordinary conditions, although 
the potential for slightly super-luminal propagation is noteworthy 
\cite{Leonard:2012fs}, and was predicted long ago \cite{Deser:1957zz,
DeWitt:1960fc}.

Although it is nice to finally be able to include quantum gravitational 
corrections to Maxwell's equations on flat space background, we could always 
have inferred physics from scattering amplitudes. The real necessity for our 
method is for studying quantum gravitational corrections to electrodynamics 
in cosmology. These effects can be significant, especially during the epoch 
of primordial inflation. For example, when the simplest de Sitter background
gauge \cite{Tsamis:1992xa,Woodard:2004ut} is employed to compute single
graviton loop corrections to the vacuum polarization \cite{Leonard:2013xsa}
one finds corrections to the Coulomb potential \cite{Glavan:2013jca}, and
to the photon field strength \cite{Wang:2014tza} which become nonperturbatively
strong at large distances and late times. When the vacuum polarization is 
computed in a much more complicated, 1-parameter family of gauges 
\cite{Glavan:2015ura}, one finds the same time dependence for the photon 
field strength, but with a different numerical coefficient \cite{Glavan:2016bvp},
signaling a slight gauge dependence which must be eliminated to infer reliable
results. First order corrections to the graviton propagator in the de Sitter 
generalization of the gauge (\ref{gauge}) have been derived recently 
\cite{Glavan:2019msf}. This should facilitate extending the current work
to de Sitter background.
\chapter{Perturbative Quantum Gravity Induced Scalar Coupling to Electromagnetism} \label{perturbative-scalar-coupling}
\counterwithin{algorithm}{chapter}
\section{Introduction}

There has been much recent interest in searching for exotic processes which
might be induced by quantum gravity \cite{Piscicchia:2022xra,Piscicchia:2022eod}.
In particular, it has been suggested \cite{Calmet:2022bin} that quantum gravitationally
induced scalar couplings to electromagnetism might be detected by planned atom 
interferometers such as MAGIS \cite{Abe:2021magis}, AION \cite{Badurina:2019hst,
Badurina:2021rgt} and AEDGE \cite{AEDGE:2019nxb}. Conventional wisdom has it that
perturbative quantum gravity can at best generate couplings of dimension eight, and
that couplings of dimensions 5 and 6 could only be induced, with unknown coefficients,
by nonperturbative effects \cite{Calmet:2019frv} such as gravitational instantons 
\cite{Perry:1978fd,Chen:2021jcb} and wormholes \cite{Gilbert:1989nq}.

In this chapter we point out that there is a completely perturbative mechanism 
through which quantum gravity induces a dimension six coupling of a massive
scalar with a precisely calculable coefficient. The mechanism is simple: assuming 
that the scalar is constant in space and time, and that the potential energy 
from its mass dominates the stress-energy, the background geometry will be de 
Sitter with a Hubble parameter which depends in a precise way on the scalar. 
Unlike the graviton propagator in flat space, the coincidence limit of a
graviton propagator on de Sitter background goes like the square of the 
Hubble parameter in any gauge \cite{Tsamis:1992xa,Woodard:2004ut,Miao:2011fc,
Mora:2012zi,Glavan:2019msf}. Hence integrating out pairs of graviton fields 
from the Heisenberg operator Maxwell equation (the Hartree approximation) 
induces couplings of the desired form with precisely computable coefficients.

This short chapter contains only three sections, of which this Introduction is 
the first. We present the calculation in Section \ref{sec:calculation}. Our 
conclusions comprise Section \ref{sec:conclusion}.

\section{Calculation}\label{sec:calculation}

Consider a massive, uncharged scalar field which is coupled to electromagnetism
and gravity,
\begin{equation}
\mathcal{L} = -\frac{1}{2}\partial_{\mu} \varphi \partial_{\nu} \varphi 
g^{\mu\nu} \sqrt{-g} - \frac{1}{2} m^2 \varphi^2 \sqrt{-g} -\frac{1}{4}
F_{\mu\nu} F_{\rho\sigma} g^{\mu\rho} g^{\nu\sigma} \sqrt{-g} -
\frac{R \sqrt{-g}}{16\pi G} \; . \label{eq:lagrangian}
\end{equation}
The corresponding Einstein equation is 
\begin{equation}
G_{\mu\nu} = 8\pi G \left[\partial_{\mu} \varphi \partial_{\nu} \varphi - 
\frac{1}{2} g_{\mu\nu} g^{\rho\sigma} \partial_{\rho} \varphi \partial_{\sigma}
\varphi - \frac{1}{2} m^2 \varphi^2 g_{\mu\nu} + F_{\mu\rho} F_{\nu\sigma}
g^{\rho\sigma} - \frac{1}{4} g_{\mu\nu} g^{\alpha\beta} g^{\rho\sigma}
F_{\alpha\rho} F_{\beta\sigma} \right] . \label{eq:einstein}
\end{equation}
As discussed in the introduction, we assume that $\varphi=\varphi_0$ is a 
constant and also set $A_\mu=0$ to get,
\begin{equation}
G_{\mu\nu} = 8 \pi G \times -\frac{1}{2} m^2 \varphi_0^2 g_{\mu\nu} \; .
\label{eq:einsteinReduced}
\end{equation}
The unique, maximally symmetric solution is de Sitter with Hubble constant,
\begin{equation}
H = \sqrt{\frac{4\pi G}{3} \times m^2 \varphi_0^2} \; . \label{eq:deSitterH}
\end{equation}
where, $H$ is the Hubble constant. This shows that a constant scalar triggers
a phase of de Sitter inflation. Because de Sitter is conformally flat, there is 
no classical effect on electromagnetism in conformal coordinates. However, we
will see that the breaking of conformal invariance by gravity does induce a
quantum effect.

Consider the Maxwell equation in a general metric $g_{\mu\nu}$,
\begin{equation}
\partial_{\nu} \Bigl[\sqrt{-g} \, g^{\mu\rho} g^{\nu\sigma} F_{\sigma\rho}
\Bigr] = J^{\mu} \; , \label{eq:maxwell}
\end{equation}
where $F_{\mu\nu} \equiv \partial_{\mu} A_{\nu} - \partial_{\nu} A_{\mu}$ is the 
field strength tensor and $J^{\mu}$ is the current density. We write the quantum
metric $g_{\mu\nu}$ in terms of the Minkowski metric $\eta_{\mu\nu}$,
\begin{equation}
g_{\mu\nu} \equiv a^2 \tilde{g}_{\mu\nu} = a^2 \Bigl[\eta_{\mu\nu} + \kappa 
h_{\mu\nu} \Bigr] \; , \label{eq:gmunuPerturb}
\end{equation}
where $\kappa^2 \equiv 16 \pi G$ is the loop counting parameter, $a(\eta)
\equiv -1/H\eta$ is the scale factor at conformal time $\eta$ and $h_{\mu\nu}$ 
is the graviton field. Graviton indices are raised and lowered with the Minkowski
metric: $h^{\mu}_{~\nu} \equiv \eta^{\mu\rho} h_{\rho\nu}$, $h^{\mu\nu} \equiv
\eta^{\mu\rho} \eta^{\nu\sigma} h_{\rho\sigma}$. The inverse and determinant of 
the conformally transformed metric are,
\begin{eqnarray}
\tilde{g}^{\alpha\beta} & = & \eta^{\alpha\beta} - \kappa h^{\alpha\beta} + 
\kappa^2 h^{\alpha\rho} h^{\beta}_{~\rho} - \cdots \; , \label{eq:gInverse} \\
\sqrt{-\tilde{g}} & = & 1 + \frac{1}{2} \kappa h + \kappa^2 \left(\frac{1}{8}
h^2 - \frac{1}{4} h^{\rho\sigma} h_{\rho\sigma}\right) + \cdots \label{eq:detg}
\end{eqnarray}
Here $h$ is the trace of the graviton field $h \equiv \eta^{\mu\nu} h_{\mu\nu}$.

To facilitate dimensional regularization we formulate the theory in $D$ spacetime
dimensions. The term inside the square bracket of equation (\ref{eq:maxwell}) can
be expressed in terms of the conformally transformed metric as $\sqrt{-g} \, 
g^{\mu\rho} g^{\nu\sigma} F_{\sigma\rho} \equiv a^{D-4} \sqrt{-\tilde{g}} \, 
\tilde{g}^{\mu\rho} \tilde{g}^{\nu\sigma} F_{\sigma\rho}$. The terms involving 
$\tilde{g}_{\mu\nu}$ can be expanded as,
\begin{eqnarray}
\lefteqn{\tilde{g}^{\mu\rho} \tilde{g}^{\nu\sigma} \sqrt{-\tilde{g}} =
\eta^{\mu\rho} \eta^{\nu\sigma} + \kappa \Bigl[\frac{1}{2} h \eta^{\mu\rho}
\eta^{\nu\sigma} - h^{\mu\rho} \eta^{\nu\sigma} - \eta^{\mu\rho} h^{\nu\sigma}
\Bigr] + \kappa^2 \Biggl[\Bigl(\frac{1}{8} h^2 \!-\! \frac{1}{4} 
h^{\alpha\beta} h_{\alpha\beta} \Bigr) \eta^{\mu\rho} \eta^{\nu\sigma} }
\nonumber \\
& & \hspace{3cm} - \frac{1}{2} h h^{\mu\rho} \eta^{\nu\sigma} \!-\! \frac{1}{2} 
h\eta^{\mu\rho} h^{\nu\sigma} \!+\! h^{\mu\alpha} h^{\rho}_{~\alpha} 
\eta^{\nu\sigma} \!+\! \eta^{\mu\rho} h^{\nu\alpha} h^{\sigma}_{~\alpha} \!+\!
h^{\mu\rho} h^{\nu\sigma} \Biggr] \!+\! \mathcal{O}(\kappa^3) \; . \qquad 
\label{eq:expansion}
\end{eqnarray}
Using the Hartree approximation \cite{Hartree:1928,Mora:2013ypa}, we can replace the 
terms proportional to $\kappa$ by zero and the terms proportional to $\kappa^2$ by 
the coincidence limit of the graviton propagator $i[\mbox{}_{\mu\nu} 
\Delta_{\rho\sigma}](x;x')$,
\begin{eqnarray}
\lefteqn{ \tilde{g}^{\mu\rho} \tilde{g}^{\nu\sigma} \sqrt{-\tilde{g}} \longrightarrow
\eta^{\mu\rho} \eta^{\nu\sigma} + \kappa^2 \Biggl[ \Bigl(\frac{1}{8} 
i\Bigl[\mbox{}^{\alpha}_{~\alpha} \Delta^{\beta}_{~\beta}\Bigr] - \frac{1}{4}
i \Bigl[\mbox{}^{\alpha\beta} \Delta_{\alpha\beta}\Bigr] \Bigr) \eta^{\mu\rho}
\eta^{\nu\sigma} - \frac{1}{2} i \Bigl[\mbox{}^{\alpha}_{~\alpha} \Delta^{\mu\rho}
\Bigr] \eta^{\nu\sigma} } \nonumber \\
& & \hspace{2.5cm} - \frac{1}{2} i \Bigl[\mbox{}^{\alpha}_{~\alpha} \Delta^{\nu\sigma} 
\Bigr] \eta^{\mu\rho} + i \Bigl[\mbox{}^{\mu\alpha} \Delta^{\rho}_{~\alpha}\Bigr] 
\eta^{\nu\sigma} + \eta^{\mu\rho} i \Bigl[\mbox{}^{\nu\alpha} \Delta^{\sigma}_{~\alpha}
\Bigr] + i \Bigl[\mbox{}^{\mu\rho} \Delta^{\nu\sigma}\Bigr] \Biggr] + 
\mathcal{O}(\kappa^4) \; . \qquad \label{eq:kappaSqr}
\end{eqnarray}

The graviton is of course gauge dependent but its coincidence limit on de Sitter is
proportional to $H^{D-2}$ in all gauges \cite{Tsamis:1992xa,Woodard:2004ut,Miao:2011fc,
Mora:2012zi,Glavan:2019msf}. In the simplest gauge \cite{Tsamis:1992xa,Woodard:2004ut}
it consists of a sum of three constant tensor factors, each multiplied by a different
scalar propagator,
\begin{equation}
i\Bigl[\mbox{}_{\mu\nu} \Delta_{\alpha\beta}\Bigr](x;x') = \sum_{I=A,B,C}
\Bigl[\mbox{}_{\mu\nu} T^I_{\alpha\beta}\Bigr] \times i\Delta_I(x;x') \; .
\label{eq:gPropagator}
\end{equation}
The constant tensor factors are,
\begin{eqnarray}
\Bigl[\mbox{}_{\mu\nu} T^A_{\alpha\beta}\Bigr] & = & 2 \, \bar{\eta}_{\mu (\alpha} 
\bar{\eta}_{\beta )\nu} - \frac{2}{D-3} \, \bar{\eta}_{\mu\nu} \bar{\eta}_{\alpha\beta}
\quad , \quad \Bigl[\mbox{}_{\mu\nu} T^B_{\alpha\beta}\Bigr] = - 4 
\delta^{0}_{~(\mu} \bar{\eta}_{\nu)(\alpha} \delta^{0}_{~\beta)} \; , \\
\Bigl[\mbox{}_{\mu\nu} T^C_{\alpha\beta}\Bigr] & = & \frac{2}{(D-2)(D-3)} \Bigl[
\bar{\eta}_{\mu\nu} + (D-3) \delta^{0}_{~\mu} \delta^{0}_{~\nu} \Bigr] \Bigl[
\bar{\eta}_{\alpha\beta} + (D-3) \delta^{0}_{~\alpha} \delta^{0}_{~\beta}\Bigr] \; .
\qquad \label{eq:gPropTensorFactors}
\end{eqnarray}
where parenthesized indices are symmetrized and $\bar{\eta}_{\mu\nu} \equiv 
\eta_{\mu\nu} + \delta^{0}_{~\mu} \delta^{0}_{~\nu}$ is the purely spatial part of 
the Minkowski metric. The three scalar propagators correspond to masses $M_A^2 = 0$,
$M_B^2 = (D-2) H^2$ and $M_C^2 = 2 (D-3) H^2$. They obey the propagator equations,
\begin{equation}
\mathcal{D}_I i\Delta_I(x;x') = i \delta^D(x-x') \; , \label{eq:scalarEquation}
\end{equation}
where the various kinetic operators are,
\begin{equation}
\mathcal{D}_I \equiv \partial_{\mu} \Bigl( a^{D-2} \eta^{\mu\nu} \partial_{\nu} \Bigr)
- M_I^2 a^{D} \; . \label{eq:opeerators}
\end{equation}
The coincidence limits of the three scalar propagators are \cite{Glavan:2019msf},
\begin{eqnarray}
i\Delta_A(x;x) & = & k \Bigl[-\pi \cot\Bigl(\frac{D\pi}{2}\Bigr) + 2\ln(a) \Bigr]
\; , \label{eq:scalarA} \\
i \Delta_B(x;x)& = & -\frac{k}{D-2}\Bigl\vert_{D \rightarrow 4} =
-\frac{H^2}{16\pi^2} \; , \label{eq:scalarB} \\
i\Delta_C(x;x) & = & \frac{k}{(D-3)(D-2)}\Bigl\vert_{D \rightarrow 4} =
+\frac{H^2}{16\pi^2} \; , \label{eq:scalarC}
\end{eqnarray}
where the constant $k$ is, 
\begin{equation}
k \equiv \frac{H^{D-2}}{(4\pi)^{\frac{D}2}} \frac{\Gamma{(D-1)}}{\Gamma(\frac{D}2)}
\Bigl\vert_{D \rightarrow 4} = \frac{H^2}{8\pi^2} \; . \label{eq:kDefine}
\end{equation}

By employing the relations (\ref{eq:gPropagator}-\ref{eq:gPropTensorFactors}) 
and (\ref{eq:scalarA}-\ref{eq:scalarC}) which define the coincident graviton
propagator in expression (\ref{eq:kappaSqr}), and then substituting into the 
left hand side of Maxwell's equation (\ref{eq:maxwell}), we obtain the order
$\kappa^2$ correction,
\begin{eqnarray}
\lefteqn{\partial_{\nu} \Bigl[\sqrt{-g} \, g^{\mu\rho} g^{\nu\sigma} 
F_{\sigma\rho} \Bigr]_{\kappa^2} = \frac{\kappa^2 H^2}{4\pi^2} \Biggl[ \tfrac{1}{4}
\Bigl(\partial_{\nu} F^{\nu}_{~0} \delta^{\mu}_{~0} + \partial_0 F_{0}^{~\mu}
\Bigr) + \partial_{\nu} \Bigl\{ (F_{0}^{~\mu} \delta_{0}^{~\nu} \!+\!
F^{\nu}_{~0} \delta_{0}^{~\mu} \!-\! 2 F^{\mu\nu} ) \ln(a) \Bigr\} \Biggr]}
\nonumber \\
& & \hspace{0.5cm} -\kappa^2 k \pi \cot(\tfrac{D \pi}{2}) \partial_{\nu} 
\Biggl[ a^{D-4} \Bigl\{ (\tfrac{D^2 - 4 D + 1}{D - 3})
\Bigl( F_{0}^{~\mu} \delta_{0}^{~\nu} \!+\! F^{\nu}_{~0} \delta_{0}^{~\mu} 
\Bigr) + (\tfrac{D^3 - 12 D^2 + 31 D - 4}{4 (D - 3)}) F^{\mu\nu} \Bigr\} \Biggr] 
. \qquad \label{eq:maxwellFinal}
\end{eqnarray}
Renormalization is facilitated by expressing the divergent part in terms of 
the purely spatial components $\bar{F}^{\mu\nu} \equiv \bar{\eta}^{\mu\rho} 
\bar{\eta}^{\nu\sigma} F_{\rho\sigma}$ of the field strength tensor,
\begin{eqnarray}
\lefteqn{\partial_{\nu} \Bigl[\sqrt{-g} \, g^{\mu\rho} g^{\nu\sigma} 
F_{\sigma\rho} \Bigr]_{\kappa^2} = \frac{\kappa^2 H^2}{4 \pi^2} \Biggl[
\tfrac{1}{4} \partial_{\nu} \Bigl\{ F^{\mu\nu} - \bar{F}^{\mu\nu} \Bigr\} - 
\partial_{\nu} \Bigl\{ (F^{\mu\nu} \!+\! \bar{F}^{\mu\nu} ) \ln(a) \Bigr\} 
\Biggr]} \nonumber \\
& & \hspace{4.3cm} + \kappa^2 k \pi \cot(\tfrac{D \pi}{2}) \partial_{\nu} 
\Biggl[ a^{D-4} \Bigl\{ -\tfrac{D (D - 5)}{4} F^{\mu\nu} + 
(\tfrac{D^2 - 4 D + 1}{D - 3}) \bar{F}^{\mu\nu} \Bigr\} \Biggr] . \qquad
\label{eq:maxwellFinal1}
\end{eqnarray}
The cotangent is divergent as $D \rightarrow 4$,
\begin{equation}
\pi \cot(\tfrac{D \pi}{2}) = \frac{2}{D - 4} + O(D \!-\! 4) \; .
\end{equation}
The divergences on the final line of (\ref{eq:maxwellFinal1}) can be eliminated
by the counterterm,
\begin{equation}
\Delta\mathcal{L} = \tfrac{\kappa^2}{4} (\tfrac{\mu}{H})^{D-4} k \pi 
\cot(\tfrac{D \pi}{2}) \Bigl\{ -\tfrac{D (D - 5)}{4} F_{\alpha\beta}
F_{\rho\sigma} + (\tfrac{D^2 - 4 D + 1}{D - 3}) \bar{F}_{\alpha\beta} 
\bar{F}_{\rho\sigma} \Bigr\} \frac{R g^{\alpha\rho} g^{\beta\sigma} \sqrt{-g}}{
(D \!-\! 1) D H^2} \; . \label{eq:counterterm}
\end{equation}
Note the factor of $(\mu/H)^{D-4}$ required to cancel the $D$-dependence
in the factor of $H^{D-2}$ evident in expression (\ref{eq:kDefine}) for the 
constant $k$. Note also that the need for a noninvariant counterterm arises
from the avoidable breaking of de Sitter invariance in the simplest gauge
\cite{Tsamis:1992xa,Woodard:2004ut} and from the unavoidable time-ordering
of interactions \cite{Glavan:2015ura}.

Combining the variation of the counterterm (\ref{eq:counterterm}) with the
primitive contribution (\ref{eq:maxwellFinal1}), and then taking the unregulated
limit gives,
\begin{equation}
\partial_{\nu} \Bigl[\sqrt{-g} \, g^{\mu\rho} g^{\nu\sigma} 
F_{\sigma\rho} \Bigr]_{\kappa^2} = \frac{\kappa^2 H^2}{4 \pi^2} \Biggl[
\tfrac{1}{4} \partial_{\nu} \Bigl\{ F^{\mu\nu} \!-\! \bar{F}^{\mu\nu} \Bigr\} - 
\partial_{\nu} \Bigl\{ (F^{\mu\nu} \!+\! \bar{F}^{\mu\nu} ) 
\ln\Bigl(\frac{\mu a}{H}\Bigr) \Bigr\} \Biggr] . \label{eq:maxwellFinal2}
\end{equation}
Note the $\mu$-dependence against the scale factor $a(\eta)$ in the logarithm
at the end. This is a vestige of renormalization. Substituting for the Hubble 
constant from expression (\ref{eq:deSitterH}), and recalling that the 
loop-counting parameter is $\kappa^2 = 16 \pi G$, results in the final 
dimension six coupling to Maxwell's equation,
\begin{equation}
\partial_{\nu} \Bigl[\sqrt{-g} \, g^{\mu\rho} g^{\nu\sigma} 
F_{\sigma\rho} \Bigr]_{\kappa^2} = \tfrac{16}{3} G^2 m^2 \phi^2 \Biggl[
\tfrac{1}{4} \partial_{\nu} \Bigl\{ F^{\mu\nu} \!-\! \bar{F}^{\mu\nu} \Bigr\} - 
\partial_{\nu} \Bigl\{ (F^{\mu\nu} \!+\! \bar{F}^{\mu\nu} ) 
\ln\Bigl(\frac{\sqrt{3} \, \mu a}{\sqrt{4\pi G} \, m \phi}\Bigr) \Bigr\} 
\Biggr] . \label{eq:maxwellFinal3}
\end{equation}
Note that the scalar could have as easily been placed inside the $\partial_{\nu}$
--- as it would have been in varying the counterterm (\ref{eq:counterterm}) ---
because the computation was made assuming $\phi$ and the induced $H$ were 
constant. Although a finite renormalization could have eliminated the term 
proportional to $F^{\mu\nu} - \bar{F}^{\mu\nu}$ in (\ref{eq:maxwellFinal3}), the 
logarithm of $\phi$ multiplying the other term is a genuine prediction of the 
theory, with a specific coefficient which we have just computed.

\section{Conclusion}\label{sec:conclusion}

The usual way a constant scalar background engenders quantum corrections
is by giving some field a mass so that the coincidence limit of that field's
propagator depends on the scalar. That cannot happen in perturbative quantum
gravity because the graviton remains massless to all orders. However, 
constant scalars can also contribute by changing a field strength 
\cite{Miao:2021gic}. In our case, a constant scalar background changes the 
cosmological constant, and the graviton propagator in de Sitter background 
depends upon this cosmological constant \cite{Tsamis:1992xa,Woodard:2004ut,
Miao:2011fc,Mora:2012zi,Glavan:2019msf}. We have exploited this mechanism to 
compute the dimension six coupling (\ref{eq:maxwellFinal3}) to 
electromagnetism. Similar results could be obtained for couplings to any
other low energy field.

Three comments are in order. First, our computation depended on the
scalar being constant throughout spacetime. Although this is not a 
reasonable assumption, setting the scalar to be constant is the correct
way to compute nonderivative couplings, which should remain valid in the
resulting low energy effective field theory, even when the assumption of
constancy is relaxed. Our second comment is that we have also assumed the
scalar potential energy dominates the stress energy, which is also not
reasonable for a weak scalar. We believe the induced coupling must still
be present in a realistic cosmological background because it must be 
there in the large field limit. It should be possible to verify this
expectation by employing the same techniques which have recently been 
used to compute cosmological Coleman-Weinberg potentials 
\cite{Kyriazis:2019xgj,Sivasankaran:2020dzp,Katuwal:2021kry,
Katuwal:2022szw}. Third, although the graviton propagator is gauge 
dependent, dimensional analysis requires its coincidence limit on de 
Sitter background to go like $H^{D-2}$, a fact which is confirmed in all 
known gauges \cite{Tsamis:1992xa,Woodard:2004ut,Miao:2011fc,Mora:2012zi,
Glavan:2019msf}. A recently developed formalism \cite{Miao:2017feh,
Katuwal:2020rkv}, based on the S-matrix, can be used to remove gauge 
dependence from the effective field equations.

\chapter{Inflaton Effective Potential from Photons for General $\epsilon$} 
\label{coleman-weinberg}

\counterwithin{algorithm}{chapter}
\renewcommand*{\thealgorithm}{\thechapter-\arabic{algorithm}} 

\def\gtwid{\mathrel{\raise.3ex\hbox{$>$\kern-.75em\lower1ex\hbox{$\sim$}}}}
\def\ltwid{\mathrel{\raise.3ex\hbox{$<$\kern-.75em\lower1ex\hbox{$\sim$}}}}
\def\square{\kern1pt\vbox{\hrule height 1.2pt\hbox{\vrule width 1.2pt\hskip 3pt
   \vbox{\vskip 6pt}\hskip 3pt\vrule width 0.6pt}\hrule height 0.6pt}\kern1pt}

\section{Introduction}

No one knows what caused primordial inflation but the data \cite{Akrami:2018odb}
are consistent with a minimally coupled, complex scalar inflaton\footnote{This chapter has been adapted from a published article in Phys. Rev. D\cite{Katuwal:2021kry}.} $\varphi$,
\begin{equation}
\mathcal{L} = -\partial_{\mu} \varphi \partial_{\nu} \varphi^* g^{\mu\nu} 
\sqrt{-g} - V(\varphi \varphi^*) \sqrt{-g} \; .
\end{equation}
If the inflaton couples only to gravity the loop corrections to its effective
potential come only from quantum gravity and are suppressed by powers of the
loop-counting parameter $G H^2 \ltwid 10^{-10}$, where $G$ is Newton's constant
and $H$ is the Hubble parameter during inflation. In that case the classical
evolution suffers little disturbance but reheating is very slow. 

Efficient reheating requires coupling the inflaton to normal matter such as 
electromagnetism with a non-infinitesimal charge $q$,
\begin{eqnarray}
\lefteqn{\mathcal{L} = - \Bigl( \partial_{\mu} - i q A_{\mu}\Bigr) \varphi 
\Bigl( \partial_{\nu} + i q A_{\nu}\Bigr) \varphi^* g^{\mu\nu} \sqrt{-g} }
\nonumber \\
& & \hspace{5cm} - V(\varphi \varphi^*) \sqrt{-g} - \frac14 F_{\mu\nu} 
F_{\rho\sigma} g^{\mu\rho} g^{\nu\sigma} \sqrt{-g} \; . \qquad 
\label{Lagrangian1}
\end{eqnarray}
But the price of efficient reheating is significant one loop corrections to 
the inflaton effective potential \cite{Green:2007gs}. For large fields these
corrections approach the Coleman-Weinberg form of flat space $\Delta V
\longrightarrow \frac{3}{16 \pi^2} (q^2 \varphi \varphi^*)^2 \ln(q^2 \varphi 
\varphi^*/s^2)$, where $s$ is the renormalization scale \cite{Coleman:1973jx}.
However, cosmological Coleman-Weinberg potentials generally depend in a 
complicated way on the geometry of inflation \cite{Miao:2015oba},
\begin{equation}
ds^2 = a^2 \Bigl[-d\eta^2 + d\vec{x} \!\cdot\! d\vec{x} \Bigr] \qquad 
\Longrightarrow \qquad H \equiv \frac{\partial_0 a}{a^2} \;\; , \;\; 
\epsilon(t) \equiv -\frac{\partial_0 H}{a H^2} \; . \label{geometry1}
\end{equation}
For the special case of de Sitter (with constant $H$ and $\epsilon = 0$)
the result takes the form \cite{Allen:1983dg,Prokopec:2007ak,Miao:2019bnq},
\begin{equation}
\Delta V\Bigl\vert_{\epsilon=0} = \frac{3 H^4}{16 \pi^2} \Biggl\{ b\Bigl(
\frac{q^2 \varphi \varphi^*}{H^2}\Bigr) + \Bigl(\frac{q^2 \varphi \varphi^*}{H^2}
\Bigr) \ln\Bigl( \frac{H^2}{s^2}\Bigr) + \frac12 \Bigl( 
\frac{q^2 \varphi \varphi^*}{H^2}\Bigr)^2 \ln\Bigl( \frac{H^2}{s^2}\Bigr) 
\Biggr\} , \label{deSitterA}
\end{equation}
where the function $b(z)$ (whose $z$ and $z^2$ terms depend on renormalization 
conventions) is,
\begin{eqnarray}
\lefteqn{ b(z) = \Bigl(-1 + 2 \gamma\Bigr) z + \Bigl(-\frac32 + \gamma\Bigr) z^2}
\nonumber \\
& & \hspace{2cm} + \int_{0}^{z} \!\!\!\! dx \, (1 \!+\! x) \Biggl[ \psi\Bigl(
\frac32 \!+\! \frac12 \sqrt{1 \!-\! 8 x} \, \Bigr) + \psi\Bigl( \frac32 \!-\!
\frac12 \sqrt{1 \!-\! 8 x} \, \Bigr) \Biggr\} . \qquad \label{deSitterB} 
\end{eqnarray}

Cosmological Coleman-Weinberg potentials are problematic because they make 
large corrections which cannot be completely subtracted using allowed local 
counterterms \cite{Miao:2015oba}. The classical evolution of inflation is 
subject to unacceptable modifications when partial subtractions are restricted 
to just functions of the inflaton \cite{Liao:2018sci}, or functions of the 
inflaton and the Ricci scalar \cite{Miao:2019bnq}. No other local subtractions
are permitted \cite{Woodard:2006nt} but it has been suggested that an acceptably
small distortion of classical inflation might result from cancellations between
the effective potentials induced by fermions and by bosons \cite{Miao:2020zeh}. 
The purpose of this chapter is to facilitate study of this scheme by developing an 
accurate approximation for extending the de Sitter results 
(\ref{deSitterA}-\ref{deSitterB}) to a general cosmological geometry 
(\ref{geometry1}).

As before on flat space \cite{Coleman:1973jx}, and on de Sitter background
\cite{Prokopec:2007ak}, we define the derivative of the one loop effective 
potential through the equation,
\begin{equation}
\Delta V'(\varphi \varphi^*) = \delta \xi R + \frac12 \delta \lambda \varphi
\varphi^* + q^2 g^{\mu\nu} i\Bigl[ \mbox{}_{\mu} \Delta_{\nu}\Bigr](x;x) \; .
\label{DeltaVdef}
\end{equation}
Here $i [\mbox{}_{\mu} \Delta_{\nu}](x;x')$ is the massive photon propagator 
in Lorentz gauge \cite{Tsamis:2006gj},
\begin{equation}
\Bigl[ \square_{\mu}^{~\nu} - R_{\mu}^{~\nu} - M^2 \delta_{\mu}^{~\nu}\Bigr] 
i\Bigl[\mbox{}_{\nu} \Delta_{\rho}\Bigr](x;x') = 
\frac{g_{\mu \rho} \, i \delta^D(x \!-\! x')}{\sqrt{-g}} + \partial_{\mu} 
\partial'_{\rho} i\Delta_t(x;x') \; , \label{propeqn}
\end{equation}
where $\square_{\mu}^{~\nu}$ is the covariant vector d'Alembertian, $M^2 
\equiv 2 q^2 \varphi \varphi^*$ is the photon mass-squared, and 
$i\Delta_t(x;x')$ is the propagator of a massless, minimally coupled (MMC) 
scalar. We regulate the ultraviolet by working in $D$ spacetime dimensions.

In section 2 we express the photon propagator as an exact spatial Fourier 
mode sum involving massive temporal and spatially transverse vectors, along
with gradients of the MMC scalar. Section 3 begins by converting the various
mode equations to a dimensionless form, then these are approximated. Each
approximation is checked against explicit numerical evolution, both for the
simple quadratic potential, which is excluded by the lower bound on the 
tensor-to-scalar ratio \cite{Aghanim:2018eyx}, and for a plateau potential 
\cite{Starobinsky:1980te} that is in good agreement with all data. In section 
4 our approximations are applied to relation (\ref{DeltaVdef}) to compute the 
one loop effective potential. This consists of a local part which depends on 
the instantaneous geometry and a numerically smaller nonlocal part which
depends on the past geometry. Exact expressions are obtained, as well as 
expansions in the large field and small field regimes. Our conclusions are
given in section 5.

\section{Photon Mode Sum}

The purpose of this section is to express the Lorentz gauge propagator for 
a massive photon as a spatial Fourier mode sum. We begin by expressing the
right hand side of the propagator equation (\ref{propeqn}) as mode sum. Then
the various transverse vector modes are introduced. Next these modes are
combined so as to enforce the propagator equation. The section closes by
checking the de Sitter and flat space correspondence limits.

\subsection{Lessons from the Propagator Equation}

If we exploit Lorentz gauge, the $\mu = 0$ component of (\ref{propeqn}) reads,
\begin{eqnarray}
\lefteqn{ -\frac1{a^2} \Bigl[ -\partial^2 + (D\!-\!2) \partial_0 a H + a^2 M^2
\Bigr] i \Bigl[\mbox{}_0 \Delta_{\rho}\Bigr](x;x') } \nonumber \\
& & \hspace{5.5cm} = -\frac{\delta^0_{~\rho} i\delta^D(x \!-\! x')}{a^{D-2}} +
\partial_0 \partial'_{\rho} i\Delta_t(x;x') \; , \qquad \label{mutime}
\end{eqnarray}
where $\partial^2 \equiv \eta^{\mu\nu} \partial_{\mu} \partial_{\nu}$ is the 
flat space d'Alembertian. The $\mu = m$ component of equation (\ref{propeqn}) 
reads,
\begin{eqnarray}
\lefteqn{ -\frac1{a^2} \Biggl\{ \Bigl[ -\partial^2 + (D\!-\!4) a H \partial_0 
+ a^2 M^2 \Bigr] i \Bigl[\mbox{}_m \Delta_{\rho}\Bigr](x;x') + 2 a H \partial_m
i \Bigl[\mbox{}_0 \Delta_{\rho}\Bigr](x;x') \Biggr\} } \nonumber \\
& & \hspace{5.5cm} = \frac{\delta_{m \rho} i\delta^D(x \!-\! x')}{a^{D-2}} +
\partial_m \partial'_{\rho} i\Delta_t(x;x') \; . \qquad \label{muspace} 
\end{eqnarray}
We begin by writing the right hand sides of expressions (\ref{mutime}) and
(\ref{muspace}) as Fourier mode sums.

The MMC scalar propagator $i\Delta_t(x;x')$ can be expressed as a Fourier mode 
sum over functions $t(\eta,k)$ whose wave equation and Wronskian are,
\begin{equation}
\Bigl[\partial_0^2 + (D\!-\! 2) a H \partial_0 + k^2\Bigr] t(\eta,k) = 0
\qquad , \qquad t \!\cdot\! \partial_0 t^* - \partial_0 t \!\cdot\! t^* = 
\frac{i}{a^{D-2}} \; . \label{MMCeqn}
\end{equation}
Although no closed form solution exists to the $t(\eta,k)$ wave equation for
a general scale factor, relations (\ref{MMCeqn}) do define a unique solution 
when combined with the early time asymptotic form,
\begin{equation}
k \gg a H \qquad \Longrightarrow \qquad t(\eta,k) \longrightarrow 
\frac{e^{-i k \eta}}{\sqrt{2 k a^{D-2}}} \; . \label{asformt}
\end{equation}
Up to infrared corrections \cite{Janssen:2008px}, which are irrelevant owing to
the derivatives in expressions (\ref{propeqn}) and (\ref{mutime}), the Fourier
mode sum for $i\Delta_t(x;x')$ is,
\begin{eqnarray}
\lefteqn{ i\Delta_t(x;x') = \int \!\! \frac{d^{D-1}k}{(2\pi)^{D-1}} \Biggl\{ 
\theta(\Delta \eta) \, t(\eta,k) t^*(\eta',k) e^{i \vec{k} \cdot \Delta \vec{x}}
} \nonumber \\
& & \hspace{6cm} + \theta(-\Delta \eta) \, t^*(\eta,k) t(\eta',k) 
e^{-i\vec{k} \cdot \Delta \vec{x}} \Biggr\} , \qquad \label{MMCprop}
\end{eqnarray}
where $\Delta \eta \equiv \eta - \eta'$ and $\Delta \vec{x} \equiv \vec{x} - 
\vec{x}'$. Acting $\partial_0 \partial_{\rho}'$ on (\ref{MMCprop}) produces 
a term proportional to $\delta^0_{~\rho} \delta(\Delta \eta)$, which the 
Wronskian (\ref{MMCeqn}) and the change of variable $\vec{k} \rightarrow -\vec{k}$ 
allows us to recognize as a $D$-dimensional delta function,
\begin{eqnarray}
\lefteqn{ \partial_0 \partial_{\rho}' i\Delta_t(x;x') = \!\int \!\! 
\frac{d^{D-1}k}{(2\pi)^{D-1}} \Biggl\{ \delta^0_{~\rho} \delta(\Delta \eta)
\Bigl[t \!\cdot\! \partial_0 t^* \!-\! \partial_0 t \!\cdot\! t^*\Bigr] 
e^{i \vec{k} \cdot \Delta \vec{x}} + \theta(\Delta \eta) \partial_0 
\partial_{\rho}' } \nonumber \\
& & \hspace{0.7cm} \times \Bigl[ t(\eta,k)
 t^*(\eta',k) e^{i \vec{k} \cdot \Delta \vec{x}} \Bigr] + 
\theta(-\Delta \eta) \partial_0 \partial_{\rho}' \Bigl[t^*(\eta,k) t(\eta',k) 
e^{-i \vec{k} \cdot \Delta \vec{x}} \Bigr] \Biggr\} , \qquad \\
& & \hspace{-0.5cm} = \frac{\delta^0_{~\rho} i \delta^D(x \!-\! x')}{a^{D-2}}
+ \int \!\! \frac{d^{D-1}k}{(2\pi)^{D-1}} \Biggl\{ \theta(\Delta \eta) \,
T_{0}(x,\vec{k}) T^*_{\rho}(x',\vec{k}) \nonumber \\
& & \hspace{7cm} + \theta(-\Delta \eta) \, 
T^*_0(x,\vec{k}) T_{\rho}(x',\vec{k}) \Biggr\} . \qquad \label{d0drho}
\end{eqnarray}
Here we define $T_{\mu}(x,\vec{k}) \equiv \partial_{\mu} [ t(\eta,k) 
e^{i \vec{k} \cdot \vec{x}} \,]$.

Substituting (\ref{d0drho}) in the right hand side of (\ref{mutime}) gives,
\begin{eqnarray}
\lefteqn{ -\frac1{a^2} \Bigl[ -\partial^2 + (D\!-\!2) \partial_0 a H + a^2 M^2
\Bigr] i \Bigl[\mbox{}_0 \Delta_{\rho}\Bigr](x;x') } \nonumber \\
& & \hspace{0.2cm} = \! \int \!\! \frac{d^{D-1}k}{(2\pi)^{D-1}} \Biggl\{ 
\theta(\Delta \eta) \, T_{0}(x,\vec{k}) T^*_{\rho}(x',\vec{k}) + 
\theta(-\Delta \eta) \, T^*_{0}(x,\vec{k}) T_{\rho}(x',\vec{k}) \Biggr\} . 
\qquad \label{proptime}
\end{eqnarray}
The corresponding expression for (\ref{muspace}) is,
\begin{eqnarray}
\lefteqn{ -\frac1{a^2} \Biggl\{ \Bigl[ -\partial^2 + (D\!-\!4) a H \partial_0 
+ a^2 M^2 \Bigr] i \Bigl[\mbox{}_m \Delta_{\rho}\Bigr](x;x') + 2 a H \partial_m
i \Bigl[\mbox{}_0 \Delta_{\rho}\Bigr](x;x') \Biggr\} } \nonumber \\
& & \hspace{0.5cm} = \! \int \!\! \frac{d^{D-1}k}{(2\pi)^{D-1}} \Biggl\{
\frac{\delta_{m \rho} i\delta(\Delta \eta) e^{i \vec{k} \cdot \Delta \vec{x}}}{
a^{D-2}} + \theta(\Delta \eta) \, T_{m}(x,\vec{k}) T^*_{\rho}(x',\vec{k}) 
\nonumber \\
& & \hspace{6.9cm} + \theta(-\Delta \eta) \, T^*_{m}(x,\vec{k}) 
T_{\rho}(x',\vec{k}) \Biggr\} . \qquad \label{propspace}
\end{eqnarray}
The right hand sides of (\ref{proptime}) and (\ref{propspace}) are the Fourier 
mode sums that will guide us in constructing the photon propagator.

\subsection{Transverse Vector Mode Functions}

In the cosmological geometry (\ref{geometry1}) a transverse (Lorentz gauge) 
vector field $F_{\mu}(x)$ obeys,
\begin{equation}
0 = D^{\mu} F_{\mu}(x) = \frac1{a^2} \Bigl[ -\Bigl( \partial_0 \!+\! (D\!-\!2)
a H \Bigr) F_0 + \partial_i F_i \Bigr] \equiv \frac1{a^2} \Bigl[ -\mathcal{D}
F_0 + \partial_i F_i \Bigr] \; . \label{transversality}
\end{equation}
We seek to express the photon propagator as a Fourier mode sum over a linear 
combination of transverse vector mode functions. Expressions 
(\ref{proptime}-\ref{propspace}) imply that one of these must be the gradient 
of a MMC scalar plane wave,
\begin{equation}
T_{\mu}(x,\vec{k}) \equiv \partial_{\mu} \Bigl[ t(\eta,k) e^{i \vec{k} \cdot 
\vec{x}} \Bigr] \; . \label{Tdef}
\end{equation}
Its transversality follows from the MMC mode equation (\ref{MMCeqn}),
\begin{equation}
-\mathcal{D} T_0 + \partial_i T_i = -\Bigl[\partial_0^2 \!+\! (D\!-\!2) a H 
\partial_0 \!+\! k^2 \Bigr] t(\eta,k) e^{i \vec{k} \cdot \vec{x}} = 0 \; .
\end{equation}

In $D$ spacetime dimensions there are $D-2$ purely spatial and transverse 
massive vector modes of the form,
\begin{equation}
V_{\mu}(x,\vec{k},\lambda,M) \equiv \epsilon_{\mu}(\vec{k},\lambda) 
\!\times\! v(\eta,k) e^{i \vec{k} \cdot \vec{x}} \qquad , \qquad \epsilon_0 
= 0 = k_i \epsilon_i \; .
\end{equation}
The polarization vectors $\epsilon_{\mu}(\vec{k},\lambda)$ are the same as those
of flat space, and their polarization sum is,
\begin{equation}
\sum_{\lambda} \epsilon_{\mu}(\vec{k},\lambda) \epsilon^*_{\rho}(\vec{k},\lambda)
= \left( \begin{matrix}0 & 0 \cr 0 & \delta_{mr} - \widehat{k}_m \widehat{k}_r\end{matrix}\right)
\equiv \overline{\Pi}_{\mu\rho}(\vec{k}) \; . \label{polsum1}
\end{equation}
The wave equation and Wronskian of $v(\eta,k)$ are,
\begin{equation}
\Bigl[ \partial_0^2 + (D \!-\! 4) a H \partial_0 + k^2 + a^2 M^2\Bigr] 
v(\eta,k) = 0 \quad , \quad v \cdot \partial_0 v^* - \partial_0 v \cdot
v^* = \frac{i}{a^{D-4}} \; . \label{spaceeqn}
\end{equation}
Relations (\ref{spaceeqn}) define a unique solution when coupled with the 
form for asymptotically early times,
\begin{equation}
k \gg \Bigl\{ a H, a M\Bigr\} \qquad \Longrightarrow \qquad v(\eta,k)
\longrightarrow \frac{a e^{-ik \eta}}{\sqrt{2 k a^{D-2}}} \; . \label{asformv}
\end{equation}

The spatially transverse vector modes $V_{\mu}(x,\vec{k},\lambda,M)$ 
represent dynamical photons. There is also a single temporal-longitudinal 
mode which represents the constrained part of the electromagnetic field.
It is a combination of $T_{\mu}(x,\vec{k})$ with a transverse vector 
formed from the $\mu = 0$ component $u(\eta,k,M)$ of a massive vector,
\begin{equation}
\Bigl[ \partial_0^2 + (D\!-\!2) \partial_0 a H + k^2 + a^2 M^2\Bigr] 
u(\eta,k,M) = 0 \;\; , \;\; u \cdot \partial_0 u^* - \partial_0 u
\cdot u^* = \frac{i}{a^{D-2}} \; . \label{tempeqn}
\end{equation} 
Relations (\ref{tempeqn}) define a unique solution when combined with the
early time asymptotic form,
\begin{equation}
k \gg \Bigl\{ a H, a M\Bigr\} \qquad \Longrightarrow \qquad u(\eta,k) 
\longrightarrow \frac{e^{-ik \eta}}{\sqrt{2 k a^{D-2}}} \; . \label{asformu}
\end{equation}
One converts $u(\eta,k,M)$ to a transverse vector $U_{\mu}(x,\vec{k},M)$,
\begin{equation}
U_{\mu}(x,\vec{k},M) \equiv \overline{\partial}_{\mu} \Bigl[ u(\eta,k)
e^{i \vec{k} \cdot \vec{x}} \Bigr] \; ,
\end{equation}
where the differential operator $\overline{\partial}_{\mu}$ has the $3+1$
decomposition,
\begin{equation}
\overline{\partial}_0 \equiv \sqrt{-\nabla^2} \longrightarrow k \qquad , 
\qquad \overline{\partial}_i \equiv -\frac{\partial_i \mathcal{D}}{
\sqrt{-\nabla^2}} \longrightarrow -i\widehat{k}_i \mathcal{D} \; .
\end{equation}

\subsection{Enforcing the Propagator Equation}

We have seen that the photon propagator $i[\mbox{}_{\rho} \Delta_{\rho}](x;x')$ 
is the spatial Fourier integral of contributions from the three transverse 
vector modes, each having the general form of constants times,
\begin{equation}
\mathcal{F}_{\mu\rho}(x;x') = \theta(\Delta \eta) F_{\mu}(x) F^*_{\rho}(x')
+ \theta(-\Delta \eta) F^*_{\mu}(x) F_{\rho}(x') \;\; , \;\;
F_{\mu} \in \Bigl\{ T_{\mu} , U_{\mu} , V_{\mu}\Bigr\} \; . \label{genform}
\end{equation}
We might anticipate that the spatially transverse modes contribute with unit
amplitude but the MMC scalar and temporal photon modes must be multiplied by 
the square of an inverse mass to even have the correct dimensions. The
multiplicative factors are chosen to enforce the propagator equation
(\ref{propeqn}).

To check the temporal components (\ref{proptime}) of the propagator equation we
must compute,
\begin{equation}
-\frac1{a^2} \Bigl[-\partial^2 + (D \!-\! 2) \partial_0 a H + a^2 M^2\Bigr]
\mathcal{F}_{0\rho}(x;x') \; . \label{proptimeF}
\end{equation}
To check the spatial components (\ref{propspace}) we need,
\begin{equation}
-\frac1{a^2} \Bigl[-\partial^2 + (D \!-\! 4) a H \partial_0 + a^2 M^2\Bigr]
\mathcal{F}_{m\rho}(x;x') -\frac1{a^2} \!\times\! 2 a H \partial_m 
\mathcal{F}_{0\rho}(x;x') \; . \label{propspaceF}
\end{equation}
The factors of $\partial_0$ in the differential operators of 
(\ref{proptimeF}-\ref{propspaceF}) can act on the theta functions or on the mode 
functions. When all derivatives act on the MMC contribution, the result is $-M^2$ 
times the original mode function,
\begin{eqnarray}
\lefteqn{-\frac1{a^2} \Bigl[-\partial^2 + (D \!-\! 2) \partial_0 a H + a^2 M^2\Bigr]
T_{0}(x) = -M^2 T_{0}(x) \; , } \label{MMCtime} \\
\lefteqn{-\frac1{a^2} \Bigl[-\partial^2 + (D \!-\! 4) a H \partial_0 + a^2 M^2\Bigr]
T_{m}(x) } \nonumber \\
& & \hspace{5.9cm} -\frac1{a^2} \!\times\! 2 a H \partial_m T_{0}(x) = -M^2 T_{m}(x) 
\; . \qquad \label{MMCspace}
\end{eqnarray}
This suggests that the MMC contribution enters the mode sum with a multiplicative
factor of $-M^{-2}$. No further information comes from acting the full differential 
operators on the other modes,
\begin{eqnarray}
-\frac1{a^2} \Bigl[-\partial^2 + (D \!-\! 2) \partial_0 a H + a^2 M^2\Bigr]
U_{0}(x) = 0 \; , \label{temptime} \\
-\frac1{a^2} \Bigl[-\partial^2 + (D \!-\! 4) a H \partial_0 + a^2 M^2\Bigr]
U_{m}(x) -\frac1{a^2} \!\times\! 2 a H \partial_m U_{0}(x) = 0 \; , 
\label{tempspace} \\
-\frac1{a^2} \Bigl[-\partial^2 + (D \!-\! 2) \partial_0 a H + a^2 M^2\Bigr]
V_{0}(x) = 0 \; , \label{transtime} \\
-\frac1{a^2} \Bigl[-\partial^2 + (D \!-\! 4) a H \partial_0 + a^2 M^2\Bigr]
V_{m}(x) -\frac1{a^2} \!\times\! 2 a H \partial_m V_{0}(x) = 0 \; .
\label{transspace}
\end{eqnarray}

It remains to check what happens when one or two factors of $\partial_0$ from
the differential operators in (\ref{proptimeF}-\ref{propspaceF}) act on the 
factors of $\theta(\pm \Delta \eta)$. A single conformal time derivative gives,
\begin{equation}
\partial_0 \mathcal{F}_{\mu\rho}(x;x') = \theta(\Delta \eta) \partial_0 F_{\mu}
F^*_{\rho} + \theta(-\Delta \eta) \partial_0 F^*_{\mu} F_{\rho} +
\delta(\Delta \eta) \Bigl[ F_{\mu} F^*_{\rho} - F^*_{\mu} F_{\rho}\Bigr] \; .
\label{firstetader}
\end{equation}
If we change the Fourier integration variable $\vec{k}$ to $-\vec{k}$ in the 
second of the delta function terms, the result for the MMC modes is,
\begin{eqnarray}
T_{\mu} T^*_{\rho} - T^*_{\mu} T_{\rho} \Bigl\vert_{\vec{k} \rightarrow -\vec{k}} 
&\!\!\!\!\! = \!\!\!\!\!& \left( \begin{matrix} [\partial_0 t \partial_0 t^* - 
\partial_0 t^* \, \partial_0 t] & -i k_r [\partial_0 t \, t^* - \partial_0 t^* 
\, t] \cr i k_m [t \, \partial t^* - t^* \partial_0 t] & k_m k_r [t \, t^* - t^* t]
\end{matrix} \right) e^{i \vec{k} \cdot \Delta \vec{x}} . \qquad \\
&\!\!\!\!\! = \!\!\!\!\!& \left( \begin{matrix} 0 & -k_r \cr -k_m & 0\end{matrix} \right) 
\frac{e^{i \vec{k} \cdot \Delta \vec{x}}}{a^{D-2}} . \qquad \label{firstMMC}
\end{eqnarray}
The temporal photon modes make exactly the same contribution,
\begin{eqnarray}
\lefteqn{U_{\mu} U^*_{\rho} - U^*_{\mu} U_{\rho} \Bigl\vert_{\vec{k} \rightarrow 
-\vec{k}} } \nonumber \\
& & \hspace{1cm} = \left( \begin{matrix} k^2 [u \, u^* - u^* u] & i k_r [u \, \mathcal{D}
u^* - u^* \mathcal{D} u] \cr -i k_m [\mathcal{D} u \, u^* - \mathcal{D} u^* u] & 
\widehat{k}_m \widehat{k}_r [\mathcal{D} u \, \mathcal{D} u^* - \mathcal{D} u^* \,
\mathcal{D} u] \end{matrix} \right) e^{i \vec{k} \cdot \Delta \vec{x}} . \qquad \\
& & \hspace{1cm} = \left( \begin{matrix} 0 & -k_r \cr -k_m & 0\end{matrix} \right) 
\frac{e^{i \vec{k} \cdot \Delta \vec{x}}}{a^{D-2}} . \qquad \label{firsttemp}
\end{eqnarray}
Canceling (\ref{firsttemp}) against (\ref{firstMMC}) --- whose multiplicative 
coefficient is $-M^{-2}$ --- fixes the multiplicative coefficient for the temporal 
photons as $+ M^{-2}$. The delta function term in (\ref{firstetader}) vanishes for 
the spatially transverse modes.

We turn now to second derivative which come from $-\partial^2 = \partial_0^2 -
\nabla^2$,
\begin{eqnarray}
\lefteqn{ \partial_0^2 \mathcal{F}_{\mu\rho}(x;x') = \theta(\Delta \eta) \, 
\partial_0^2 F_{\mu}(x) \, F^*_{\rho}(x') + \theta(-\Delta \eta) \, \partial_0^2 
F^*_{\mu}(x) \, F_{\rho}(x') } \nonumber \\
& & \hspace{1.5cm} + \delta(\Delta \eta) \Bigl[ \partial_0 F_{\mu} F^*_{\rho} -
\partial_0 F^*_{\mu} F_{\rho} \Bigr] + \partial_0 \Biggl\{ \delta(\Delta \eta) 
\Bigl[ F_{\mu} F^*_{\rho} - F^*_{\mu} F_{\rho}\Bigr] \Biggr\} . \qquad 
\label{secondetader}
\end{eqnarray}
We have already arranged for the cancellation of the final term in 
(\ref{secondetader}). For the new delta function term the MMC modes give,
\begin{eqnarray}
\lefteqn{\partial_0 T_{\mu} T^*_{\rho} - \partial_0 T^*_{\mu} T_{\rho} 
\Bigr\vert_{\vec{k} \rightarrow -\vec{k}} } \nonumber \\
& & \hspace{1.8cm} = \left( \begin{matrix} [\partial_0^2 t \, \partial_0 t^* - 
\partial_0^2 t^* \partial_0 t] & -i k_r [\partial_0^2 t \, t^* - \partial_0^2 
t^* t] \cr i k_m [ \partial_0 t \, \partial_0 t^* - \partial_0 t^* \partial_0 t] 
& k_m k_r [ \partial_0 t \, t^* - \partial_0 t^* t]\end{matrix}\right) 
e^{i \vec{k} \cdot \Delta \vec{x}} \; , \qquad \\
& & \hspace{1.8cm} = -i \left( \begin{matrix} k^2 & i k_r (D\!-\! 2) a H \cr 0 & k_m k_r\end{matrix}
\right) \frac{e^{i \vec{k} \cdot \Delta \vec{x}}}{a^{D-2}} \; , 
\label{secondMMCS} \qquad
\end{eqnarray}
where we have used $\partial_0^2 t = -[(D\!-\!2) a H \partial_0 + k^2] t$. The
corresponding contribution for the temporal modes is,
\begin{eqnarray}
\lefteqn{\partial_0 U_{\mu} U^*_{\rho} - \partial_0 U^*_{\mu} U_{\rho} 
\Bigr\vert_{\vec{k} \rightarrow -\vec{k}} } \nonumber \\
& & \hspace{-0.1cm} = \left( \begin{matrix} k^2 [\partial_0 u \, u^* - \partial_0 u^* u] 
\!&\! i k_r [\partial_0 u \, \mathcal{D} u^* - \partial_0 u^* \mathcal{D} u] \cr 
-i k_m [\partial_0 \mathcal{D} u \, u^* - \partial_0 \mathcal{D} u^* u] \!&\! 
\widehat{k}_r \widehat{k}_m [ \partial_0 \mathcal{D} u \, \mathcal{D} u^* - 
\partial_0 \mathcal{D} u^* \mathcal{D} u]\end{matrix}\right) e^{i\vec{k} \cdot \Delta \vec{x}} 
, \qquad \\
& & \hspace{-0.1cm} = -i \left( \begin{matrix} k^2 & i k_r (D\!-\!2) a H \cr 0 & 
\widehat{k}_m \widehat{k}_r (k^2 + a^2 M^2) \end{matrix}\right) 
\frac{e^{i\vec{k} \cdot \Delta \vec{x}}}{a^{D-2}} \; , \label{secondtemp} \qquad
\end{eqnarray}
where we have used $\partial_0 \mathcal{D} u_0 = -(k^2 + a^2 M^2) u_0$. And each 
of the spatially transverse modes gives,
\begin{eqnarray}
\partial_0 V_{\mu} V^*_{\rho} - \partial_0 V^*_{\mu} V_{\rho} 
\Bigr\vert_{\vec{k} \rightarrow -\vec{k}} & = & \left( \begin{matrix} 0 & 0 \cr 0 & 
\epsilon_m \epsilon_r^* [ \partial_0 v \, v* - \partial_0 v^* v]\end{matrix}\right) 
e^{i\vec{k} \cdot \Delta \vec{x}} \; , \qquad \\
& = & -i \left( \begin{matrix} 0 & 0 \cr 0 & \epsilon_m \epsilon_r^*\end{matrix}\right)
\frac{e^{i \vec{k} \cdot \Delta \vec{x}}}{a^{D-4}} \; . \qquad \label{secondtrans} 
\end{eqnarray}

The second conformal time derivatives in both expression (\ref{proptimeF}) 
and the corresponding spatial relation (\ref{propspaceF}) come in the form 
$-\frac1{a^2} \times \partial_0^2$. Including the multiplicative factors, we see 
that the temporal delta functions which are induced consist of $\frac1{a^2 M^2}$ 
times (\ref{secondMMCS}) minus the same factor times (\ref{secondtemp}), plus the 
polarization sum (\ref{polsum1}) over (\ref{secondtrans}),
\begin{eqnarray}
\lefteqn{ \frac{i}{M^2} \left( \begin{matrix} k^2 & i k_r (D\!-\!2) a H \cr 0 & k_m k_r\end{matrix}
\right) \frac{e^{i \vec{k} \cdot \Delta \vec{x}}}{a^{D}} - \frac{i}{M^2} \left( 
\begin{matrix} k^2 & i k_r (D\!-\!2) a H \cr 0 & \widehat{k}_m \widehat{k}_r (k^2 + a^2 M^2)\end{matrix}
\right) \frac{e^{i \vec{k} \cdot \Delta \vec{x}}}{a^{D}} } \nonumber \\
& & \hspace{2.5cm} -i \left( \begin{matrix} 0 & 0 \cr 0 & \delta_{mr} - \widehat{k}_m
\widehat{k}_r \end{matrix} \right) \frac{e^{i \vec{k} \cdot \Delta \vec{x}}}{a^{D-2}} = -i
\left( \begin{matrix} 0 & 0 \cr 0 & \delta_{mr}\end{matrix} \right) 
\frac{e^{i \vec{k} \cdot \Delta \vec{x}}}{a^{D-2}} \; . \qquad \label{deltaterms} 
\end{eqnarray}
With $-\frac1{M^2}$ times expressions (\ref{MMCtime}-\ref{MMCspace}) we see that 
the propagator equations (\ref{proptime}-\ref{propspace}) are obeyed by the
Fourier mode sum,
\begin{eqnarray}
\lefteqn{ i \Bigl[\mbox{}_{\mu} \Delta_{\rho} \Bigr](x;x') = \!\!\int\!\!\! 
\frac{d^{D-1}k}{(2\pi)^{D-1}} \Biggl\{ \!\theta(\Delta \eta) \!\Biggl[ 
\frac{U_{\mu}(x,\vec{k},M) U^*_{\rho}(x',\vec{k},M) \!-\! T_{\mu}(x,\vec{k}) 
T^*_{\rho}(x',\vec{k})}{M^2} } \nonumber \\
& & \hspace{1cm} + \overline{\Pi}_{\mu\rho}(\vec{k}) v(\eta,k) v^*(\eta',k) 
e^{i \vec{k} \cdot \Delta \vec{x}} \Biggr] + \theta(-\Delta \eta) \Biggl[ 
\frac{U^*_{\mu}(x,\vec{k},M) U_{\rho}(x',\vec{k},M)}{M^2} \nonumber \\
& & \hspace{2.5cm} - \frac{T^*_{\mu}(x,\vec{k}) T_{\rho}(x',\vec{k})}{M^2} + 
\overline{\Pi}_{\mu\rho}(\vec{k}) v^*(\eta,k) v(\eta',k) 
e^{-i \vec{k} \cdot \Delta \vec{x}} \Biggr] \Biggr\} . \qquad \label{propagator}
\end{eqnarray}
Note that the $U_{\mu}(x,\vec{k},M)$ and $T_{\mu}(x,\vec{k})$ modes combine
to form a vector integrated propagator analogous to the scalar ones
introduced in \cite{Miao:2011fc}.

The photon propagator can also be expressed as the sum of three bi-vector 
differential operator acting on a scalar propagator,
\begin{eqnarray}
\lefteqn{i\Bigl[\mbox{}_{\mu} \Delta_{\rho}\Bigr](x;x') = \frac1{M^2} \Bigl[
-\eta_{\mu\rho} + \overline{\Pi}_{\mu\rho}\Bigr] \frac{i \delta^D(x \!-\! x')}{
a^{D-2}} } \nonumber \\
& & \hspace{2cm} + \frac1{M^2} \Bigl[ \overline{\partial}_{\mu} 
\overline{\partial}_{\rho}' i\Delta_u(x;x') - \partial_{\mu} \partial_{\rho}' 
i\Delta_t(x;x')\Bigr] + \overline{\Pi}_{\mu\rho} i\Delta_v(x;x') \; . \qquad
\label{scalarsum}
\end{eqnarray}
The Fourier mode sum for the MMC scalar propagator $i\Delta_t(x;x')$ was given
in expression (\ref{MMCprop}). The mode sum for the temporal propagator
$i\Delta_u(x;x')$ comes from replacing $t(\eta,k)$ with $u(\eta,k)$ in 
(\ref{MMCprop}), and the mode sum for the transverse spatial propagator 
$i\Delta_v(x;x')$ is obtained by replacing $t(\eta,k)$ with $v(\eta,k)$. The
resulting lowest order (free) field strength correlators are,
\begin{eqnarray}
\lefteqn{\Bigl\langle \Omega \Bigl\vert T^*\Bigl[ F_{0j}(x) F_{0\ell}(x')\Bigr]
\Bigr\vert \Omega \Bigr\rangle = \frac{\partial_j \partial_{\ell}}{\nabla^2}
\frac{i \delta^D(x \!-\! x')}{a^{D-4}} } \nonumber \\
& & \hspace{3.9cm} + a^2 {a'}^2 M^2 \frac{\partial_j \partial_{\ell}}{\nabla^2}
i\Delta_u(x;x') + \overline{\Pi}_{j\ell} \partial_0 \partial_0' i \Delta_v(x;x') 
\; , \qquad \\
\lefteqn{\Bigl\langle \Omega \Bigl\vert T^*\Bigl[ F_{0j}(x) F_{k\ell}(x')\Bigr]
\Bigr\vert \Omega \Bigr\rangle = \Bigl[\delta_{jk} \partial_{\ell} \!-\! 
\delta_{j\ell} \partial_k\Bigr] \partial_0 i \Delta_v(x;x') \; , } \\
\lefteqn{\Bigl\langle \Omega \Bigl\vert T^*\Bigl[ F_{ij}(x) F_{k\ell}(x')\Bigr]
\Bigr\vert \Omega \Bigr\rangle } \nonumber \\
& & \hspace{3cm} = - \Bigl[\delta_{ik} \partial_j \partial_{\ell} \!-\! 
\delta_{kj} \partial_{\ell} \partial_i \!+\! \delta_{j\ell} \partial_i 
\partial_k \!-\! \delta_{\ell i} \partial_k \partial_j \Bigr] i\Delta_v(x;x') 
\; . \qquad
\end{eqnarray}
The $T^*$-ordering symbol in these correlators indicates that the derivatives in 
forming the field strength tensor, $F_{\mu\nu}(x) \equiv \partial_{\mu} A_{\nu}(x) 
- \partial_{\nu} A_{\mu}(x)$, are taken outside the time-ordering symbol.

An important simplification is,
\begin{equation}
T_{\mu}(x,\vec{k}) = -i \lim_{M \rightarrow 0} U_{\mu}(x,\vec{k},M) \; . 
\label{MMCtotemp}
\end{equation}
Comparing equations (\ref{MMCtime}) with (\ref{temptime}), and (\ref{MMCspace}) 
with (\ref{tempspace}), shows that both sides of relation (\ref{MMCtotemp}) obey
the same wave equation for $M = 0$. That they are identical follows from 
$t(\eta,k)$ and $u(\eta,k)$ having the same asymptotic forms (\ref{asformt}) and 
(\ref{asformu}). Relation (\ref{MMCtotemp}) is of great importance because it
guarantees that the propagator has no $\frac1{M^2}$ pole.

\subsection{The de Sitter Limit}

In the limit of $\epsilon = 0$ the mode functions have closed form 
solutions,\footnote{In the phase factors for $u(\eta,k,M)$ and $v(\eta,k,M)$ one 
must regard $\nu_b$ as a real number, even if $M^2 > \frac14 (D-3)^2 H^2$.}
\begin{eqnarray}
t(\eta,k) &\!\!\! \longrightarrow \!\!\!& e^{\frac{i \pi}{2} (\nu_A + \frac12)} 
\sqrt{\frac{\pi}{4 H a^{D-1}}} \!\times\! H^{(1)}_{\nu_A}\Bigl( \frac{k}{Ha}\Bigr) 
\; , \qquad \\
u(\eta,k,M) &\!\!\! \longrightarrow \!\!\!& e^{\frac{i \pi}{2} (\nu_b + \frac12)} 
\sqrt{\frac{\pi}{4 H a^{D-1}}} \!\times\! H^{(1)}_{\nu_b}\Bigl( \frac{k}{Ha}\Bigr) 
\; , \qquad \\
v(\eta,k,M) & \longrightarrow & e^{\frac{i \pi}{2} (\nu_b + \frac12)} 
\sqrt{\frac{\pi}{4 H a^{D-3}}} \!\times\! H^{(1)}_{\nu_b}\Bigl( \frac{k}{Ha}\Bigr) 
\; , \qquad 
\end{eqnarray}
where the indices are,
\begin{equation}
\nu_A \equiv \Bigl( \frac{D\!-\!1}{2}\Bigr) \qquad , \qquad \nu_b \equiv 
\sqrt{ \Bigl(\frac{D\!-\!3}{2}\Bigr)^2 \!-\! \frac{M^2}{H^2} } \; .
\label{indices}
\end{equation}
The Fourier mode sums for the three propagators can be mostly expressed in terms
of the de Sitter length function $y(x;x')$,
\begin{equation}
y(x;x') \equiv \Bigl\Vert \vec{x} \!-\! \vec{x}' \Bigr\Vert^2 - \Bigl(
\vert \eta \!-\! \eta'\vert \!-\! i \varepsilon \Bigr)^2 \; . \label{ydef}
\end{equation}
The de Sitter limit of the temporal photon propagator is a Hypergeometric 
function,
\begin{equation}
i\Delta_u(x;x') \longrightarrow \frac{H^{D-2}}{(4\pi)^{\frac{D}2}} 
\frac{\Gamma(\nu_a \!+\! \nu_b) \Gamma(\nu_A \!-\! \nu_b)}{\Gamma(\frac{D}2)}
\mbox{}_{2}F_{1}\Bigl(\nu_A \!+\! \nu_b, \nu_A \!-\! \nu_b , \frac{D}2 ;1 \!-\!
\frac{y}{4}\Bigr) \equiv b(y) \; . \label{dSu}
\end{equation}
The de Sitter limit of the spatially transverse photon propagator is closely
related,
\begin{equation}
i\Delta_v(x;x') \longrightarrow a a' b(y) \; . \label{dSv}
\end{equation}
However, infrared divergences break de Sitter invariance in the MMC scalar 
propagator \cite{Vilenkin:1982wt,Linde:1982uu,Starobinsky:1982ee}. The result
for the noncoincident propagator takes the form \cite{Onemli:2002hr,Onemli:2004mb},
\begin{equation}
i\Delta_t(x;x') \longrightarrow A(y) + \frac{H^{D-2}}{(4\pi)^{\frac{D}2}}
\frac{\Gamma(D \!-\! 1)}{\Gamma(\frac{D}2)} \ln(a a') \; , \label{dSt}
\end{equation}
where we only need derivatives of the function $A(y)$ \cite{Miao:2009hb},
\begin{eqnarray}
A'(y) & = & \frac12 (2 \!-\! y) B'(y) - \frac12 (D \!-\! 2) B(y) \; , 
\label{Adef} \\
B(y) & \equiv & \frac{\Gamma(D \!-\! 2) \Gamma(1)}{\Gamma(\frac{D}2)} 
\, \mbox{}_{2}F_1\Bigl(D \!-\! 2, 1,\frac{D}2 ; 1 \!-\! \frac{y}{4}\Bigr) \; .
\label{Bdef}
\end{eqnarray}
It is useful to note that the functions $B(y)$ and $b(y)$ obey,
\begin{eqnarray}
0 & = & (4 y \!-\! y^2) B''(y) + D (2 \!-\! y) B'(y) - (D\!-\!2) B(y) \; , \\
0 & = & (4 y \!-\! y^2) b''(y) + D (2 \!-\! y) b'(y) - (D\!-\!2) b(y) 
- \frac{M^2}{H} b(y) \; . 
\end{eqnarray}

A direct computation of the photon propagator on de Sitter background gives
\cite{Tsamis:2006gj},
\begin{eqnarray}
\lefteqn{ i\Bigl[\mbox{}_{\mu} \Delta_{\rho}\Bigr](x;x') \longrightarrow -
\frac{\partial^2 y}{\partial x^{\mu} \partial {x'}^{\rho}} \Bigl[ (4 y \!-\! y^2)
\frac{\partial}{\partial y} + (D \!-\! 1) (2 \!-\! y) \Bigr] \Bigl[ 
\frac{b'(y) \!-\! B'(y)}{2 M^2} \Bigr] } \nonumber \\
& & \hspace{3.5cm} + \frac{\partial y}{\partial x^{\mu}} 
\frac{\partial y}{\partial {x'}^{\rho}} \Bigl[ (2 \!-\! y) 
\frac{\partial}{\partial y} - (D \!-\! 1) \Bigr] \Bigl[ 
\frac{b'(y) \!-\! B'(y)}{2 M^2} \Bigr] . \qquad \label{directdS}
\end{eqnarray}
To see that the de Sitter limit of our mode sum (\ref{scalarsum}) agrees 
with (\ref{directdS}) we substitute the de Sitter limits (\ref{dSt}), (\ref{dSu}) 
and (\ref{dSv}) and make some tedious reorganizations. This is simplest for the 
MMC contribution,
\begin{eqnarray}
\lefteqn{ \frac{ \delta^0_{~\mu} \delta^0_{~\rho} i \delta^D(x \!-\! x')}{M^2 
a^{D-2}} - \frac{ \partial_{\mu} \partial_{\rho}' i\Delta_t(x;x')}{M^2} 
\longrightarrow - \frac{\partial^2 y}{\partial x^{\mu} \partial {x'}^{\rho}} 
\frac{A'}{M^2} - \frac{\partial y}{\partial x^{\mu}} \frac{\partial y}{\partial 
{x'}^{\rho}} \frac{A''}{M^2} , } \\
& & \hspace{-0.6cm} = - \frac{\partial^2 y}{\partial x^{\mu} \partial {x'}^{\rho}}
\Bigl[ \frac{(2 \!-\! y) B' \!-\! (D \!-\! 2) B}{2 M^2} \Bigr] - 
\frac{\partial y}{\partial x^{\mu}} \frac{\partial y}{\partial {x'}^{\rho}}
\Bigl[ \frac{(2 \!-\! y) B'' \!-\! (D \!-\! 1) B'}{2 M^2} \Bigr] , \qquad \\
& & \hspace{-0.6cm} = \frac{\partial^2 y}{\partial x^{\mu} \partial {x'}^{\rho}}
\Bigl[ \frac{(4 y \!-\! y^2) B'' \!+\! (D \!-\! 1) (2 \!-\! y) B'}{2 M^2} \Bigr]
\nonumber \\
& & \hspace{5.9cm} - \frac{\partial y}{\partial x^{\mu}} 
\frac{\partial y}{\partial {x'}^{\rho}} \Bigl[ \frac{(2 \!-\! y) B'' \!-\! 
(D \!-\! 1) B'}{2 M^2} \Bigr] . \qquad
\end{eqnarray}

Each tensor component of the temporal photon contribution requires a separate 
treatment. The case of $\mu = 0 = \rho$ gives,
\begin{eqnarray}
\lefteqn{ \overline{\partial}_{0} \overline{\partial}_{0}' \,
\frac{i \Delta_u(x;x')}{M^2} \longrightarrow -\nabla^2 \frac{b(y)}{M^2} =
-\nabla^2 y \, \frac{b'}{M^2} - \partial_i y \, \partial_i y \, \frac{b''}{M^2} } 
\\
& & \hspace{-0.5cm} = \frac{a a' H^2}{M^2} \Biggl\{ -2 (D \!-\! 1)b' + 4 \Bigl[
2 \!-\! y \!-\! \frac{a}{a'} \!-\! \frac{a'}{a}\Bigr] b'' \Biggr\} , \\
& & \hspace{-0.5cm} = \frac{a a' H^2}{2 M^2} \Biggl\{ \Bigl[-(2 \!-\! y) + 2 \Bigl( 
\frac{a}{a'} \!+\! \frac{a'}{a}\Bigr) \Bigr] \Bigl[- (4 y \!-\! y^2) b'' - (D \!-\! 1)
(2 \!-\! y) b' \Bigr] \nonumber \\
& & \hspace{1.5cm} + \Bigl[8 \!-\! 4 y \!+\! y^2 - 2 (2 \!-\! y) \Bigl( \frac{a}{a'} 
\!+\! \frac{a'}{a} \Bigr) \Bigr] \Bigl[ (2 \!-\! y) b'' - (D \!-\! 1) b'\Bigr] 
\Biggr\} , \qquad \\
& & \hspace{-0.5cm} = -\frac{\partial^2 y}{\partial x^0 \partial {x'}^0} 
\Bigl[ (4y \!-\! y^2) \frac{\partial}{\partial y} + (D\!-\!1) (2 \!-\! y)\Bigr] 
\frac{b'}{2 M^2} \nonumber \\
& & \hspace{5cm} + \frac{\partial y}{\partial x^0} \frac{\partial y}{\partial {x'}^0} 
\Bigl[ (2 \!-\! y) \frac{\partial}{\partial y} - (D \!-\!1)\Bigr] \frac{b'}{2 M^2} . 
\qquad 
\end{eqnarray}
For $\mu = 0$ and $\rho = r$ we have,
\begin{eqnarray}
\lefteqn{ \overline{\partial}_{0} \overline{\partial}_{r}' \,
\frac{i \Delta_u(x;x')}{M^2} \longrightarrow \partial_r \mathcal{D}' \frac{b(y)}{M^2} 
= \partial_r \mathcal{D}' y \, \frac{b'}{M^2} + \partial_r y \, \partial_0' y \, 
\frac{b''}{M^2} } \\
& & \hspace{-0.5cm} = \frac{a {a'}^2 H^3 \Delta x^r}{M^2} \Biggl\{ 2 (D \!-\! 1) b'
- 2 (2 \!-\! y) b'' + 4 \frac{a}{a'} b''\Biggr\} , \\
& & \hspace{-0.5cm} = \frac{a^2 a' H^3 \Delta x^r}{M^2} \Biggl\{ \Bigl[ (4 y \!-\! y^2)
b'' + (D \!-\! 1) (2 \!-\! y) b'\Bigr] \nonumber \\
& & \hspace{5cm} + \Bigl[2 \!-\! y - 2 \frac{a'}{a}\Bigr] \Bigl[ (2 \!-\! y) b'' -
(D \!-\! 1) b'\Bigr] \Biggr\} , \qquad \\
& & \hspace{-0.5cm} = -\frac{\partial^2 y}{\partial x^0 \partial {x'}^r} 
\Bigl[ (4y \!-\! y^2) \frac{\partial}{\partial y} + (D\!-\!1) (2 \!-\! y)\Bigr] 
\frac{b'}{2 M^2} \nonumber \\
& & \hspace{5cm} + \frac{\partial y}{\partial x^0} \frac{\partial y}{\partial {x'}^r} 
\Bigl[ (2 \!-\! y) \frac{\partial}{\partial y} - (D \!-\!1)\Bigr] \frac{b'}{2 M^2} . 
\qquad 
\end{eqnarray}
And the result for $\mu = m$ and $\rho = 0$ is,
\begin{eqnarray}
\lefteqn{ \overline{\partial}_{m} \overline{\partial}_{0}' \,
\frac{i \Delta_u(x;x')}{M^2} \longrightarrow -\partial_m \mathcal{D} \frac{b(y)}{M^2} 
= -\partial_m \mathcal{D} y \, \frac{b'}{M^2} - \partial_m y \, \partial_0 y \, 
\frac{b''}{M^2} } \\
& & \hspace{-0.5cm} = -\frac{a^2 a' H^3 \Delta x^m}{M^2} \Biggl\{ 2 (D \!-\! 1) b'
- 2 (2 \!-\! y) b'' + 4 \frac{a'}{a} b''\Biggr\} , \\
& & \hspace{-0.5cm} = -\frac{a {a'}^2 H^3 \Delta x^m}{M^2} \Biggl\{ \Bigl[ (4 y \!-\! y^2)
b'' + (D \!-\! 1) (2 \!-\! y) b'\Bigr] \nonumber \\
& & \hspace{5cm} + \Bigl[2 \!-\! y - 2 \frac{a}{a'}\Bigr] \Bigl[ (2 \!-\! y) b'' -
(D \!-\! 1) b'\Bigr] \Biggr\} , \qquad \\
& & \hspace{-0.5cm} = -\frac{\partial^2 y}{\partial x^m \partial {x'}^0} 
\Bigl[ (4y \!-\! y^2) \frac{\partial}{\partial y} + (D\!-\!1) (2 \!-\! y)\Bigr] 
\frac{b'}{2 M^2} \nonumber \\
& & \hspace{5cm} + \frac{\partial y}{\partial x^m} \frac{\partial y}{\partial {x'}^0} 
\Bigl[ (2 \!-\! y) \frac{\partial}{\partial y} - (D \!-\!1)\Bigr] \frac{b'}{2 M^2} . 
\qquad 
\end{eqnarray}

The case of $\mu = m$ and $\rho = r$ requires the most intricate analysis. It begins
with the observation,
\begin{equation}
\frac{\partial_m \partial_r}{\nabla^2} \frac{i \delta^D(x \!-\! x')}{a^{D-2}} + 
\overline{\partial}_m \overline{\partial}_r' \frac{i \Delta_u(x;x')}{M^2} 
\longrightarrow \frac{\partial_m \partial_r}{\nabla^2} \, \frac{\mathcal{D} 
\mathcal{D}' b(y)}{M^2} \; . \label{spaceu}
\end{equation}
This component combines with the contribution from spatially transverse photons,
\begin{equation}
\overline{\Pi}_{mr} i\Delta_v(x;x') \longrightarrow \Bigl( \delta_{mr} - 
\frac{\partial_m \partial_r}{\nabla^2} \Bigr) a a' b(y) \; . \label{spacev}
\end{equation} 
The $\partial_m \partial_r/\nabla^2$ terms from expressions (\ref{spaceu}) and
(\ref{spacev}) give,
\begin{eqnarray}
\lefteqn{ \mathcal{D} \mathcal{D}' b(y) - a a' M^2 b(y) = a a' H^2 \Biggl\{
\Bigl[8 \!-\! 4 y \!+\! y^2 - 2 (2 \!-\! y) \Bigl( \frac{a}{a'} \!+\! \frac{a'}{a}
\Bigr) \Bigr] b'' } \nonumber \\
& & \hspace{0.3cm} + \Bigl[-(2 D \!-\! 3) (2 \!-\! y) + 2 (D \!-\! 1) \Bigl(
\frac{a}{a'} \!+\! \frac{a'}{a}\Bigr) \Bigr] b' + \Bigl[ (D \!-\! 2)^2 -
\frac{M^2}{H^2}\Bigr] b \Biggr\} , \qquad \\ 
& & \hspace{-0.5cm} = a a' H^2 \Biggl\{ 2 (2 \!-\! y)^2 b'' - 3 (D\!-\!1) 
(2 \!-\! y) b' + (D\!-\!2) (D\!-\! 1) b \nonumber \\
& & \hspace{5cm} + 2 \Bigl( \frac{a}{a'} \!+\! \frac{a'}{a}\Bigr) \Bigl[-
(2 \!-\! y) b'' + (D \!-\!1) b'\Bigr] \Biggr\} , \qquad \\
& & \hspace{-0.5cm} = \frac12 \nabla^2 I\Bigl[-(2 \!-\! y) b' + (D \!-\!2) b\Bigr]
\; , \qquad \label{longitudinal}
\end{eqnarray}
where $I[f(y)]$ represents the indefinite integral of $f(y)$ with respect to $y$.

Substituting relation (\ref{longitudinal}) in (\ref{spaceu}) and (\ref{spacev})
gives,
\begin{eqnarray}
\lefteqn{ \frac{\partial_m \partial_r}{\nabla^2} \frac{i \delta^D(x \!-\! x')}{a^{D-2}}
+ \overline{\partial}_m \overline{\partial}_r' \, \frac{i \Delta_u(x;x')}{M^2}
+ \overline{\Pi}_{mr} i\Delta_v(x;x') } \nonumber \\
& & \hspace{3.5cm} \longrightarrow a a' \delta_{mr} b(y) + \frac{\partial_m 
\partial_r}{2 M^2} I\Bigl[-(2 \!-\! y) b' + (D\!-\!2) b\Bigr] ,  \qquad \\
& & \hspace{-0.5cm} = \frac{a a' H^2}{M^2} \Biggl\{ \delta_{mr} \Bigl[(4 y \!-\!y^2) 
b'' + (D\!-\!1) (2 \!-\! y) b' \Bigr] \nonumber \\
& & \hspace{4cm} + 2 a a' H^2 \Delta x^m \Delta x^r 
\Bigl[-(2 \!-\! y) b'' + (D \!-\! 1) b'\Bigr] \Biggr\} , \qquad \\
& & \hspace{-0.5cm} = -\frac{\partial^2 y}{\partial x^m \partial {x'}^r} 
\Bigl[ (4y \!-\! y^2) \frac{\partial}{\partial y} + (D\!-\!1) (2 \!-\! y)\Bigr] 
\frac{b'}{2 M^2} \nonumber \\
& & \hspace{5cm} + \frac{\partial y}{\partial x^m} \frac{\partial y}{\partial {x'}^r} 
\Bigl[ (2 \!-\! y) \frac{\partial}{\partial y} - (D \!-\!1)\Bigr] \frac{b'}{2 M^2} . 
\qquad 
\end{eqnarray}
This completes our demonstration that the de Sitter limit of our propagator agrees
with the direct calculation (\ref{directdS}). It should also be noted that taking 
$H \rightarrow 0$ in the de Sitter limit gives the well known flat space result 
\cite{Tsamis:2006gj}, so we have really checked two correspondence limits.

\section{Approximating the Amplitudes}

The results of the previous section are exact but they rely upon mode functions
$t(\eta,k)$, $u(\eta,k,M)$ and $v(\eta,k,M)$ for which no explicit solution
is known in a general cosmological geometry (\ref{geometry1}). The purpose of 
this section is to develop approximations for the amplitudes (norm-squares) of 
these mode functions. We begin converting all the dependent and independent 
variables to dimensionless form. Then approximations are developed for each of 
the three amplitudes, checked against numerical evolution for the inflationary 
geometry of a simple quadratic potential which reproduces the scalar amplitude 
and spectral index but gives too large a value for the tensor-to-scalar ratio. 
The section closes by demonstrating that our approximations remain valid for 
the plateau potentials which agree with current data. 

\subsection{Dimensionless Formulation}

Time scales vary so much during cosmology that it is desirable to change the
independent variable from conformal time $\eta$ to the number of e-foldings since
the start of inflation $n$,
\begin{equation}
n \equiv \ln\Bigl[ \frac{a(\eta)}{a_i}\Bigr] \qquad \Longrightarrow \qquad 
\partial_0 = a H \partial_n \quad , \quad \partial_0^2 = a^2 H^2 \Bigl[ 
\partial_n^2 + (1 - \epsilon) \partial_n \Bigr] \; . \label{ndef}
\end{equation}
We convert the wave number $k$ and the mass $M$ to dimensionless parameters
using factors of $8 \pi G$,
\begin{equation}
\kappa \equiv \sqrt{8\pi G} \, k \qquad , \qquad \mu \equiv \sqrt{8\pi G} \, M
\; . \label{kmudef}
\end{equation}
And the dimensionless Hubble parameter, inflaton and classical potential are,
\begin{equation}
\chi(n) \equiv \sqrt{8\pi G} \, H(\eta) \;\; , \;\; \psi(n) \equiv 
\sqrt{8\pi G} \, \varphi(\eta) \;\; , \;\; U(\psi \psi^*) \equiv (8\pi G)^2
V(\varphi \varphi^*) \; . \label{chipsiUdef}
\end{equation}
The first slow roll parameter is already dimensionless and we consider it to be
a function of $n$,
\begin{equation}
\epsilon(n) \equiv -\frac{\chi'}{\chi} \; . \label{epsilondef}
\end{equation}

In terms of these dimensionless variables the nontrivial Einstein equations
are,
\begin{eqnarray}
\frac12 (D\!-\!2) (D\!-\!1) \chi^2 & = & \chi^2 \psi' {\psi'}^* + U(\psi \psi^*) 
\; , \label{Einstein1} \\
-\frac12 (D\!-\!2) \Bigl(D \!-\!1 \!-\! 2\epsilon\Bigr) \chi^2 & = & \chi^2
\psi' {\psi'}^* - U(\psi \psi^*) \; . \label{Einstein2}
\end{eqnarray}
The dimensionless inflaton evolution equation is,
\begin{equation}
\chi^2 \Bigl[ \psi'' + (D \!-\! 1 \!-\! \epsilon) \psi'\Bigr] + \psi \,
U'(\psi\psi^*) = 0 \; . \label{inflaton1}
\end{equation}
This can be expressed entirely in terms of $\psi$ and its derivatives,
\begin{equation}
\psi'' + \Bigl(D \!-\! 1 \!-\! \frac{2 \psi' {\psi'}^*}{D \!-\! 2} \Bigr) 
\Biggl[ \psi' + \frac{(D \!-\! 2) U'(\psi \psi^*) \psi}{2 U(\psi \psi^*)}
\Biggr] = 0 \; . \label{imflaton2}
\end{equation}

Although our analytic approximations apply for any model of inflation,
comparing them with exact numerical results of course requires an explicit
model of inflation. It is simplest to carry out most of the analysis using 
a quadratic model with $U(\psi) = c^2 \psi \psi^*$. Applying the slow 
roll approximation gives analytic expressions for the scalar, the 
dimensionless Hubble parameter and the first slow roll parameter,
\begin{equation}
\psi(n) \simeq \sqrt{\psi_0^2 \!-\! 2 n} \quad , \quad \chi(n) \simeq 
\frac{c}{\sqrt{3}} \sqrt{\psi_0^2 \!-\! 2 n} \quad , \quad \epsilon(n) 
\simeq \frac{1}{\psi_0^2 \!-\! 2 n} \; , \label{slowroll1}
\end{equation}
Note also that $\chi(n) \simeq \chi_0 \sqrt{1 - 2 n/\psi_0^2}$. By starting 
from $\psi_0 = 10.6$ one gets somewhat over 50 e-foldings of inflation. 
Setting $c = 7.126 \times 10^{-6}$ makes this model consistent with the 
observed values of the scalar spectral index and the scalar amplitude 
\cite{Aghanim:2018eyx}, but the model's tensor-to-scalar ratio is about 
three times larger than the 95\% confidence upper limit. Although we exploit
the simple slow roll results (\ref{slowroll1}) of this phenomenologically
excluded model to develop approximations, the section closes with a 
demonstration that our analytic approximations continue to apply for 
viable models.

We define the dimensionless MMC scalar amplitude,
\begin{equation}
\mathcal{T}(n,\kappa) \equiv \ln\Bigl[ \frac{\vert t(\eta,k)\vert^2}{\sqrt{8\pi G}}
\Bigr] \; . \label{scTdef}
\end{equation}
Following the procedure of \cite{Romania:2011ez,Romania:2012tb,Brooker:2015iya} we
convert the mode equation and Wronskian (\ref{MMCeqn}) into the nonlinear relation,
\begin{equation}
\mathcal{T}'' + \frac12 {\mathcal{T}'}^2 + (D \!-\! 1 \!-\! \epsilon) \mathcal{T}'
+ \frac{2 \kappa^2 e^{-2n}}{\chi^2} - \frac{e^{-2(D-1) n - 2 \mathcal{T}}}{2 \chi^2}
= 0 \; . \label{scTeqn}
\end{equation}
The asymptotic relation (\ref{asformt}) implies the initial conditions needed for
equation (\ref{scTeqn}) to produce a unique solution,
\begin{equation}
\mathcal{T}(0,\kappa) = -\ln(2 \kappa) \qquad , \qquad \mathcal{T}'(0,\kappa) =
-(D\!-\!2) \; . \label{scTinitial}
\end{equation}

The temporal photon and spatially transverse photon amplitudes are defined
analogously,
\begin{equation}
\mathcal{U}(n,\kappa,\mu) \equiv \ln\Bigl[ \frac{\vert u(\eta,k,M)\vert^2}{
\sqrt{8\pi G}}\Bigr] \qquad , \qquad 
\mathcal{V}(n,\kappa,\mu) \equiv \ln\Bigl[ \frac{\vert v(\eta,k,M)\vert^2}{
\sqrt{8\pi G}}\Bigr] \; . \label{scUVdef}
\end{equation}
Applying the same procedure \cite{Romania:2011ez,Romania:2012tb,Brooker:2015iya}
to the temporal photon mode equation and Wronskian (\ref{tempeqn}) gives,
\begin{eqnarray}
\lefteqn{\mathcal{U}'' + \frac12 {\mathcal{U}'}^2 + (D \!-\! 1 \!-\! \epsilon) 
\mathcal{U}' } \nonumber \\
& & \hspace{2cm} + \frac{2 \kappa^2 e^{-2n}}{\chi^2} + 2 (D\!-\!2) (1 \!-\!
\epsilon) + \frac{2 \mu^2}{\chi^2} - \frac{e^{-2 (D-1) n - 2 \mathcal{U}}}{
2 \chi^2} = 0 \; . \qquad \label{scUeqn}
\end{eqnarray}
And the initial conditions follow from (\ref{asformu}),
\begin{equation}
\mathcal{U}(0,\kappa,\mu) = -\ln(2 \kappa) \qquad , \qquad 
\mathcal{U}'(0,\kappa,\mu) = -(D \!-\! 2) \; . \label{scUinitial}
\end{equation}
The analogous transformation of the spatially transverse photon mode equation
and Wronskian (\ref{spaceeqn}) produces,
\begin{equation}
\mathcal{V}'' + \frac12 {\mathcal{V}'}^2 + (D \!-\! 3 \!-\! \epsilon) 
\mathcal{V}' + \frac{2 \kappa^2 e^{-2n}}{\chi^2} + \frac{2 \mu^2}{\chi^2} 
- \frac{e^{-2 (D-3) n - 2 \mathcal{V}}}{2 \chi^2} = 0 \; . \label{scVeqn}
\end{equation}
The initial conditions associated with (\ref{asformv}) are,
\begin{equation}
\mathcal{V}(0,\kappa,\mu) = -\ln(2 \kappa) \qquad , \qquad 
\mathcal{V}'(0,\kappa,\mu) = -(D \!-\! 4) \; . \label{scVinitial}
\end{equation}

\subsection{Massless, Minimally Coupled Scalar}

\begin{figure}[H]
\centering
\begin{subfigure}[b]{0.33\textwidth}
\centering
\includegraphics[width=\textwidth]{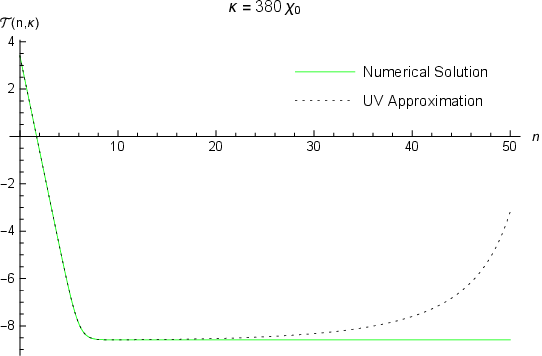}
\caption{$n_{\kappa} \simeq 6.0$}
\end{subfigure}
\begin{subfigure}[b]{0.33\textwidth}
\centering
\includegraphics[width=\textwidth]{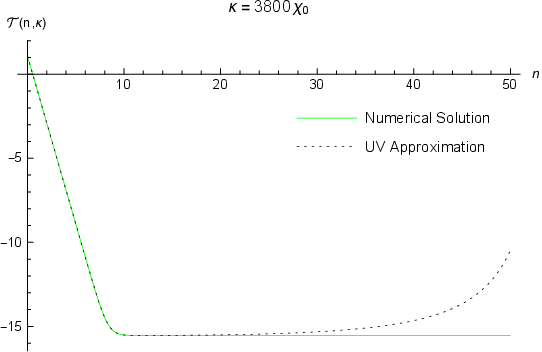}
\caption{$n_{\kappa} \simeq 8.3$}
\end{subfigure}
\begin{subfigure}[b]{0.33\textwidth}
\centering
\includegraphics[width=\textwidth]{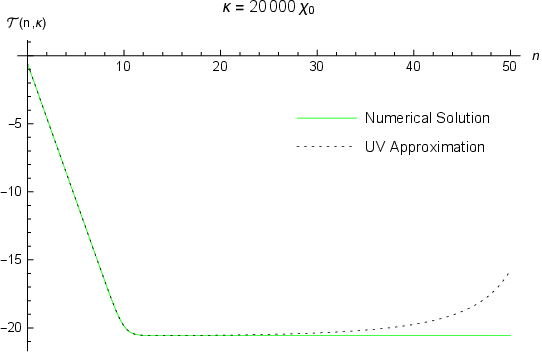}
\caption{$n_{\kappa} \simeq 10.0$}
\end{subfigure}%
\caption[MMC scalar amplitude and its UV approximation]{\footnotesize Plots the massless, minimally coupled scalar amplitude 
$\mathcal{T}(n,\kappa)$ (in solid green) and the (black dashed) ultraviolet 
approximation (\ref{T1def}) versus the e-folding $n$ for three different 
values of $\kappa$.}
\label{MMCSUV}
\end{figure}

The MMCS amplitude is controlled by the relation between the physical wave
number $\kappa e^{-n}$ and the Hubble parameter $\chi(n)$. In the sub-horizon
regime of $\kappa > \chi(n) e^{n}$ the amplitude falls off roughly like
$\mathcal{T}(n,\kappa) \simeq -\ln(2 \kappa) - (D-2) n$, whereas it approaches
a constant in the super-horizon regime of $\kappa < \chi(n) e^{n}$. (The 
e-folding of first horizon crossing is $n_{\kappa}$ such that $\kappa =
\chi(n_{\kappa}) e^{n_{\kappa}}$.) Figure~\ref{MMCSUV} shows that both the 
sub-horizon regime, and also the initial phases of the super-horizon regime, 
are well described by the constant $\epsilon$ solution \cite{Brooker:2015iya},
\begin{equation}
\mathcal{T}_{1}(n,\kappa) \equiv \ln\Biggl[ \frac{\frac{\pi}{2} z(n,\kappa)}{2
\kappa e^{(D-2) n}} \Bigl\vert H^{(1)}_{\nu_t(n)}\Bigl( z(n,\kappa)\Bigr) 
\Bigr\vert^2 \Biggr] . \label{T1def}
\end{equation}
Here the ratio $z(n,\kappa)$ and the MMCS index $\nu_t(n)$ are,
\begin{equation}
z(n,\kappa) \equiv \frac{\kappa e^{-n}}{[1 \!-\! \epsilon(n)] \chi(n)} \qquad , 
\qquad \nu_t(n) \equiv \frac12 \Bigl( \frac{D \!-\! 1 \!-\! \epsilon(n)}{1 \!-\! 
\epsilon(n)} \Bigr) \; . \label{znudef}
\end{equation}

Of course expression (\ref{T1def}) is an approximation to the exact result.
Because we propose to use this to compute the divergent coincidence limit of
the propagator it is important to see how well $\mathcal{T}_1(n,\kappa)$ 
captures the ultraviolet behavior of $\mathcal{T}(n,\kappa)$. Because 
(\ref{T1def}) is exact for constant first slow roll parameter, the deviation 
must involve derivatives of $\epsilon(n)$. It turns out to fall off like 
$\kappa^{-4}$ \cite{Brooker:2015iya},
\begin{equation}
\mathcal{T}(n,\kappa) - \mathcal{T}_1(n,\kappa) = \Bigl( \frac{D\!-\!2}{16}
\Bigr) \Bigl[ (D + 5 - 7 \epsilon) \epsilon' + \epsilon''\Bigr] \Bigl( 
\frac{\chi e^n}{\kappa}\Bigr)^4 + O\Biggl( \Bigl(\frac{\chi e^n}{\kappa}
\Bigr)^6 \Biggr) . \label{Tasymp}
\end{equation}
We will see in section 4 that this suffices for an exact description of the
ultraviolet. 

\begin{figure}[H]
\centering
\begin{subfigure}[b]{0.33\textwidth}
\centering
\includegraphics[width=\textwidth]{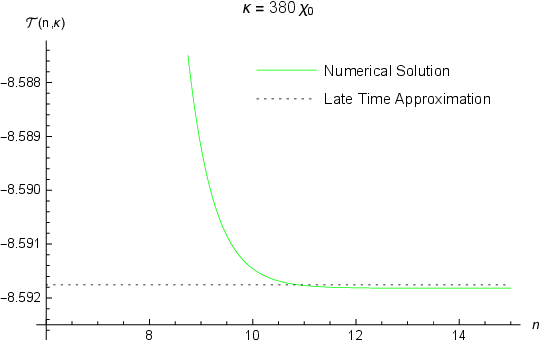}
\caption{$n_{\kappa} \simeq 6.0$}
\end{subfigure}
\begin{subfigure}[b]{0.33\textwidth}
\centering
\includegraphics[width=\textwidth]{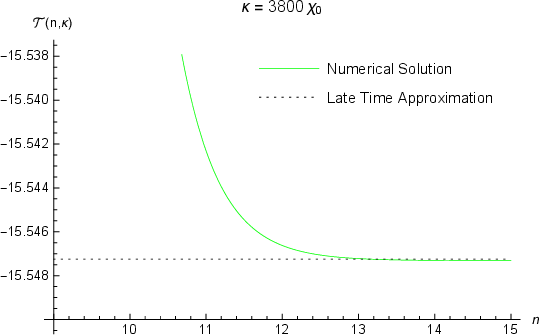}
\caption{$n_{\kappa} \simeq 8.3$}
\end{subfigure}
\begin{subfigure}[b]{0.33\textwidth}
\centering
\includegraphics[width=\textwidth]{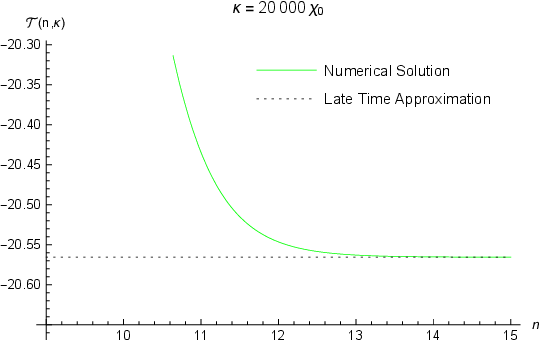}
\caption{$n_{\kappa} \simeq 10.0$}
\end{subfigure}%
\caption[MMC scalar amplitude and its late time approximations]{\footnotesize Plots the massless, minimally coupled scalar amplitude 
$\mathcal{T}(n,\kappa)$ (in solid green) and the (black dashed) late time 
approximation (\ref{T2def}) versus the e-folding $n$ for three different 
values of $\kappa$.}
\label{MMCSlate}
\end{figure}

The discrepancy between $\mathcal{T}(n,\kappa)$ and $\mathcal{T}_1(n,\kappa)$
that is evident at late times in Figure~\ref{MMCSUV} is due to evolution of
the first slow roll parameter $\epsilon(n)$. Figure~\ref{MMCSlate} shows that
the asymptotic late time phase is captured with great accuracy by the form, 
\begin{equation}
\mathcal{T}_2(n,\kappa) = \ln\Biggl[ \frac{\chi^2(n_{\kappa})}{2 \kappa^3} 
\times C\Bigl( \epsilon(n_{\kappa})\Bigr) \Biggr] \; , \label{T2def}
\end{equation}
where the nearly unit correction factor $C(\epsilon)$ is,
\begin{equation}
C(\epsilon) \equiv \frac1{\pi} \Gamma^2\Bigl( \frac12 + \frac1{1 \!-\! \epsilon}
\Bigr) [ 2 (1 \!-\! \epsilon)]^{\frac{2}{1-\epsilon}} \; . \label{Cdef1}
\end{equation}
\noindent Expression (\ref{T2def}) is exact for constant $\epsilon(n)$. When
the first slow roll parameter evolves there are very small nonlocal corrections
whose form is known \cite{Brooker:2017kjd} but whose net contribution is 
negligible for smooth potentials.

\subsection{Temporal Photon}

\begin{figure}[H]
\centering
\begin{subfigure}[b]{0.33\textwidth}
\centering
\includegraphics[width=\textwidth]{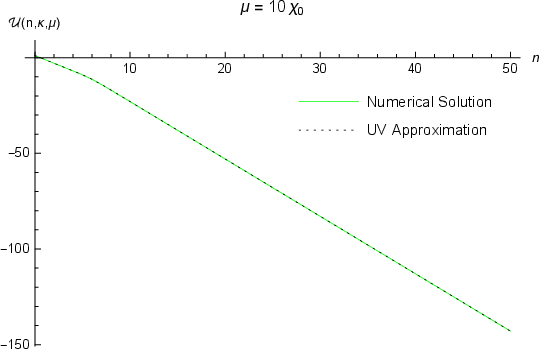}
\caption{$n_{\mu} < 0$}
\end{subfigure}
\begin{subfigure}[b]{0.33\textwidth}
\centering
\includegraphics[width=\textwidth]{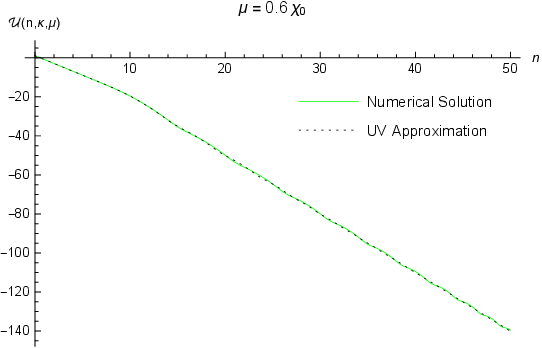}
\caption{$n_{\mu} < 0$}
\end{subfigure}
\begin{subfigure}[b]{0.33\textwidth}
\centering
\includegraphics[width=\textwidth]{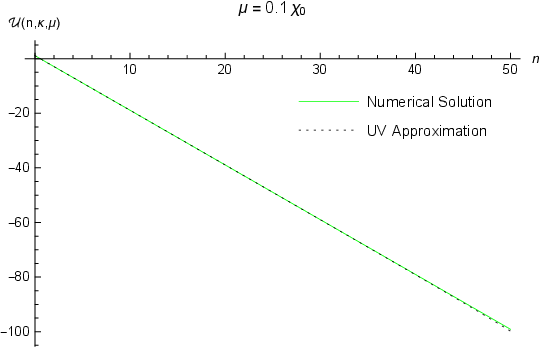}
\caption{$n_{\mu} > 50$}
\end{subfigure}%
\caption[Temporal amplitudes and its UV approximations for different values of mass $\mu$]{\footnotesize Plots the temporal amplitude $\mathcal{U}(n,\kappa,\mu)$ 
and the ultraviolet approximation (\ref{U1def}) versus the e-folding $n$ for 
$\kappa = 3800 \chi_0$ (with $n_{\kappa} \simeq 8.3$) and three different values 
of $\mu$ with outside the range of inflation.}
\label{TempUV}
\end{figure}

The temporal photon amplitude is very similar to the massive scalar which 
was the subject of a previous study \cite{Kyriazis:2019xgj}. Like that system,
the functional form of the amplitude is controlled by two key events:
\begin{enumerate}
\item{First horizon crossing at $n_{\kappa}$ such that $\kappa e^{-n_{\kappa}}
= \chi(n_{\kappa})$; and}
\item{Mass domination at $n_{\mu}$ such that $\mu = \frac12 
\chi(n_{\mu})$.\footnote{The quadratic slow roll approximation 
(\ref{slowroll1}) gives $n_{\mu} \simeq \frac12 \psi_0^2 [1 -
(2 \mu/\chi_0)^2]$.}}
\end{enumerate}
The ultraviolet is well approximated by the form that applies for constant
$\epsilon(n)$ and $\mu \propto \chi(n)$ \cite{Janssen:2009pb},
\begin{equation}
\mathcal{U}_{1}(n,\kappa,\mu) \equiv \ln\Biggl[ \frac{\frac{\pi}{2} 
z(n,\kappa)}{2 \kappa e^{(D-2) n}} \Bigl\vert H^{(1)}_{\nu_u(n,\mu)}
\Bigl( z(n,\kappa)\Bigr) \Bigr\vert^2 \Biggr] , \label{U1def}
\end{equation}
where the temporal index is,
\begin{equation}
\nu^2_u(n,\mu) \equiv \frac14 \Bigl( \frac{D \!-\! 3 \!+\! \epsilon(n)}{1 
\!-\! \epsilon(n)} \Bigr)^2 \!\!- \frac{\mu^2}{[1 \!-\! \epsilon(n)]^2 
\chi^2(n)} . \label{nuudef}
\end{equation}
Figure~\ref{TempUV} shows that the ultraviolet approximation is excellent
when matter domination comes either before or after inflation.

The ultraviolet regime is $\kappa e^{-n} \gg \{\chi(n),\mu\}$. To see how well
the ultraviolet approximation captures this regime we substitute the difference
into the exact evolution equation (\ref{scUeqn}) and expand in powers of $e^n 
\chi(n)/\kappa$ to find \cite{Kyriazis:2019xgj},
\begin{eqnarray}
\lefteqn{\mathcal{U}(n,\kappa,\mu) - \mathcal{U}_1(n,\kappa,\mu) = \Biggl\{ 
\Bigl( 5 \epsilon - 3 \epsilon^2\Bigr) \frac{\mu^2}{4 \chi^2} } \nonumber \\
& & \hspace{1.8cm} + \Bigl( \frac{D\!-\!2}{16}\Bigr) \Bigl[ (D - 9 + 7 \epsilon) 
\epsilon' - \epsilon''\Bigr] \Biggr\} \Bigl( \frac{\chi e^n}{\kappa}\Bigr)^4 \!+ 
O\Biggl( \Bigl(\frac{\chi e^n}{\kappa}\Bigr)^6 \Biggr) . \qquad \label{Uasymp}
\end{eqnarray}
This is suffices to give an exact result for the ultraviolet so we that can take 
the unregulated limit of $D = 4$ for the approximations which pertain for $n > 
n_{\kappa}$.

The various terms in equation (\ref{scUeqn}) behave differently before and after
first horizon crossing. Evolution before first horizon crossing is controlled by
the 4th and 7th terms,
\begin{equation}
\frac{2 \kappa^2 e^{-2n}}{\chi^2} - \frac{e^{-2(D-1)n - 2\mathcal{U}}}{2 \chi^2}
\simeq 0 \qquad \Longrightarrow \qquad \mathcal{U} \simeq -\ln(2\kappa) - 
(D\!-\!2) n \; . \label{scUUV}
\end{equation}
After first horizon crossing these terms rapidly redshift into insignificance.
We can take the unregulated limit ($D=4$), and equation (\ref{scUeqn}) becomes,
\begin{equation}
\mathcal{U}'' + \frac12 {\mathcal{U}'}^2 + (3 \!-\! \epsilon) \mathcal{U}'
+ 4 (1 \!-\! \epsilon) + \frac{2 \mu^2}{\chi^2} \simeq 0 \; . \label{scUlate}
\end{equation}
This is a nonlinear, first order equation for $\mathcal{U}'$. Following
\cite{Kyriazis:2019xgj} we make the ansatz,
\begin{equation}
\mathcal{U}' \simeq \alpha + \beta \tanh(\gamma) \; . \label{ansatz}
\end{equation}
Substituting (\ref{ansatz}) in (\ref{scUlate}) gives,
\begin{eqnarray}
\lefteqn{ \Bigl({\rm Eqn.\ \ref{scUlate}} \Bigr) = \alpha' + \frac12 \alpha^2
+ \frac12 \beta^2 + (3 \!-\! \epsilon) \alpha + 4 (1 \!-\! \epsilon) +
\frac{2 \mu^2}{\chi^2} } \nonumber \\
& & \hspace{2.7cm} + \Bigl[ (3 \!-\! \epsilon \!+\! \alpha) \beta + \beta'\Bigr]
\tanh(\gamma) + \beta \Bigl( \gamma' - \frac12 \beta\Bigr) {\rm sech}^2(\gamma) 
\; . \qquad \label{almost}
\end{eqnarray}Ansatz (\ref{ansatz}) does not quite solve (\ref{scUlate}), but 
the following choices reduce the residue to terms of order $\epsilon \times 
\tanh(\gamma)$,
\begin{equation}
\alpha = -3 \qquad , \qquad \frac14 \beta^2 = \frac14 + \frac{\epsilon}{2} - 
\frac{\mu^2}{\chi^2} \qquad , \qquad \gamma' = \frac12 \beta \; . 
\label{ansatzparams}
\end{equation}

\begin{figure}[H]
\centering
\begin{subfigure}[b]{0.33\textwidth}
\centering
\includegraphics[width=\textwidth]{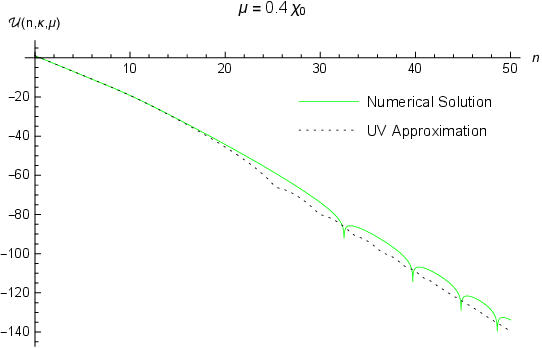}
\end{subfigure}
\begin{subfigure}[b]{0.33\textwidth}
\centering
\includegraphics[width=\textwidth]{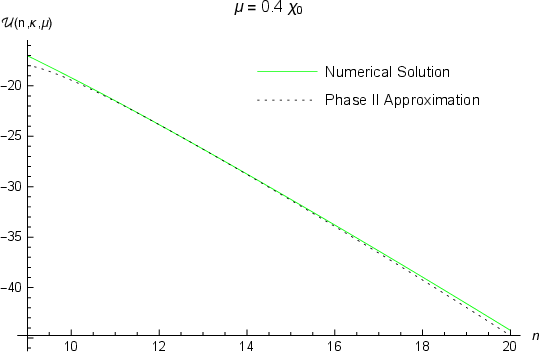}
\end{subfigure}
\begin{subfigure}[b]{0.33\textwidth}
\centering
\includegraphics[width=\textwidth]{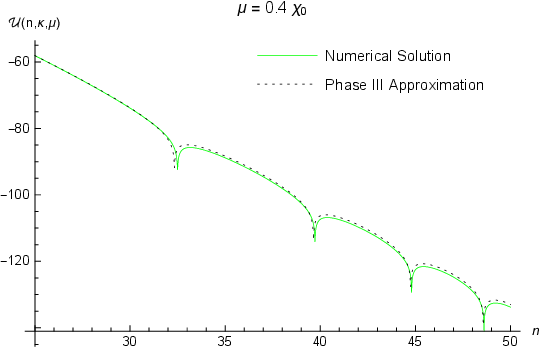}
\end{subfigure}
\caption[Temporal amplitudes and its different approximations for $\mu=0.4\chi_0$]{\footnotesize Plots the temporal amplitude 
$\mathcal{U}(n,3800 \chi_0, 0.4 \chi_0)$ and the three approximations: 
(\ref{U1def}), (\ref{U2def}) and (\ref{U3def}). For $\kappa = 3800 \chi_0$
horizon crossing occurs at $n_{\kappa} \simeq 8.3$; for $\mu = 0.4 \chi_0$ mass
domination occurs at $n_{\mu} \simeq 20.2$.}
\label{Temp3phasesA}
\end{figure}

\begin{figure}[H]
\centering
\begin{subfigure}[b]{0.33\textwidth}
\centering
\includegraphics[width=\textwidth]{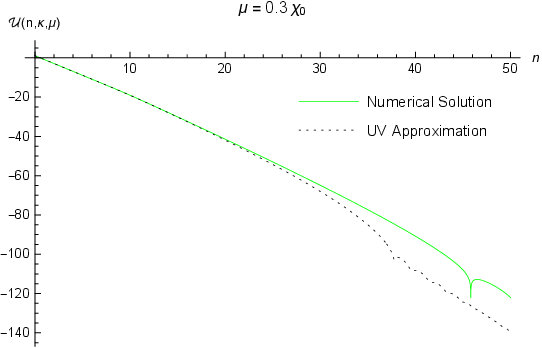}
\end{subfigure}
\begin{subfigure}[b]{0.33\textwidth}
\centering
\includegraphics[width=\textwidth]{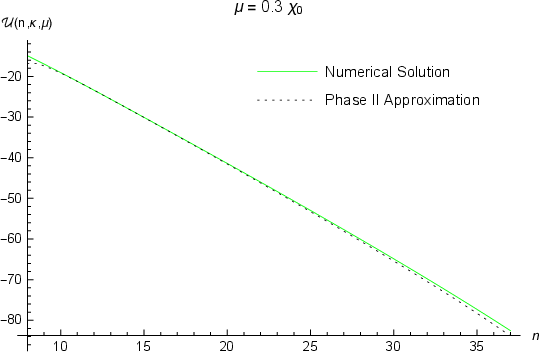}
\end{subfigure}
\begin{subfigure}[b]{0.33\textwidth}
\centering
\includegraphics[width=\textwidth]{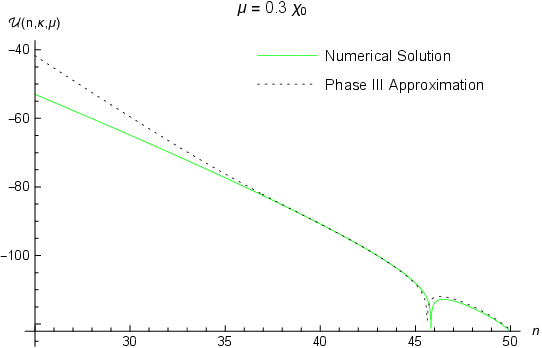}
\end{subfigure}
\caption[Temporal amplitudes and its different approximations for $\mu=0.3\chi_0$]{\footnotesize Plots the temporal amplitude 
$\mathcal{U}(n,3800 \chi_0, 0.3 \chi_0)$ and the three approximations: 
(\ref{U1def}), (\ref{U2def}) and (\ref{U3def}). For $\kappa = 3800 \chi_0$
horizon crossing occurs at $n_{\kappa} \simeq 8.3$; for $\mu = 0.3 \chi_0$ mass
domination occurs at $n_{\mu} \simeq 36.0$.}
\label{Temp3phasesB}
\end{figure}

Figures~\ref{Temp3phasesA} and \ref{Temp3phasesB} show how 
$\mathcal{U}(n,\kappa,\mu)$ behaves when mass domination comes after first horizon
crossing and before the end of inflation.
\noindent First comes a phase of slow decline followed by a period of
oscillations. From (\ref{ansatz}) with (\ref{ansatzparams}) we see that these
phases are controlled by a ``frequency'' defined as,
\begin{equation}
\omega^2_u(n,\mu) \equiv \frac14 + \frac{\epsilon(n)}{2} - 
\frac{\mu^2}{\chi^2(n)} \equiv -\Omega^2_u(n,\mu) \; . \label{tempomega}
\end{equation}
During the phase of slow decline $\omega^2_u(n,\mu) > 0$. Integrating 
(\ref{ansatz}) with (\ref{ansatzparams}) for this case gives,
\begin{eqnarray}
\lefteqn{\mathcal{U}_{2}(n,\kappa,\mu) = \mathcal{U}_{2} - 3 (n \!-\! n_2) + 
2 \ln\Biggl[\cosh\Bigl( \int_{n_2}^{n} \!\!\! dn' \omega_u(n',\mu)\Bigr) } 
\nonumber \\
& & \hspace{5cm} + \Bigl( \frac{3 \!+\! \mathcal{U}_{2}'}{2 \omega_u(n_2,\mu)} 
\Bigr) \sinh\Bigl( \int_{n_2}^{n} \!\!\! dn' \omega_u(n',\mu)\Bigr) \Biggr] ,
\qquad \label{U2def}
\end{eqnarray}
where $n_2 \equiv n_{\kappa} + 4$. The oscillatory phase is characterized by 
$\omega^2_u(n,\mu) < 0$. Integrating (\ref{ansatz}) with (\ref{ansatzparams})
for this case produces,
\begin{eqnarray}
\lefteqn{\mathcal{U}_{3}(n,\kappa,\mu) = \mathcal{U}_{3} - 3 (n \!-\! n_3) + 
2 \ln\Biggl[ \Biggl\vert\cos\Bigl( \int_{n_3}^{n} \!\!\! dn' \Omega_u(n',\mu)\Bigr) } 
\nonumber \\
& & \hspace{5cm} + \Bigl( \frac{3 \!+\! \mathcal{U}_{3}'}{2 \Omega_u(n_3,\mu)} 
\Bigr) \sin\Bigl( \int_{n_3}^{n} \!\!\! dn' \Omega_u(n',\mu)\Bigr) \Biggr\vert 
\Biggr] , \qquad \label{U3def}
\end{eqnarray}
where $n_3 \equiv n_{\mu} + 4$. Figures~\ref{Temp3phasesA} and 
\ref{Temp3phasesB} show that these approximations are excellent.

\begin{figure}[H]
\centering
\begin{subfigure}[b]{0.33\textwidth}
\centering
\includegraphics[width=\textwidth]{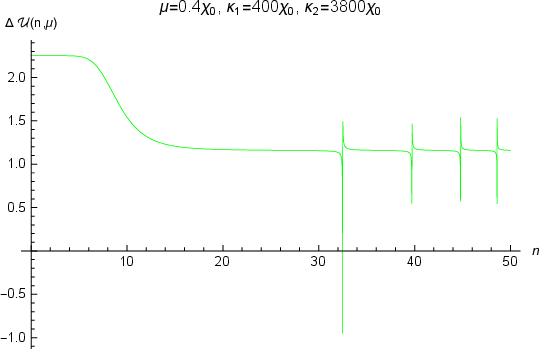}
\end{subfigure}
\begin{subfigure}[b]{0.33\textwidth}
\centering
\includegraphics[width=\textwidth]{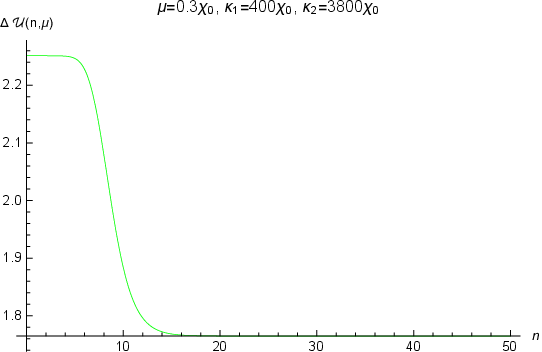}
\end{subfigure}
\begin{subfigure}[b]{0.33\textwidth}
\centering
\includegraphics[width=\textwidth]{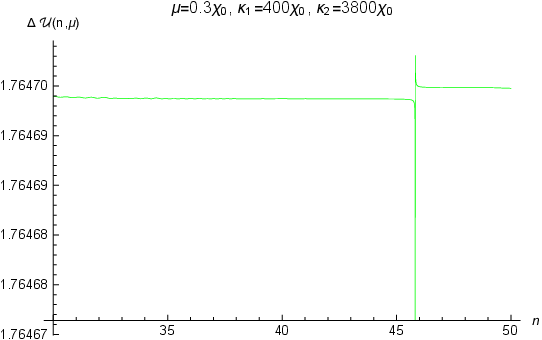}
\end{subfigure}
\caption[Difference of the temporal amplitude for different values of $\kappa$]{\footnotesize Plots the difference of the temporal amplitude 
$\Delta \mathcal{U} \equiv \mathcal{U}(n,\kappa_1,\mu) -
\mathcal{U}(n,\kappa_2,\mu)$ for $\kappa_1 = 400 \chi_0$ and $\kappa_2
= 3800 \chi_0$ with $\mu$ chosen so that all three approximations 
(\ref{U1def}), (\ref{U2def}) and (\ref{U3def}) are necessary.}
\label{Tempkdep}
\end{figure}

It is worth noting that the approximations (\ref{U2def}) and (\ref{U3def})
depend on $\kappa$ principally through the integration constants $\mathcal{U}_2
\equiv \mathcal{U}(n_2,\kappa,\mu)$ and $\mathcal{U}_3 \equiv 
\mathcal{U}(n_3,\kappa,\mu)$. Figure~\ref{Tempkdep} shows the difference
$\mathcal{U}(n,400 \chi_0,\mu) - \mathcal{U}(n,3800 \chi_0,\mu)$ for the same
two choices of $\mu$ in Figures~\ref{Temp3phasesA} and \ref{Temp3phasesB}. One
can see that the difference freezes into a constant after first horizon 
crossing to better than five significant figures!

\subsection{Spatially Transverse Photons}

\begin{figure}[H]
\centering
\begin{subfigure}[b]{0.33\textwidth}
\centering
\includegraphics[width=\textwidth]{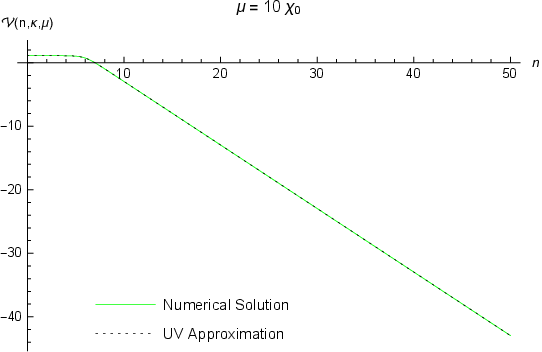}
\caption{$n_{\mu} < 0$}
\end{subfigure}
\begin{subfigure}[b]{0.33\textwidth}
\centering
\includegraphics[width=\textwidth]{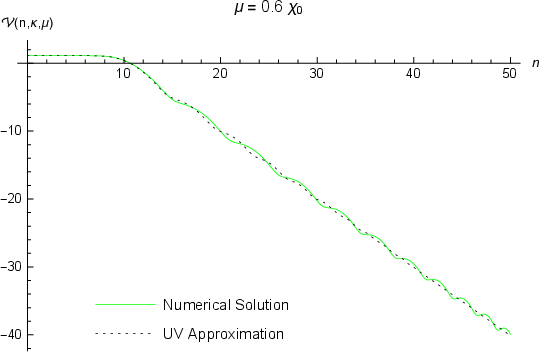}
\caption{$n_{\mu} < 0$}
\end{subfigure}
\begin{subfigure}[b]{0.33\textwidth}
\centering
\includegraphics[width=\textwidth]{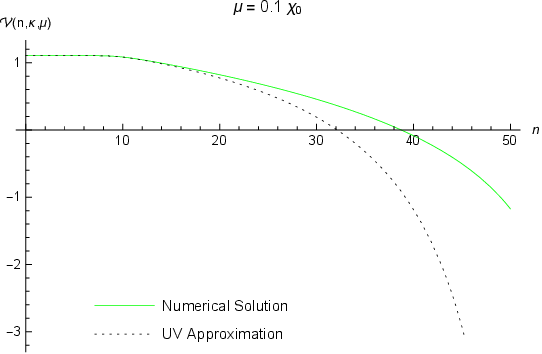}
\caption{$n_{\mu} > 50$}
\end{subfigure}%
\caption[Transverse amplitudes and its UV approximations for different values of $\mu$]{\footnotesize Plots the transverse amplitude $\mathcal{V}(n,\kappa,\mu)$ 
and the ultraviolet approximation (\ref{V1def}) versus the e-folding $n$ for 
$\kappa = 3800 \chi_0$ (with $n_{\kappa} \simeq 8.3$) and three different values 
of $\mu$ with outside the range of inflation.}
\label{TransUV}
\end{figure}

The general considerations for the amplitude of spatially transverse photons 
are similar to those for temporal photons. Before first horizon crossing it 
is the 4th and last terms of equation (\ref{scVeqn}) which control the 
evolution,
\begin{equation}
\frac{2 \kappa^2 e^{-2n}}{\chi^2} - \frac{e^{-2(D-3)n - 2\mathcal{V}}}{
2 \chi^2} \simeq 0 \qquad \Longrightarrow \qquad \mathcal{V} \simeq 
-\ln(2\kappa) - (D \!-\! 4) n \; . \label{UVscVapprox}
\end{equation}
A more accurate approximation is,
\begin{equation}
\mathcal{V}_{1}(n,\kappa,\mu) \equiv \ln\Biggl[ \frac{\frac{\pi}{2} 
z(n,\kappa)}{2 \kappa e^{(D-4) n}} \Bigl\vert H^{(1)}_{\nu_v(n,\mu)}
\Bigl( z(n,\kappa)\Bigr) \Bigr\vert^2 \Biggr] , \label{V1def}
\end{equation}
where $z(n,\kappa)$ is the same as (\ref{znudef}) and the transverse index is,
\begin{equation}
\nu^2_v(n,\mu) \equiv \frac14 \Bigl( \frac{D \!-\! 3 \!-\! \epsilon(n)}{
1 \!-\! \epsilon(n)} \Bigr)^2 - \frac{\mu^2}{[1 \!-\! \epsilon(n)]^2 
\chi^2(n)} \; . \label{nuvdef}
\end{equation}
Note the slight (order $\epsilon$) difference between $\nu^2_u(n,\mu)$ and 
$\nu^2_v(n,\mu)$. Figure~\ref{TransUV} shows that (\ref{V1def}) is excellent
up to several e-foldings after first horizon crossing, and throughout inflation
for $n_{\mu} < 0$.
\noindent Expression (\ref{V1def}) also models the ultraviolet to high precision,
\begin{eqnarray}
\lefteqn{\mathcal{V}(n,\kappa,\mu) - \mathcal{V}_1(n,\kappa,\mu) = \Biggl\{ 
\Bigl( 5 \epsilon - 3 \epsilon^2\Bigr) \frac{\mu^2}{4 \chi^2} } \nonumber \\
& & \hspace{1.8cm} + \Bigl( \frac{D\!-\!4}{16}\Bigr) \Bigl[ (D + 3 - 7 \epsilon) 
\epsilon' + \epsilon''\Bigr] \Biggr\} \Bigl( \frac{\chi e^n}{\kappa}\Bigr)^4 \!+ 
O\Biggl( \Bigl(\frac{\chi e^n}{\kappa}\Bigr)^6 \Biggr) . \qquad \label{Vasymp}
\end{eqnarray}

\begin{figure}[H]
\centering
\begin{subfigure}[b]{0.33\textwidth}
\centering
\includegraphics[width=\textwidth]{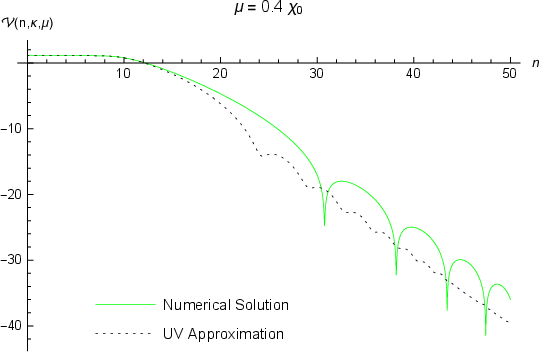}
\end{subfigure}
\begin{subfigure}[b]{0.33\textwidth}
\centering
\includegraphics[width=\textwidth]{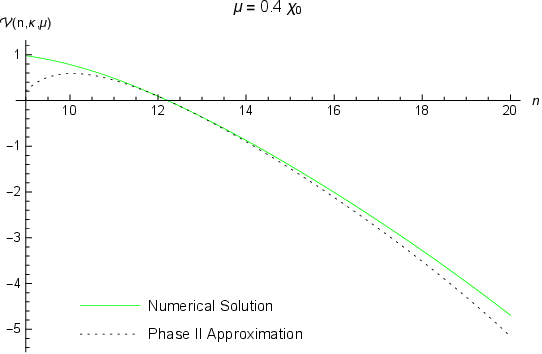}
\end{subfigure}
\begin{subfigure}[b]{0.33\textwidth}
\centering
\includegraphics[width=\textwidth]{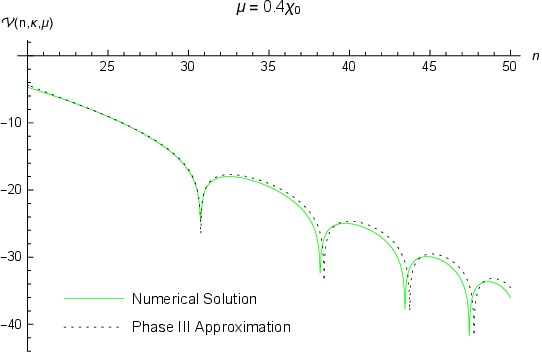}
\end{subfigure}
\caption[Transverse amplitude and its approximations for $\mu=0.4\chi_0$]{\footnotesize Plots the transverse amplitude 
$\mathcal{V}(n,3800 \chi_0, 0.4 \chi_0)$ and the three approximations: 
(\ref{V1def}), (\ref{V2def}) and (\ref{V3def}). For $\kappa = 3800 \chi_0$
horizon crossing occurs at $n_{\kappa} \simeq 8.3$; for $\mu = 0.4 \chi_0$ mass
domination occurs at $n_{\mu} \simeq 20.2$.}
\label{Trans3phasesA}
\end{figure}

Figure~\ref{Trans3phasesA} shows $\mathcal{V}(n,\kappa,\mu)$ for the case
where $n_{\mu}$ happens after first horizon crossing and before the end of
inflation. One sees the same phases of slow decline after first horizon 
crossing, followed by oscillations.
\noindent The second and third phases can be understood by noting that the two 
terms of expression (\ref{UVscVapprox}) redshift into insignificance after 
first horizon crossing. We can also set $D=4$ so that equation (\ref{scVeqn}) 
degenerates to,
\begin{equation}
\mathcal{V}'' + \frac12 {\mathcal{V}'}^2 + (1 \!-\! \epsilon) \mathcal{V}'
+ \frac{2 \mu^2}{\chi^2} \simeq 0 \; .
\end{equation}
The same ansatz (\ref{ansatz}) applies to this regime, with the parameter
choices,
\begin{equation}
\alpha = -1 \quad , \quad \frac14 \beta^2 = \frac14 - \frac{\mu^2}{\chi^2} 
\equiv \omega^2_v \equiv -\Omega^2_v \quad , \quad \gamma' = \frac12 \beta \; .
\label{Vparams}
\end{equation}
Just as there was an order $\epsilon$ difference between the temporal and 
transverse indices --- expressions (\ref{znudef}) and (\ref{nuvdef}), respectively 
--- so too there is an order $\epsilon$ difference between $\omega^2_u(n,\mu)$ and
$\omega^2_v(n,\kappa)$. 

\begin{figure}[H]
\centering
\begin{subfigure}[b]{0.33\textwidth}
\centering
\includegraphics[width=\textwidth]{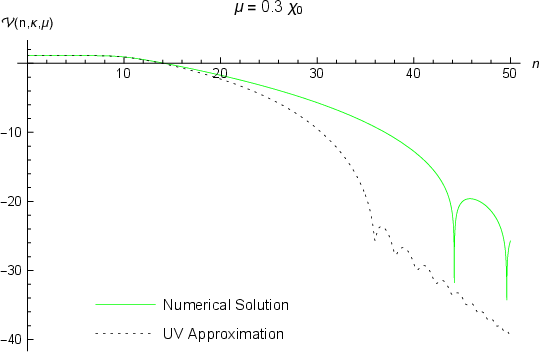}
\end{subfigure}
\begin{subfigure}[b]{0.33\textwidth}
\centering
\includegraphics[width=\textwidth]{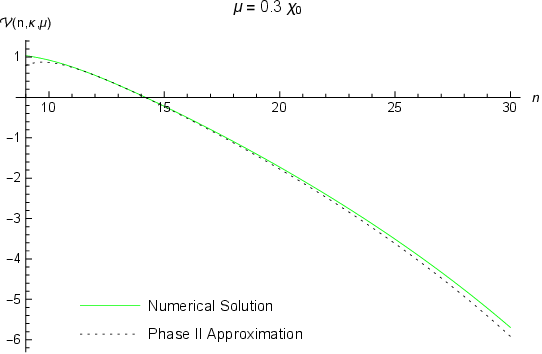}
\end{subfigure}
\begin{subfigure}[b]{0.33\textwidth}
\centering
\includegraphics[width=\textwidth]{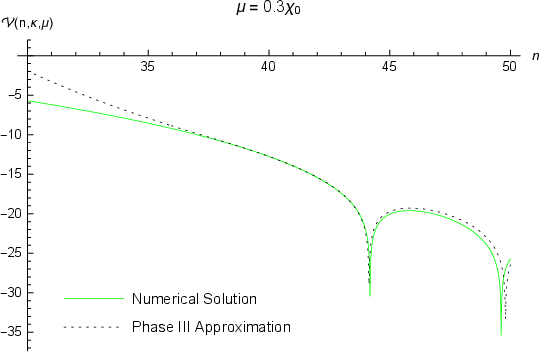}
\end{subfigure}
\caption[Transverse amplitude and its approximations for $\mu=0.3\chi_0$]{\footnotesize Plots the transverse amplitude 
$\mathcal{V}(n,3800 \chi_0, 0.3 \chi_0)$ and the three approximations: 
(\ref{V1def}), (\ref{V2def}) and (\ref{V3def}). For $\kappa = 3800 \chi_0$
horizon crossing occurs at $n_{\kappa} \simeq 8.3$; for $\mu = 0.3 \chi_0$ mass
domination occurs at $n_{\mu} \simeq 36.0$.}
\label{Trans3phasesB}
\end{figure}

Integrating (\ref{ansatz}) with (\ref{Vparams}) for $\omega^2_v(,\mu) > 0$ gives,
\begin{eqnarray}
\lefteqn{\mathcal{V}_{2}(n,\kappa,\mu) = \mathcal{V}_{2} - (n \!-\! n_2) + 
2 \ln\Biggl[\cosh\Bigl( \int_{n_2}^{n} \!\!\! dn' \omega_v(n',\mu)\Bigr) } 
\nonumber \\
& & \hspace{5cm} + \Bigl( \frac{1 \!+\! \mathcal{V}_{2}'}{2 \omega_v(n_2,\mu)} 
\Bigr) \sinh\Bigl( \int_{n_2}^{n} \!\!\! dn' \omega_v(n',\mu)\Bigr) \Biggr] ,
\qquad \label{V2def}
\end{eqnarray}
where $n_2 \equiv n_{\kappa} + 4$. Integrating (\ref{ansatz}) with (\ref{Vparams})
for $\omega^2_v(n,\mu) < 0$ results in,
\begin{eqnarray}
\lefteqn{\mathcal{V}_{3}(n,\kappa,\mu) = \mathcal{V}_{3} - (n \!-\! n_3) + 
2 \ln\Biggl[ \Biggl\vert\cos\Bigl( \int_{n_3}^{n} \!\!\! dn' \Omega_v(n',\mu)\Bigr) } 
\nonumber \\
& & \hspace{5cm} + \Bigl( \frac{1 \!+\! \mathcal{V}_{3}'}{2 \Omega_v(n_3,\mu)} 
\Bigr) \sin\Bigl( \int_{n_3}^{n} \!\!\! dn' \Omega_v(n',\mu)\Bigr) \Biggr\vert 
\Biggr] , \qquad \label{V3def}
\end{eqnarray}
where $n_3 \equiv n_{\mu} + 4$. Figures~\ref{Trans3phasesA} and \ref{Trans3phasesB}
demonstrate that the (\ref{V2def}) and (\ref{V3def}) approximations are excellent.

\begin{figure}[H]
\centering
\begin{subfigure}[b]{0.33\textwidth}
\centering
\includegraphics[width=\textwidth]{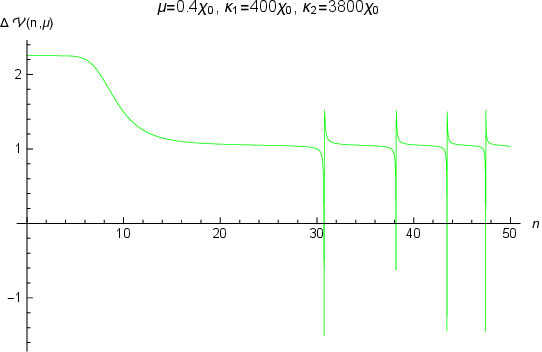}
\end{subfigure}
\begin{subfigure}[b]{0.33\textwidth}
\centering
\includegraphics[width=\textwidth]{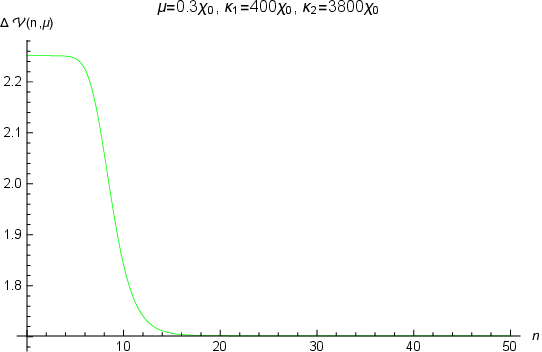}
\end{subfigure}
\begin{subfigure}[b]{0.33\textwidth}
\centering
\includegraphics[width=\textwidth]{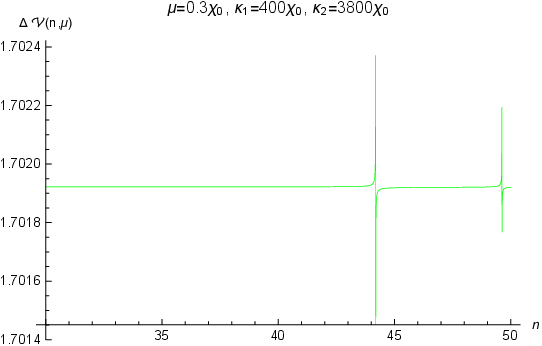}
\end{subfigure}
\caption[Difference of the transverse amplitudes for different values of $\kappa$]{\footnotesize Plots the difference of the transverse amplitude 
$\Delta \mathcal{V} \equiv \mathcal{V}(n,\kappa_1,\mu) -
\mathcal{V}(n,\kappa_2,\mu)$ for $\kappa_1 = 400 \chi_0$ and $\kappa_2 =
3800 \chi_0$ with $\mu$ chosen so that all three approximations (\ref{V1def}), 
(\ref{V2def}) and (\ref{V3def}) are necessary.}
\label{Transkdep}
\end{figure}

Finally, we note that from Figure~\ref{Transkdep} that $\mathcal{V}'(n,\kappa,\mu)$
is nearly independent of $\kappa$ after first horizon crossing.
\noindent One consequence for the (\ref{V2def}) and (\ref{V3def}) 
approximations is that only the integration constants $\mathcal{V}_2$ and
$\mathcal{V}_3$ depend on $\kappa$.

\subsection{Plateau Potentials}

We chose the quadratic dimensionless potential $U(\psi\psi^*) = c^2 \psi \psi^*$ for 
detailed studies because it gives simple, analytic expressions (\ref{slowroll1}) in
the slow roll approximation for the dimensionless Hubble parameter $\chi(n)$ and the 
first slow roll parameter $\epsilon(n)$. Setting $c \simeq 7.126 \times 10^{-6}$ makes
this model consistent with the observed values for the scalar amplitude and the 
scalar spectral index \cite{Aghanim:2018eyx}. On the other hand, the model's large
prediction of $r \simeq 0.14$ is badly discordant with limits on the tensor-to-scalar 
ratio \cite{Aghanim:2018eyx}. We shall therefore briefly consider how our analytic
approximations fare when used with the plateau potentials currently consistent with
observation. 

\begin{figure}[H]
\centering
\begin{subfigure}[b]{0.33\textwidth}
\centering
\includegraphics[width=\textwidth]{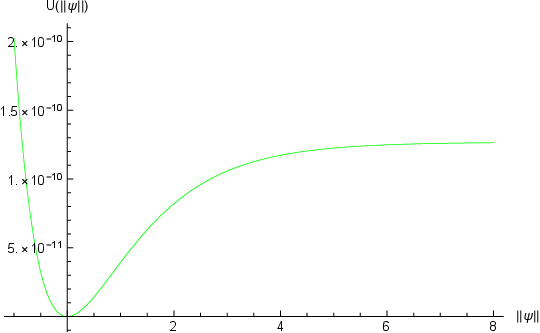}
\end{subfigure}
\begin{subfigure}[b]{0.33\textwidth}
\centering
\includegraphics[width=\textwidth]{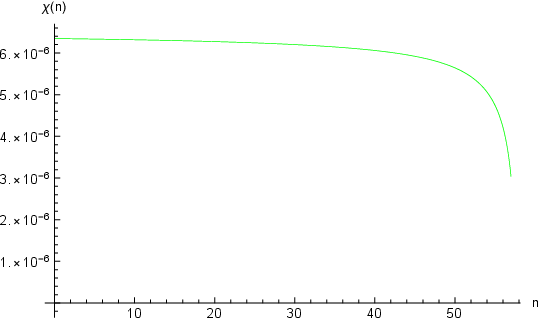}
\end{subfigure}
\begin{subfigure}[b]{0.33\textwidth}
\centering
\includegraphics[width=\textwidth]{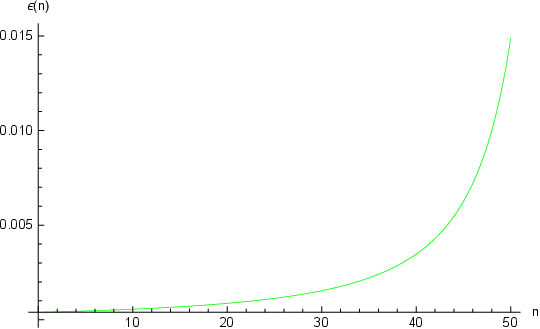}
\end{subfigure}
\caption[Potential and geometry for the Einstein-frame representation 
of Starobinsky's inflation]{\footnotesize{Potential and geometry for the Einstein-frame representation 
of Starobinsky's original model of inflation \cite{Starobinsky:1980te}. The left shows
the dimensionless potential $U(\psi\psi^*)$ (\ref{StaroU}); the middle plot gives the
dimensionless Hubble parameter $\chi(n)$ and the right hand plot depicts the first
slow roll parameter $\epsilon(n)$. Inflation was assumed to start from $\psi_0 = 4.6$.}}
\label{StarobinskyA}
\end{figure}

The best known plateau potential is the Einstein-frame version of Staro\-binsky's
famous $R + R^2$ model \cite{Starobinsky:1980te}. Expressing the dimensionless 
potential for this model in our notation gives \cite{Brooker:2016oqa},
\begin{equation}
U(\psi\psi^*) = \frac34 M^2 \Bigl( 1 - e^{-\sqrt{\frac23} \, \vert\psi\vert}\Bigr)^2 
\qquad , \qquad M = 1.3 \times 10^{-5} \; . \label{StaroU}
\end{equation}
Somewhat over 50 e-foldings of inflation result if one starts from $\psi_0 = 4.6$, 
and the choice of $M = 1.3 \times 10^{-5}$ makes the model consistent with observation 
\cite{Aghanim:2018eyx}. Figure~\ref{StarobinskyA} shows why $r = 16 \epsilon$ is 
so small for this model: its dimensionless Hubble parameter $\chi(n)$ is nearly constant.  

\begin{figure}[H]
\centering
\begin{subfigure}[b]{0.5\textwidth}
\centering
\includegraphics[width=\textwidth]{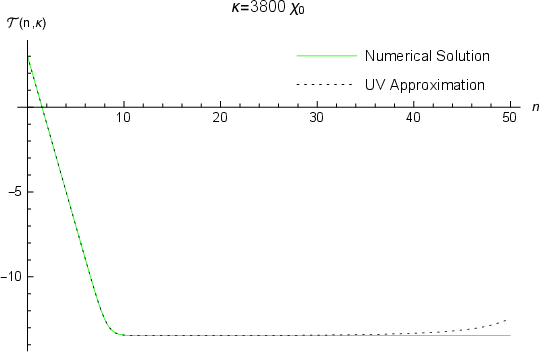}
\end{subfigure}
\begin{subfigure}[b]{0.5\textwidth}
\centering
\includegraphics[width=\textwidth]{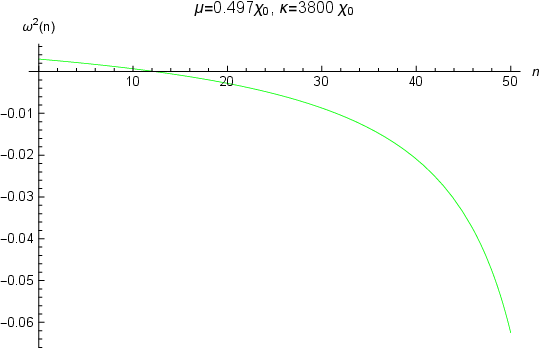}
\end{subfigure}
\caption[Amplitude of the MMC scalar for the Strobinsky's potential]{\footnotesize The left hand plot shows the amplitude 
$\mathcal{T}(n,\kappa)$ of the massless, minimally coupled scalar
for $\kappa = 3800 \chi_0$, which corresponds to $n_{\kappa} \simeq
8.3$. The right hand graph shows the frequency $\omega_u^2(n,\mu)
\simeq \omega^2_v(n,\mu)$ for $\mu = 0.497 \chi_0$ which passes 
through zero at $n_{\mu} \simeq 12$.} 
\label{StarobinskyB}
\end{figure}
\begin{figure}[H]
\centering
\begin{subfigure}[b]{0.5\textwidth}
\centering
\includegraphics[width=\textwidth]{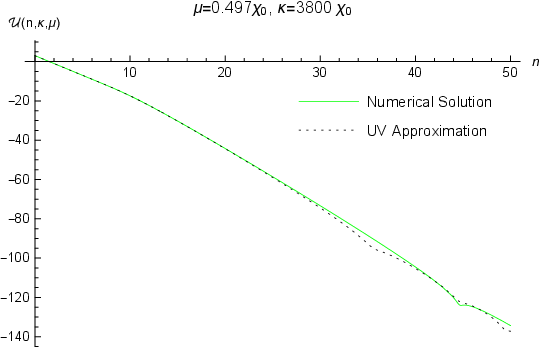}
\end{subfigure}
\begin{subfigure}[b]{0.5\textwidth}
\centering
\includegraphics[width=\textwidth]{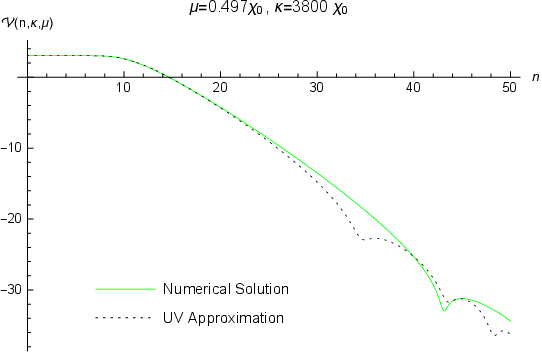}
\end{subfigure}
\caption[Temporal and spatially transverse amplitudes for the Starobinsky potential]{\footnotesize Plots of the temporal amplitude 
$\mathcal{U}(n,\kappa,\mu)$ (left) and the spatially transverse 
amplitude $\mathcal{V}(n,\kappa,\mu)$ (right) versus $n$ for the 
Starobinsky potential (\ref{StaroU}). For each amplitude $\kappa 
= 3800 \chi_0$ (which implies $n_{\kappa} \simeq 8.3$) and $\mu 
= 0.497 \chi_0$ (which implies $n_{\mu} \simeq 12$).}
\label{StarobinskyC}
\end{figure}

\noindent All our approximations pertain for this model, but the 
general effect of $\chi(n)$ being so nearly constant is to increase
the range over which the ultraviolet approximations pertain. The left 
hand plot of Figure~\ref{StarobinskyB} shows this for the MMCS 
amplitude $\mathcal{T}(n,\kappa)$. Because $\epsilon(n)$ is so small, 
the temporal and transverse frequencies are nearly equal $\omega^2_u(n,\mu) 
\simeq \omega^2_v(n,\mu)$ and nearly constant. The right hand plot of 
Figure~\ref{StarobinskyB} shows this for a carefully chosen value of 
$\mu = 0.497 \chi_0$ which causes mass domination to occur during 
inflation. For this case we can just see the second and third phases 
occur in Figure~\ref{StarobinskyC}.

\section{Effective Potential}

The purpose of this section is to evaluate the one photon loop contribution 
to the inflaton effective potential defined by equation (\ref{DeltaVdef}).
We begin by deriving some exact results for the trace of the coincident 
propagator, and we recall that $\mathcal{T}(n,\kappa)$ can be obtained from
$\mathcal{U}(n,\kappa,0)$. Then the ultraviolet approximations (\ref{U1def})
and (\ref{V1def}) are used to derive a divergent result whose renormalization
gives the part of the effective potential that depends locally on the geometry.
We give large field and small field expansions for this local part, and we
study its dependence on derivatives of $\epsilon(n)$. The section closes with
a discussion of the nonlocal part of the effective potential which derives 
from the late time approximations (\ref{U2def}), (\ref{U3def}), (\ref{V2def})
and (\ref{V3def}).

\subsection{Trace of the Coincident Photon Propagator}

At coincidence the mixed time-space components of the photon mode sum vanish,
and factors of $\widehat{k}_m \widehat{k}_n$ average to $\delta_{mn}/(D-1)$,
\begin{eqnarray}
\lefteqn{ i \Bigl[ \mbox{}_{\mu} \Delta_{\nu}\Bigr](x;x) = \int \!\! 
\frac{d^{D-1}k}{(2\pi)^{D-1}} \, \Biggl\{ \frac1{M^2} \left( 
\begin{matrix} k^2 u u^* & 0 \cr 0 & \frac{\delta_{mn}}{D - 1} \mathcal{D} u
\mathcal{D} u^*\end{matrix}\right) } \nonumber \\
& & \hspace{1.8cm} - \frac1{M^2} \left( \begin{matrix}\partial_0 t \partial_0 t^* & 0 \cr 
0 & \frac{\delta_{mn}}{D - 1} k^2 t t^*\end{matrix}\right) + \left( \begin{matrix} 0 & 0 \cr 0 &
(\frac{D-2}{D-1}) \delta_{mn} v v^*\end{matrix}\right) \Biggr\} . \qquad \label{coincprop}
\end{eqnarray}
Its trace is,
\begin{eqnarray}
\lefteqn{g^{\mu\nu} i \Bigl[ \mbox{}_{\mu} \Delta_{\nu}\Bigr](x;x) =} \nonumber \\
& & \hspace{1cm} \int \!\! \frac{d^{D-1}k}{(2\pi)^{D-1}} \, \Biggl\{ 
\frac{ \mathcal{D} u \mathcal{D} u^* \!-\! k^2 u u^* \!+\! \partial_0 t 
\partial_0 t^* \!-\! k^2 t t^*}{a^2 M^2} + \frac{(D \!-\!2) v v^*}{a^2} \Biggr\} 
. \qquad \label{coinctrace}
\end{eqnarray}
Relation (\ref{MMCtotemp}) allows us to replace the MMCS mode function $t(\eta,k)$
with the massless limit of the temporal mode function $u_0(\eta,k) \equiv 
u(\eta,k,0)$,
\begin{equation}
\partial_0 t \partial_0 t^* = k^2 u_0 u^*_0 \qquad , \qquad k^2 t t^* = 
\mathcal{D} u_0 \mathcal{D} u^*_0 \; . \label{ttouzero}
\end{equation}
Substituting (\ref{ttouzero}) in (\ref{coinctrace}) gives,
\begin{eqnarray}
\lefteqn{g^{\mu\nu} i \Bigl[ \mbox{}_{\mu} \Delta_{\nu}\Bigr](x;x) =} \nonumber \\
& & \hspace{0.8cm} \int \!\! \frac{d^{D-1}k}{(2\pi)^{D-1}} \, \Biggl\{ 
\frac{ \mathcal{D} u \mathcal{D} u^* \!-\! \mathcal{D} u_0 \mathcal{D} u_0^*
\!-\! k^2 (u u^* \!-\! u_0 u^*_0)}{a^2 M^2} + \frac{(D \!-\!2) v v^*}{a^2} \Biggr\} 
. \qquad \label{coinctrace2}
\end{eqnarray}
This second form (\ref{coinctrace2}) is very important because it demonstrates
the absence of any $1/M^2$ pole as an exact relation, before any approximations 
are made.

The mode equation for temporal photons implies,
\begin{eqnarray}
\lefteqn{\mathcal{D} u \mathcal{D} u^* = a^2 H^2 \Bigl[ u' {u'}^* + 
(D\!-\!2) (u u^*)' + (D \!-\! 2)^2 u u^*\Bigr] \; , } \\
& & \hspace{0.8cm} = (k^2 + a^2 M^2) u u^* + \frac{a^2 H^2}{2} \Bigl( 
\partial_n \!+\! D \!-\! 1 \!-\! \epsilon \Bigr) \Bigl(\partial_n \!+\! 2 D 
\!-\! 4\Bigr) (u u^*) \; . \qquad \label{tempsimp}
\end{eqnarray}
Using relations (\ref{tempsimp}) and (\ref{coinctrace2}) allows us to express 
the trace of the coincident photon propagator in terms of three coincident 
scalar propagators,
\begin{eqnarray}
\lefteqn{g^{\mu\nu} i \Bigl[ \mbox{}_{\mu} \Delta_{\nu}\Bigr](x;x) =
i\Delta_u(x;x) + \frac{(D \!-\! 2)}{a^2} i\Delta_v(x;x) } \nonumber \\
& & \hspace{1cm} + \frac{H^2}{2 M^2} \Bigl( \partial_n \!+\! D \!-\! 1 \!-\! 
\epsilon \Bigr) \Bigl(\partial_n \!+\! 2 D \!-\! 4\Bigr) \Bigl[i\Delta_u(x;x) 
- i\Delta_{u_0}(x;x) \Bigr] \; . \qquad \label{coinctrace3}
\end{eqnarray}
The disappearance of any factors of $k^2$ from the Fourier mode sums in
(\ref{coinctrace3}), coupled with the ultraviolet expansions (\ref{Uasymp}) and
(\ref{Vasymp}), means that the phase 1 approximations $\mathcal{U}_1(n,\kappa,\mu)$
and $\mathcal{V}_1(n,\kappa,\mu)$ exactly reproduce the ultraviolet divergence
structures.

Two of the scalar propagators in expression (\ref{coinctrace3}) are,
\begin{eqnarray}
\lefteqn{i\Delta_u(x;x') \equiv \int \!\! \frac{d^{D-1}k}{(2\pi)^{D-1}} \Biggl\{ 
\theta(\Delta \eta) u(\eta,k,M) u^*(\eta',k,M) e^{i \vec{k} \cdot \Delta \vec{x}} }
\nonumber \\
& & \hspace{4.5cm} + \theta(-\Delta \eta) u^*(\eta,k,M) u(\eta',k,M) e^{-i\vec{k} 
\cdot \Delta \vec{x}} \Biggr\} , \qquad \label{tempprop} \\
\lefteqn{i\Delta_v(x;x') \equiv \int \!\! \frac{d^{D-1}k}{(2\pi)^{D-1}} \Biggl\{ 
\theta(\Delta \eta) v(\eta,k,M) v^*(\eta',k,M) e^{i \vec{k} \cdot \Delta \vec{x}} }
\nonumber \\
& & \hspace{4.5cm} + \theta(-\Delta \eta) v^*(\eta,k,M) v(\eta',k,M) e^{-i\vec{k} 
\cdot \Delta \vec{x}} \Biggr\} . \qquad \label{transprop}
\end{eqnarray}
The third scalar propagator $i\Delta_{u_0}(x;x')$ is just the $M \rightarrow 0$
limit of $i\Delta_{u}(x;x')$. The coincidence limits of each propagator can be
expressed in terms of the corresponding amplitude,
\begin{equation}
\frac{i\Delta_u(x;x)}{\sqrt{8\pi G} } = \int \!\! \frac{d^{D-1}k}{(2\pi)^{D-1}} \,
e^{\mathcal{U}(n,\kappa,\mu)} \quad , \quad \frac{i\Delta_v(x;x)}{\sqrt{8\pi G} } 
= \int \!\! \frac{d^{D-1}k}{(2\pi)^{D-1}} \, e^{\mathcal{V}(n,\kappa,\mu)} \; .
\label{coincmsums}
\end{equation}

Expression(\ref{coinctrace3}) is exact but not immediately useful because we 
lack explicit expressions for the coincident propagators (\ref{coincmsums}).
It is at this stage that we must resort to the analytic approximations developed 
in section 3. Recall that the phase 1 approximation is valid until roughly 4 
e-foldings after horizon crossing. If one instead thinks of this as a condition 
on the dimensionless wave number $\kappa \equiv \sqrt{8\pi G} \, k$ at fixed $n$, 
it means that $\kappa > \kappa_{n-4}$, where we define $\kappa_n$ as the 
dimensionless wave number which experiences horizon crossing at e-folding $n$. 
Taking as an example the temporal photon contribution we can write,
\begin{eqnarray}
\lefteqn{e^{\mathcal{U}(n,\kappa,\mu)} \simeq \theta\Bigl( \kappa \!-\! 
\kappa_{n-4}\Bigr) e^{\mathcal{U}_{1}(n,\kappa,\mu)} + \theta\Bigl( \kappa_{n-4}
- \kappa \Bigr) e^{\mathcal{U}_{2,3}(n,\kappa,\mu)} \; , } \\
& & \hspace{2cm} = e^{\mathcal{U}_{1}(n,\kappa,\mu)} + \theta\Bigl( \kappa_{n-4}
- \kappa \Bigr) \Biggl[e^{\mathcal{U}_{2,3}(n,\kappa,\mu)} -
e^{\mathcal{U}_{1}(n,\kappa,\mu)} \Biggr] . \qquad \label{ampapprox}
\end{eqnarray}
Substituting the approximation (\ref{ampapprox}) into expression 
(\ref{coincmsums}) allows us to write,
\begin{equation}
i\Delta_u(x;x) \simeq L_u(n) + N_u(n) \; , 
\label{locnonloc}
\end{equation}
where we define the local ($L$) and nonlocal ($N$) contributions as,
\begin{eqnarray}
L_u(n) &\!\!\! \equiv \!\!\!& \sqrt{8\pi G} \! \int \!\! 
\frac{d^{D-1}k}{(2\pi)^{D-1}} \, e^{\mathcal{U}_{1}(n,\kappa,\mu)} \; , 
\label{localprop} \\
N_u(n) &\!\!\! \equiv \!\!\!& \sqrt{8\pi G} \! \int \!\! 
\frac{d^{3}k}{(2\pi)^{3}} \, \theta\Bigl( \kappa_{n-4} - \kappa \Bigr) 
\Biggl[e^{\mathcal{U}_{2,3}(n,\kappa,\mu)} - e^{\mathcal{U}_{1}(n,\kappa,\mu)} 
\Biggr] . \qquad \label{nonlocalprop}
\end{eqnarray}
Note that we have taken the unregulated limit ($D = 4$) in expression 
(\ref{nonlocalprop}) because it is ultraviolet finite. The same considerations
apply as well for the coincident spatially transverse photon propagator 
$i\Delta_v(x;x')$, and for the massless limit of the temporal photon propagator 
$i\Delta_{u_0}(x;x)$.

\subsection{The Local Contribution}

The local contribution for each of the coincident propagators (\ref{coincmsums})
comes from using the phase 1 approximation (\ref{localprop}). For the temporal 
modes the amplitude is approximated by expression (\ref{U1def}), whereupon we 
change variables to $z$ using $k =
(1-\epsilon) H a z$, and then employ integral $6.574\; \# 2$ of 
\cite{Gradshteyn:1965},
\begin{equation}
L_u(n) = \frac{[(1 \!-\! \epsilon) H]^{D-2}}{(4\pi)^{\frac{D}2}}
\times \frac{\Gamma(\frac{D-1}{2} \!+\! \nu_u) \Gamma(\frac{D-1}{2} \!-\! 
\nu_u)}{\Gamma(\frac12 \!+\! \nu_u) \Gamma(\frac12 \!-\! \nu_u)} \times
\Gamma\Bigl(1 \!-\! \frac{D}2\Bigr) \; . \label{Lu}
\end{equation}
Recall that the index $\nu_u(n,\mu)$ is defined in expression (\ref{nuudef}).
Of course the massless limit is,
\begin{equation}
L_{u_0}(n) = \frac{[(1 \!-\! \epsilon) H]^{D-2}}{(4\pi)^{\frac{D}2}}
\times \frac{\Gamma(\frac{D-1}{2} \!+\! \nu_{u_0}) \Gamma(\frac{D-1}{2} \!-\! 
\nu_{u_0})}{\Gamma(\frac12 \!+\! \nu_{u_0}) \Gamma(\frac12 \!-\! \nu_{u_0})} 
\times \Gamma\Bigl(1 \!-\! \frac{D}2\Bigr) \; , \label{Lu0}
\end{equation}
where the index is,
\begin{equation}
\nu_{u_0}(n) \equiv \nu_u(n,0) = \frac12 \Bigl( \frac{D \!-\! 3 \!+\! \epsilon(n)}{
1 \!-\! \epsilon(n)} \Bigr) \; . \label{nuu0def}
\end{equation}
The phase 1 approximation (\ref{V1def}) for the transverse amplitude contains 
two extra scale factors which serve to exactly cancel the inverse scale 
factors that are evident in the transverse contribution to the trace of the 
coincident photon propagator (\ref{coinctrace3}). Hence we have,
\begin{equation}
\frac{L_v(n)}{a^2} = \frac{[(1 \!-\! \epsilon) H]^{D-2}}{
(4\pi)^{\frac{D}2}}\times \frac{\Gamma(\frac{D-1}{2} \!+\! \nu_v) 
\Gamma(\frac{D-1}{2} \!-\! \nu_v)}{\Gamma(\frac12 \!+\! \nu_v) 
\Gamma(\frac12 \!-\! \nu_v)} \times \Gamma\Bigl(1 \!-\! \frac{D}2\Bigr) 
\; , \label{Lv}
\end{equation}
where the transverse index $\nu_v(n,\mu)$ is given in (\ref{nuvdef}).

Each of the local contributions (\ref{Lu}), (\ref{Lu0}) and (\ref{Lv}) is 
proportional to the same divergent Gamma function,
\begin{equation}
\Gamma\Bigl(1 \!-\! \frac{D}2\Bigr) = \frac{2}{D \!-\! 4} + 
O\Bigl( (D\!-\!4)^0 \Bigr) \; .
\end{equation}
Each also contains a similar ratio of Gamma functions,
\begin{eqnarray}
\lefteqn{ \frac{ \Gamma(\frac{D-1}2 \!+\! \nu) \Gamma(\frac{D-1}2 \!-\! \nu)}{
\Gamma(\frac12 \!+\! \nu) \Gamma(\frac12 \!-\! \nu)} = \Bigl[ 
\Bigl(\frac{D\!-\!3}{2}\Bigr)^2 \!-\! \nu^2\Bigr] \!\times\! 
\frac{ \Gamma(\frac{D-3}2 \!+\! \nu) \Gamma(\frac{D-3}2 \!-\! \nu)}{
\Gamma(\frac12 \!+\! \nu) \Gamma(\frac12 \!-\! \nu)} \; , } \\
& & \hspace{-0.7cm} = \! \Bigl[ \Bigl(\frac{D\!-\!3}{2}\Bigr)^2 \!\!\!-\! \nu^2\Bigr]
\Biggl\{ \! 1 + \Bigl[ \psi\Bigl( \frac12 \!+\! \nu\!\Bigr) \!+\! \psi\Bigl(
\frac12 \!-\! \nu\!\Bigr) \Bigr] \Bigl( \frac{D \!-\! 4}{2}\Bigr) \!+\! O\Bigl( 
(D\!-\!4)^2\Bigr) \! \Biggr\} . \qquad 
\end{eqnarray}
These considerations allow us to break up each of the three terms in 
(\ref{coinctrace3}) into a potentially divergent part plus a manifestly finite part. 
For $i\Delta_u(x;x) \rightarrow L_u(n)$ this decomposition is,
\begin{eqnarray}
\lefteqn{ L_u = \frac{[(1 \!-\! \epsilon) H]^{D-4}}{(4 \pi)^{\frac{D}2}} 
\Biggl[ M^2 - \frac{(D \!-\! 2) H^2}{2} \Bigl( (D \!-\! 3) \epsilon -
\frac12 (D\!-\! 4) \epsilon^2\Bigr) \Biggr] \Gamma\Bigl(1 \!-\! \frac{D}2\Bigr) }
\nonumber \\
& & \hspace{1.5cm} + \frac1{16 \pi^2} \Bigl[M^2 - \epsilon H^2\Bigr] \Bigl[
\psi\Bigl( \frac12 \!+\! \nu_u\Bigr) + \psi\Bigl( \frac12 \!-\! \nu_u\Bigr) \Bigr]
+ O( D\!-\! 4) \; . \qquad \label{1stterm}
\end{eqnarray}
For $(D-2) i\Delta_v(x;x) \rightarrow (D-2) L_v(n)$ we have,
\begin{eqnarray}
\lefteqn{ (D \!-\!2) L_v = \frac{[(1 \!-\! \epsilon) H]^{D-4}}{(4 \pi)^{\frac{D}2}} 
\Biggl[ (D \!-\! 2) M^2 - \frac{(D \!-\! 2) (D \! -\! 4) H^2}{2} \Bigl( (D \!-\! 3) 
\epsilon } \nonumber \\
& & \hspace{-0.5cm} - \frac{(D\!-\! 2) \epsilon^2}{2}\Bigr) \Biggr] \Gamma\Bigl(1 
\!-\! \frac{D}2\Bigr) + \frac{2 M^2}{16 \pi^2} \Bigl[ \psi\Bigl( \frac12 \!+\! 
\nu_v\Bigr) + \psi\Bigl( \frac12 \!-\! \nu_v\Bigr) \Bigr] + O( D\!-\! 4) \; . 
\qquad \label{2ndterm}
\end{eqnarray}
And the final term in (\ref{coinctrace3}) --- the one with derivatives --- becomes,
\begin{eqnarray}
\lefteqn{ \frac{H^2}{2 M^2} \Bigl( \partial_n \!+\! D \!-\! 1 \!-\! \epsilon\Bigr)
\Bigl(\partial_n \!+\! 2 D \!-\! 4\Bigr) \Bigl[ L_u \!-\! L_{u_0}\Bigr] = \frac{H^2}{2} 
\Bigl( \partial_n \!+\! D \!-\! 1 \!-\! \epsilon\Bigr) } \nonumber \\
& & \hspace{0cm} \times \Bigl(\partial_n \!+\! 2 D \!-\! 4\Bigr) \frac{[(1 \!-\! 
\epsilon) H]^{D-4}}{(4 \pi)^{\frac{D}2}} \Gamma\Bigl(1 \!-\! \frac{D}2\Bigr) + 
\frac{H^2}{32 \pi^2} \Bigl(\partial_n \!+\! 3 \!-\! \epsilon\Bigr) \Bigl( \partial_n 
\!+\! 4\Bigr) \qquad \nonumber \\
& & \hspace{1cm} \times \Biggl\{ \psi\Bigl( \frac12 \!+\! \nu_u\Bigr) + \psi\Bigl(
\frac12 \!-\! \nu_u\Bigr) - \frac{\epsilon H^2}{M^2} \Bigl[ \psi\Bigl(\frac12 \!+\! 
\nu_u\Bigr) - \psi\Bigl(\frac12 \!+\! \nu_{u_0}\Bigr) \nonumber \\
& & \hspace{4.5cm} + \psi\Bigl(\frac12 \!-\! \nu_u\Bigr) - \psi\Bigl(\frac12 \!-\! \nu_{u_0}
\Bigr) \Bigr] \Biggr\} + O(D \!-\! 4) \; . \qquad \label{3rdterm}
\end{eqnarray}
Note that the difference $\psi(\frac12 \pm \nu_u) - \psi(\frac12 \pm \nu_{u_0})$ 
is of order $M^2$ so expression (\ref{3rdterm}) has no $1/M^2$ pole. Note also that 
the $1/\epsilon$ pole in $\psi(\frac12 - \nu_{u_0}) = \psi(\frac{-\epsilon}{1 - \epsilon})$
is canceled by an explicit multiplicative factor of $\epsilon$.

The potentially divergent terms (the ones proportional to $\Gamma(1 - \frac{D}2)$) in 
expressions (\ref{1stterm}), (\ref{2ndterm} and (\ref{3rdterm}) sum to give,
\begin{eqnarray}
\lefteqn{ (\ref{1stterm})_{\rm div} + (\ref{2ndterm})_{\rm div} + 
(\ref{3rdterm})_{\rm div} = \frac{[(1 \!-\! \epsilon) H]^{D-4}}{(4\pi)^{\frac{D}2}}
\Bigl[ (D\!-\!1) M^2 + \frac12 R\Bigr] \Gamma\Bigl(1 \!-\! \frac{D}2\Bigr) }
\nonumber \\
& & \hspace{0.7cm} + \frac{H^2}{16 \pi^2} \Bigl[3 - 12 \epsilon + 4 \epsilon^2 - 2 
\epsilon' -\frac{(6 \epsilon' \!+\! \epsilon'')}{1 \!-\! \epsilon} - \Bigl( 
\frac{\epsilon'}{1 \!-\! \epsilon}\Bigr)^2 \Bigr] + O(D \!-\! 4) \; , \qquad
\end{eqnarray}
where we recall that the $D$-dimensional Ricci scalar is $R = (D-1) (D - 2 \epsilon) 
H^2$. Comparison with expression (\ref{DeltaVdef}) for $\Delta V'(\varphi \varphi^*)$
reveals that we can absorb the divergences with the following counterterms,
\begin{equation}
\delta \xi = -\frac{\Gamma(1 \!-\! \frac{D}2) s^{D-4}}{(4\pi)^{\frac{D}2}} \times 
\frac12 q^2 \qquad , \qquad \delta \lambda = -\frac{\Gamma(1 \!-\! \frac{D}2) 
s^{D-4}}{(4\pi)^{\frac{D}2}} \times 4 (D \!-\! 1) q^4 \; , \label{cterms}
\end{equation}
where $s$ is the renormalization scale. Up to finite renormalizations, these choices
agree with previous results \cite{Miao:2015oba,Prokopec:2007ak,Miao:2019bnq}, in the 
same gauge and using the same regularization, on de Sitter background.

Substituting expressions (\ref{1stterm}), (\ref{2ndterm}), (\ref{3rdterm}) and
(\ref{cterms}) into the definition (\ref{DeltaVdef}) of $\Delta V'(\varphi \varphi^*)$
and taking the unregulated limit gives the local contribution,
\begin{eqnarray}
\lefteqn{ \Delta V'_{\rm L}(\varphi \varphi^*) = \frac{q^2 H^2}{16 \pi^2} \Biggl\{ 
\frac{(6 M^2 \!+\! R)}{2 H^2} \ln\Bigl[ \frac{(1 \!-\! \epsilon)^2 H^2}{s^2}\Bigr] 
\!+\! 3 \!-\! 12 \epsilon \!+\! 4 \epsilon^2 \!-\! 2 \epsilon' \!-\! \frac{(6 \epsilon' 
\!+\! \epsilon'')}{1 \!-\! \epsilon} } \nonumber \\
& & \hspace{-0.5cm} - \Bigl( \frac{\epsilon'}{1 \!-\! \epsilon}\Bigr)^2 
+ \frac{M^2}{H^2} \Biggl[ \psi\Bigl( \frac12 \!+\! \nu_u\Bigr) +
\psi\Bigl( \frac12 \!-\! \nu_u\Bigr) + 2 \psi\Bigl( \frac12 \!+\! \nu_v\Bigr) + 2
\psi\Bigl( \frac12 \!-\! \nu_v\Bigr) \Biggr] \nonumber \\
& & \hspace{-0.5cm} + \frac12 \Bigl[ (\partial_n \!+\! 3 \!-\! \epsilon) (\partial_n 
\!+\! 4) - 2 \epsilon \Bigr] \Bigl[ \psi\Bigl( \frac12 \!+\! \nu_u\Bigr) \!+\! 
\psi\Bigl( \frac12 \!-\! \nu_u\Bigr) \Bigr] - (\partial_n \!+\! 3 \!-\! \epsilon)
(\partial_n \!+\! 4) \nonumber \\
& & \hspace{1.8cm} \times \frac{\epsilon H^2}{2 M^2} \Bigl[ \psi\Bigl( \frac12 
\!+\! \nu_u\Bigr) \!-\! \psi\Bigl( \frac1{1 \!-\! \epsilon}\Bigr) + \psi\Bigl( 
\frac12 \!-\! \nu_u\Bigr) \!-\! \psi\Bigl( \frac{-\epsilon}{1 \!-\! \epsilon}\Bigr) 
\Bigr] \Biggr\} . \qquad \label{VLprime}
\end{eqnarray}
It is worth noting that there are no singularities at $\epsilon = 1$, or when
either $1/(1 - \epsilon)$ or $-\epsilon/(1 - \epsilon)$ become non-positive integers
\cite{Janssen:2008px}. The effective potential is obtained by integrating 
(\ref{VLprime}) with respect to $\varphi \varphi^*$. The result is best expressed 
using the variable $z \equiv q^2 \varphi \varphi^*/H^2$,
\begin{eqnarray}
\lefteqn{ \Delta V_{\rm L} = \frac{H^4}{16 \pi^2} \Biggl\{\! \Bigl[
3 z^2 \!+\! \frac{R z}{2 H^2} \Bigr]\! \ln\Bigl[ \frac{ (1 \!-\! \epsilon)^2 H^2}{s^2}
\Bigr] \!+\! \Bigl[3 \!-\! 12 \epsilon \!+\! 4 \epsilon^2 \!-\! 2 \epsilon' \!-\! 
\frac{(6 \epsilon' \!+\! \epsilon'')}{1 \!-\! \epsilon} \Bigr] z } \nonumber \\
& & \hspace{-0.5cm} - \frac{{\epsilon'}^2 z}{(1 \!-\! \epsilon)^2} + 2 \! 
\int_{0}^{z} \!\!\! dx \, x \Biggl[ \psi\Bigl( \frac12 \!+\! \alpha\! \Bigr) 
\!+\! \psi\Bigl( \frac12 \!-\! \alpha \!\Bigr) \!+\! 2 \psi\Bigl( \frac12 \!+\! 
\beta \!\Bigr) \!+\! 2 \psi\Bigl( \frac12 \!-\! \beta \! \Bigr) \Biggr] 
\nonumber \\
& & \hspace{-0.5cm} + \frac12 \Bigl[ (\partial_n \!+\! 3 \!-\! 3 \epsilon) 
(\partial_n \!+\! 4 \!-\! 2 \epsilon) - 2 \epsilon\Bigr] \! \int_{0}^{z} \!\!\! dx 
\Bigl[ \psi\Bigl( \frac12 \!+\! \alpha(x)\Bigr) + \psi\Bigl( \frac12 \!-\! \alpha(x) 
\Bigr) \Bigr] \nonumber \\
& & \hspace{-0.5cm} - (\partial_n \!+\! 3 \!-\! 3 \epsilon) (\partial_n \!+\! 4 
\!-\! 2 \epsilon) \! \int_{0}^{z} \!\! \frac{dx \epsilon}{4 x} \Bigl[ \psi\Bigl( 
\frac12 \!+\! \alpha(x) \Bigr) - \psi\Bigl( \frac1{1 \!-\! \epsilon}\Bigr) 
\nonumber \\
& & \hspace{6.5cm} + \psi\Bigl( \frac12 \!-\! \alpha(x)\Bigr) - \psi\Bigl( 
\frac{-\epsilon}{1 \!-\! \epsilon}\Bigr) \Bigr] \Biggr\} , \qquad \label{VL}
\end{eqnarray}
where the $x$-dependent indices are,
\begin{equation}
\alpha(x) \equiv \sqrt{\frac14 + \frac{\epsilon \!-\! 2 x}{(1 \!-\! \epsilon)^2}}
\qquad , \qquad \beta(x) \equiv \sqrt{ \frac14 - \frac{2 x}{(1 \!-\! \epsilon)^2}} 
\; . \label{alphabeta}
\end{equation}
Note that the term inside the square brackets on the last line of (\ref{VL})
vanishes for $x = 0$, so the integrand is well defined at $x = 0$.

\subsection{Large Field \& Small Field Expansions}

\begin{figure}[H]
\centering
\begin{subfigure}[b]{0.5\textwidth}
\centering
\includegraphics[width=\textwidth]{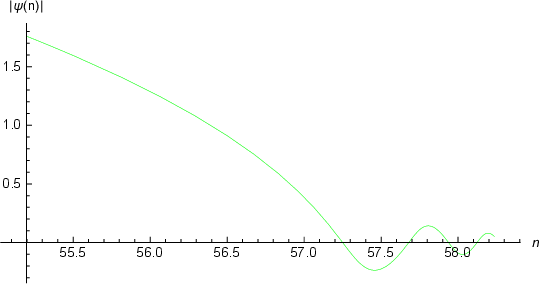}
\end{subfigure}
\begin{subfigure}[b]{0.5\textwidth}
\centering
\includegraphics[width=\textwidth]{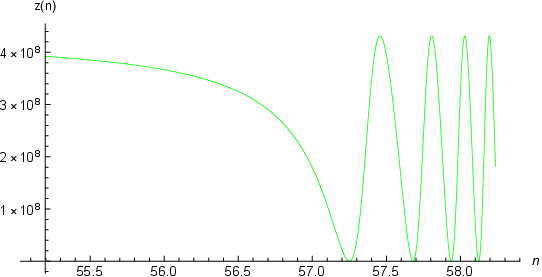}
\end{subfigure}
\caption[Inflaton field and the ratio $z \equiv q^2 \psi^2/\chi^2$ after the end of inflation]{\footnotesize Plots of the dimensionless inflaton field $\psi(n)$ 
and the ratio $z \equiv q^2 \psi^2/\chi^2$ after the end of inflation for
the quadratic potential. Here we chose $q^2 = \frac1{137}$.}
\label{psiandz}
\end{figure}

Expression (\ref{VL}) depends principally on the quantity $z = q^2 \varphi 
\varphi^*/H^2$. During inflation $z$ is typically quite large, whereas it 
touches $0$ after the end of inflation. Figure~\ref{psiandz} shows this 
for the quadratic potential, and the results are similar for the 
Starobinsky potential (\ref{StaroU}). It is therefore desirable to expand
the potential $\Delta V_L(\varphi \varphi^*)$ for large $z$ and for small $z$.

\begin{figure}[H]
\centering
\begin{subfigure}[b]{0.33\textwidth}
\centering
\includegraphics[width=\textwidth]{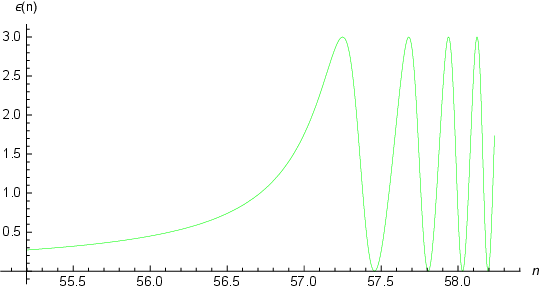}
\end{subfigure}
\begin{subfigure}[b]{0.33\textwidth}
\centering
\includegraphics[width=\textwidth]{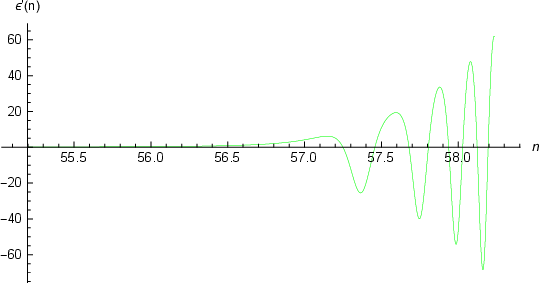}
\end{subfigure}
\begin{subfigure}[b]{0.33\textwidth}
\centering
\includegraphics[width=\textwidth]{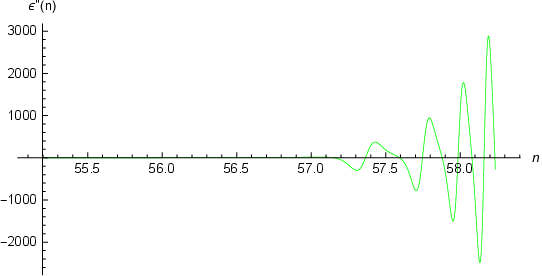}
\end{subfigure}
\caption[First slow roll and its derivatives after the end of inflation]{\footnotesize Plots of the first slow roll and its derivatives after
the end of inflation for the quadratic potential.}
\label{epsilon}
\end{figure}

The large field regime follows from the large argument expansion of the digamma
function,
\begin{equation}
\psi(x) = \ln(x) - \frac1{2 x} - \frac1{12 x^2} + \frac1{120 x^4} - \frac1{256 x^6}
+ O\Bigl( \frac1{x^8}\Bigr) \; . \label{psiexp}
\end{equation}
Substituting (\ref{psiexp}) in (\ref{VL}), and performing the various integrals
gives,
\begin{eqnarray}
\lefteqn{\Delta V_L = \frac{H^4}{16 \pi^2} \Biggl\{3 z^2 \ln\Bigl( \frac{2 q^2 
\varphi \varphi^*}{s^2}\Bigr) \!-\! \frac32 z^2 + \frac{R z}{2 H^2} \ln\Bigl( 
\frac{2 q^2 \varphi \varphi^*}{s^2}\Bigr) \!-\! (4 \!+\! 8 \epsilon \!-\! 3 
\epsilon^2) z } \nonumber \\
& & \hspace{0.5cm} -\epsilon' z - \Bigl[\frac34 \epsilon (1 \!-\! \epsilon) (2 \!-\! 
\epsilon) + \frac78 (1 \!-\! \epsilon) \epsilon' + \frac18 \epsilon''\Bigr]
\ln^2(2 z) + O\Bigl( \ln(z)\Bigr) \Biggr\} . \qquad \label{largezexp}
\end{eqnarray}
The leading contribution of (\ref{largezexp}) agrees with the famous flat space
result of Coleman and Weinberg \cite{Coleman:1973jx},
\begin{equation}
\Delta V \longrightarrow \frac{3 (q^2 \varphi \varphi^*)^2}{16 \pi^2} 
\ln\Bigl( \frac{2 q^2 \varphi \varphi^*}{s^2}\Bigr) \; .
\end{equation}
The first three terms of (\ref{largezexp}) could be subtracted using allowed 
counterterms of the form $F(\varphi \varphi^*,R)$ \cite{Woodard:2006nt}. A
prominent feature of the remaining terms is the presence of derivatives of
the first slow roll parameter. These derivatives are typically very small
during inflation but Figure~\ref{epsilon} shows that they can be quite large
after the end of inflation.

The small field expansion derives from expanding the digamma functions in
expression (\ref{VL}) in powers of $x$,
\begin{eqnarray}
\psi\Bigl( \frac12 \!+\! \alpha(x)\Bigr) & = & \psi\Bigl( \frac1{1 \!-\! \epsilon}
\Bigr) - \psi'\Bigl( \frac1{1 \!-\! \epsilon} \Bigr) \frac{2x}{1 \!-\! \epsilon^2}
+ O(x^2) \; , \\
\psi\Bigl( \frac12 \!-\! \alpha(x)\Bigr) & = & \psi\Bigl( \frac{-\epsilon}{1 \!-\! 
\epsilon}\Bigr) + \psi'\Bigl( \frac{-\epsilon}{1 \!-\! \epsilon} \Bigr) 
\frac{2x}{1 \!-\! \epsilon^2} + O(x^2) \; , \\
\psi\Bigl( \frac12 \!+\! \beta(x)\Bigr) & = & -\gamma - \frac{\pi^2}{6} \frac{2x}{
(1 \!-\! \epsilon)^2} + O(x^2) \; , \\
\psi\Bigl( \frac12 \!-\! \beta(x)\Bigr) & = & -\frac{(1 \!-\! \epsilon)^2}{2 x}
+ 1 -\gamma + \Bigl[1 \!+\! \frac{\pi^2}{6}\Bigr] \frac{2 x}{(1 \!-\! \epsilon)^2}
+ O(x^2) \; . \qquad
\end{eqnarray}
The result is,
\begin{eqnarray}
\lefteqn{ \Delta V_L = \frac{H^4}{16 \pi^2} \Biggl\{ \Biggl[ \frac{R}{2 H^2}
\ln\Bigl[ \frac{(1 \!-\! \epsilon)^2 H^2}{s^2} \Bigr] \!+\! 1 \!-\! 8 \epsilon 
\!+\! 2 \epsilon^2 \!-\! 2 \epsilon' - \frac{(6 \epsilon' \!+\! \epsilon'')}{1
\!-\! \epsilon} \!-\! \frac{{\epsilon'}^2}{(1 \!-\! \epsilon)^2} \Biggr] z }
\nonumber \\
& & \hspace{2cm} + \frac12 \Bigl[ (\partial_n \!+\! 3 \!-\! 3 \epsilon)
(\partial_n \!+\! 4 \!-\! 2 \epsilon) \!-\! 2 \epsilon\Bigr] \Bigl[
\psi\Bigl( \frac1{1 \!-\! \epsilon}\Bigr) \!+\! \psi\Bigl( \frac{-\epsilon}{1
\!-\! \epsilon}\Bigr) \Bigr] z \nonumber \\
& & \hspace{0cm} + \frac12 (\partial_n \!+\! 3 \!-\! 3 \epsilon) (\partial_n 
\!+\! 4 \!-\! 2 \epsilon) \Bigl[ \psi'\Bigl( \frac1{1 \!-\! \epsilon}\Bigr) 
\!-\! \psi'\Bigl( \frac{-\epsilon}{1 \!-\! \epsilon}\Bigr) \Bigr] 
\frac{\epsilon z}{1 \!-\! \epsilon^2} + O(z^2) \Biggr\} . \qquad 
\label{smallzexp}
\end{eqnarray}
Note that the $1/\epsilon$ pole from $\psi(\frac{-\epsilon}{1 - \epsilon})$ on
the penultimate line of (\ref{smallzexp}) cancels against the double pole from
$\psi'( \frac{-\epsilon}{1 - \epsilon})$ on the last line.

\subsection{The Nonlocal Contribution}

The nonlocal contribution to the effective potential is obtained by 
substituting the nonlocal contribution (\ref{nonlocalprop}) to each coincident
propagator in (\ref{coinctrace3}), and then into expression (\ref{DeltaVdef}),
\begin{eqnarray}
\lefteqn{ \Delta V'_{\rm N}(\varphi \varphi^*) = q^2 N_u(n) + 2 q^2 e^{-2n} 
N_v(n) } \nonumber \\
& & \hspace{4cm} + \frac{q^2 H^2}{2 M^2} \Bigl(\partial_n \!+\! 3 \!-\! 
\epsilon \Bigr) \Bigl(\partial_n \!+\! 4\Bigr) \Bigl[ N_u(n) \!-\! N_{u_0}(n)
\Bigr] . \qquad \label{nonlocal1}
\end{eqnarray}
The nonlocal contributions to the various propagators are,
\begin{eqnarray}
N_u(n) &\!\!\! = \!\!\!& \int_{0}^{\kappa_{n-4}} \!\! \frac{d\kappa \, 
\kappa^2}{32 \pi^3 G} \Biggl[e^{\mathcal{U}_{2,3}(n,\kappa,\mu)} - 
e^{\mathcal{U}_{1}(n,\kappa,\mu)} \Biggr] , \qquad \label{newNu} \\
N_{u_0}(n) &\!\!\! = \!\!\!& \int_{0}^{\kappa_{n-4}} \!\! \frac{d\kappa \, 
\kappa^2}{32 \pi^3 G} \Biggl[e^{\mathcal{U}_{2}(n,\kappa,0)} - 
e^{\mathcal{U}_{1}(n,\kappa,0)} \Biggr] , \qquad \label{newNu0} \\
N_v(n) &\!\!\! = \!\!\!& \int_{0}^{\kappa_{n-4}} \!\! \frac{d\kappa \, 
\kappa^2}{32 \pi^3 G} \Biggl[e^{\mathcal{V}_{2,3}(n,\kappa,\mu)} - 
e^{\mathcal{V}_{1}(n,\kappa,\mu)} \Biggr] . \qquad \label{newNv}
\end{eqnarray}
The nonlocal nature of these contributions derives from the integration over
$\kappa$, which can be converted to an integration over $n_{\kappa}$,
\begin{equation}
\kappa \equiv e^{n_{\kappa}} \chi(n_{\kappa}) \qquad \Longrightarrow \qquad
\frac{d\kappa}{\kappa} = \Bigl[ 1 \!-\! \epsilon(n_{\kappa})\Bigr] dn_{\kappa}
\; . \label{kappaton}
\end{equation}
After this is done, any factors of $\kappa$ depend on the earlier geometry.

A number of approximations result in huge simplification. First, note from 
Figures~\ref{Temp3phasesA} and \ref{Temp3phasesB} that the ultraviolet 
approximation (\ref{U1def}) for $\mathcal{U}(n,\kappa,\mu)$ is typically more
negative than the late time approximations (\ref{U2def}) and (\ref{U3def}).
Figures~\ref{Trans3phasesA} and \ref{Trans3phasesB} show that the same rule 
applies to $\mathcal{V}(n,\kappa,\mu)$. Hence we can write,
\begin{equation}
N_u(n) \simeq \int_{0}^{\kappa_{n-4}} \!\! \frac{d\kappa \, \kappa^2}{16 \pi^3 G} 
\, e^{\mathcal{U}_{2,3}(n,\kappa,\mu)} \qquad , \qquad 
N_v(n) \simeq \int_{0}^{\kappa_{n-4}} \!\! \frac{d\kappa \, \kappa^2}{16 \pi^3 G} 
\, e^{\mathcal{V}_{2,3}(n,\kappa,\mu)} \; . \label{simp1}
\end{equation} 
Second, because the temporal and transverse frequencies are nearly equal, we 
can write,
\begin{equation}
\omega^2_u(n,\mu) \simeq \omega^2_v(n,\mu) \qquad \Longrightarrow \qquad
\mathcal{U}(n,\kappa,\mu) \simeq \mathcal{V}(n,\kappa,\mu) - 2 n \; . \label{simp2}
\end{equation}
When the mass vanishes there is so little difference between the ultraviolet
approximation (\ref{U1def}) and its late time extension (\ref{U2def}) that we
can ignore this contribution, $N_{u_0}(n) \simeq 0$. Next, Figures~\ref{Tempkdep} 
and \ref{Transkdep} imply that the late time approximations for 
$\mathcal{U}(n,\kappa,\mu)$ and $\mathcal{V}(n,\kappa,\mu)$ inherit their $\kappa$
dependence from the ultraviolet approximation at $n \simeq n_{\kappa} + 4$, which 
is itself independent of $\mu$,
\begin{equation}
n > n_{\kappa} + 4 \qquad \Longrightarrow \qquad \mathcal{U}_{2,3}(n,\kappa,\mu)
\simeq \mathcal{U}_1(n_{\kappa} \!+\! 4,\kappa,0) + f_{2,3}(n,\mu) \; ,
\label{simp4}
\end{equation}
where $f_{2,3}(n,\mu)$ can be read off from expressions (\ref{U2def}) and (\ref{U3def})
by omitting the $\kappa$-dependent integration constants. Finally, we can use the 
slow roll form (\ref{T2def}) for the amplitude reached after first horizon crossing 
and before the mass dominates,
\begin{equation}
e^{\mathcal{U}_1(n_{\kappa} + 4,\kappa,0)} \simeq \frac{\chi^2(n_{\kappa})}{2 \kappa^3}
\times C\Bigl( \epsilon(n_{\kappa}) \Bigr) 
\; .
\end{equation}
Putting it all together gives,
\begin{eqnarray}
\lefteqn{\Delta V_N(\varphi \varphi^*) \simeq 3 q^2 \!\! \int_{0}^{n-4} 
\!\!\!\!\!\!\! d n_{\kappa} \frac{ [1 \!-\! \epsilon(n_{\kappa})] \chi^2(n_{\kappa})
C(n_{\kappa}))}{32 \pi^3 G} \times e^{f_{2,3}(n,\mu)} } \nonumber \\
& & \hspace{-0.5cm} + \frac{q^2 \chi^2(n)}{2 \mu^2} ( \partial_n \!+\! 3 \!-\!
\epsilon )( \partial_n \!+\! 4) \!\! \int_{0}^{n-4} \!\!\!\!\!\! 
d n_{\kappa} \frac{ [1 \!-\! \epsilon(n_{\kappa})] \chi^2(n_{\kappa})
C(\epsilon(n_{\kappa}))}{32 \pi^3 G} \!\times\! e^{f_{2,3}(n,\mu)} \!. 
\qquad \label{nonlocal2}
\end{eqnarray}

\section{Conclusions}

In section 2 we derived an exact, dimensionally regulated, Fourier mode 
sum (\ref{propagator}) for the Lorentz gauge propagator of a massive photon on 
an arbitrary cosmological background (\ref{geometry1}). Our result is expressed 
in terms of mode functions $t(\eta,k)$, $u(\eta,k,M)$ and $v(\eta,k,M)$ whose 
defining relations are (\ref{MMCeqn}), (\ref{tempeqn}) and (\ref{spaceeqn}), which
respectively represent massless minimally coupled scalars, massive temporal
photons, and massive spatially transverse photons. The photon propagator can 
also be expressed as a sum (\ref{scalarsum}) of bi-vector differential 
operators acting on the scalar propagators $i\Delta_t(x;x')$, $i\Delta_u(x;x')$ 
and $i\Delta_v(x;x')$ associated with the three mode functions. Because Lorentz 
gauge is an exact gauge there should be no linearization instability, even on 
de Sitter, such as occurs for Feynman gauge \cite{Kahya:2005kj,Kahya:2006ui}.

In section 3 we converted to a dimensionless form with time represented 
by the number of e-foldings $n$ since the beginning of inflation, and the wave 
number, mass and Hubble parameter all expressed in reduced Planck units, $\kappa 
\equiv \sqrt{8\pi G} \,k$, $\mu \equiv \sqrt{8\pi G} \, M$ and $\chi(n) \equiv 
\sqrt{8 \pi G} \, H(\eta)$. Analytic approximations were derived for the amplitudes
$\mathcal{T}(n,\kappa)$, $\mathcal{U}(n,\kappa,\mu)$ and $\mathcal{V}(n,\kappa,\mu)$
associated with each of the mode functions. Which approximation to use is controlled 
by first horizon crossing at $\kappa = e^{n_{\kappa}} \chi(n_{\kappa})$ and mass
domination at $\mu = \frac12 \chi(n_{\mu})$. Until shortly after first horizon
crossing we employ the ultraviolet approximations (\ref{T1def}), (\ref{U1def}) and
(\ref{V1def}). After first horizon crossing and before mass domination the 
appropriate approximations are (\ref{T2def}), (\ref{U2def}) and (\ref{V2def}).
And after mass domination (which $\mathcal{T}(n,\kappa)$ never experiences) the
amplitudes are well approximated by (\ref{U3def}) and (\ref{V3def}). The validity
of these approximations was checked against explicit numerical solutions for 
inflation driven by the simple quadratic model, and by the phenomenologically 
favored plateau model (\ref{StaroU}).

In section 4 we applied our approximations to compute the effective potential 
induced by photons coupled to a charged inflaton. Our result consists of a 
part (\ref{VL}) which depends locally on the geometry (\ref{geometry1}) and a
numerically smaller part (\ref{nonlocal2}) which depends on the past history.
The local part was expanded both for the case of large field strength 
(\ref{largezexp}), and for small field strength (\ref{smallzexp}). The 
existence of the second, nonlocal contribution, was conjectured on the 
basis of indirect arguments \cite{Miao:2015oba} that have now been explicitly 
confirmed. Another conjecture that has been confirmed is the rough validity of
extrapolating de Sitter results \cite{Allen:1983dg,Prokopec:2007ak} from the
constant Hubble parameter of de Sitter background to the time dependent one
of a general cosmological background (\ref{geometry1}). However, we now have
good approximations for the dependence on the first slow roll parameter 
$\epsilon(n)$. 

Our most important result is probably the fact that electromagnetic
corrections to the effective potential depend upon first and second derivatives
of the first slow roll parameter. One consequence is that the effective
potential from electromagnetism responds more strongly to changes in the 
geometry than for scalars \cite{Kyriazis:2019xgj} or spin one half 
fermions \cite{Sivasankaran:2020dzp}. This can be very important
during reheating (see Figure~\ref{epsilon}); it might also be significant if 
features occur during inflation. Another consequence is that there cannot be 
perfect cancellation between the positive effective potentials induced by 
bosons and the negative potentials induced by fermions \cite{Miao:2020zeh}.
Note that the derivatives of $\epsilon$ come exclusively from the 
constrained part of the photon propagator --- the $t(\eta,k)$ and $u(\eta,k)$
modes --- which is responsible for long range electromagnetic interactions. 
Dynamical photons --- the $v(\eta,k)$ modes --- produce no derivatives
at all. These statements can be seen from expression (\ref{coinctrace3}), 
which is exact, independent of any approximation.

We close with a speculation based on the correlation between the spin of 
the field and the number of derivatives it induces in the effective potential: 
scalars produce no derivatives \cite{Kyriazis:2019xgj}, spin one half fermions 
induce one derivative \cite{Sivasankaran:2020dzp}, and this paper has shown
that spin one vectors give two derivatives. It would be interesting to see if 
the progression continues for gravitinos (which ought to induce three derivatives)
and gravitons (which would induce four derivatives). Of course gravitons do not 
acquire a mass through coupling to a scalar inflaton, but they do respond to it, 
and the mode equations have been derived in a simple gauge \cite{Iliopoulos:1998wq,
Abramo:2001dc}. Until now it was not possible to do much with this system because
it can only be solved exactly for the case of constant $\epsilon(n)$, however,
we now have a reliable approximation scheme that can be used for arbitrary
$\epsilon(n)$. Further, we have a worthy object of study in the graviton 1-point
function, which defines how quantum 0-point fluctuations back-react to change
the classical geometry. At one loop order it consists of the same sort of 
coincident propagator we have studied in this paper. On de Sitter background the
result is just a constant times the de Sitter metric \cite{Tsamis:2005je}, which
must be absorbed into a renormalization of the cosmological constant if ``$H$''
is to represent the true Hubble parameter. Now suppose that the graviton 
propagator for general first slow roll parameter consists of a local part with 
up to 4th derivatives of $\epsilon(n)$ plus a nonlocal part. That sort of result
could {\it not} be absorbed into any counterterm. So perhaps there is one loop
back-reaction after all \cite{Geshnizjani:2002wp}, and de Sitter represents a 
case of unstable equilibrium?

\chapter{Reheating with Effective Potential} 
\label{reheatingChapter}
\counterwithin{algorithm}{chapter}
\section{Introduction}

Scalar-driven inflation is supported by the slow roll of the inflaton down
its potential.\footnote{This chapter has been adapted from a published article in JCAP\cite{Katuwal:2022szw}.} At the end of inflation the inflaton begins oscillating, and
its kinetic energy is transferred to ordinary matter during the process of
reheating. The efficiency of this transfer obviously depends on the way the
inflaton is coupled to ordinary matter. Ema et al. have shown that the most
efficient coupling is that of a charged inflaton to electromagnetism
\cite{Ema:2016dny}. 

What happens is that the evolution of a charged inflaton induces a 
time-dependent photon mass which oscillates around zero during reheating. 
The temporal and longitudinal components of the photon diverge as the
mass goes to zero, which makes reheating very efficient. The process has 
been previously studied by discretizing space, carrying out a finite Fourier 
transform, and then numerically evolving the nonlinear system of the inflaton 
plus electromagnetism \cite{Bezrukov:2020txg}. However, the energy transfer
is broadly distributed over so many modes that there is little point to 
including nonlinear effects in the photon field, provided that its response 
to the inflaton $0$-mode is known to all orders. In that case, one merely 
sums the contribution from each photon mode's wave vector, which can be 
accomplished by varying the inflaton effective potential. The goal of this 
chapter is to develop a good analytic approximation for the massive photon 
propagator in a time-dependent inflaton background, and then use it to 
compute the quantum-induced, effective force in equation for the inflaton 
$0$-mode. In this way reheating can be studied by numerically solving a 
nonlocal equation for the inflaton $0$-mode.

This chapter consists of five sections, of which the first is this Introduction.
In section 2 we derive a spatial Fourier mode sum for the massive photon
propagator which is valid when the mass becomes time-dependent. Section 3
develops analytic approximations for the temporal and longitudinal modes,
checking them against explicit numerical analysis for a simple model of 
inflation. In section 4 we discuss how these approximations can be used to 
estimate the quantum-induced effective force which controls the process 
of reheating. Section 5 gives our conclusions.

\section{The Massive Photon Propagator}

The purpose of this section is to generalize the massive photon propagator 
from its known form for a constant mass \cite{Katuwal:2021kry} to the case of
a time-dependent mass. The Lagrangian is,
\begin{eqnarray}
\lefteqn{\mathcal{L} = -\frac14 F_{\mu\nu} F_{\rho\sigma} g^{\mu\rho} g^{\nu\sigma} 
\sqrt{-g} } \nonumber \\
& & \hspace{2cm} - \Bigl( \partial_{\mu} \!-\! i q A_{\mu}\Bigr) \varphi \Bigl(
\partial_{\nu} \!+\! i q A_{\nu} \Bigr) \varphi^* g^{\mu\nu} \sqrt{-g} - 
V(\varphi \varphi^*) \sqrt{-g} \; , \qquad \label{Lagrangian}
\end{eqnarray}
where $\varphi$ is the inflaton and $F_{\mu\nu} \equiv \partial_{\mu} A_{\nu} - 
\partial_{\nu} A_{\mu}$ is the electromagnetic field strength. We work on a 
general homogeneous, isotropic and spatially flat geometry in $D$-dimensional, 
conformal coordinates, with Hubble parameter $H$ and first slow roll parameter 
$\epsilon$,
\begin{equation}
ds^2 = a^2 \Bigl[-d\eta^2 + d\vec{x} \!\cdot\! d\vec{x}\Bigr] \qquad , \qquad
H \equiv \frac{\partial_0 a}{a^2} \quad , \quad \epsilon \equiv -
\frac{\partial_0 H}{a H^2} \; . \label{geometry}
\end{equation}
The section first reviews the constant mass case, and then makes the 
generalizations necessary to incorporate a time-dependent mass.

\subsection{Constant Mass}

When the photon's mass is constant its propagator $i[\mbox{}_{\mu} 
\Delta_{\rho}](x;x')$ is transverse,
\begin{equation}
\partial_{\mu} \Bigl\{ \!\sqrt{-g(x)} \, g^{\mu\nu}(x) \, i\Bigl[\mbox{}_{\nu}
\Delta_{\rho}\Bigr](x;x')\Bigr\} = 0 = \partial'_{\rho} \Bigl\{\! \sqrt{-g(x')} \,
g^{\rho\sigma}(x') \, i\Bigl[\mbox{}_{\mu} \Delta_{\sigma}\Bigr](x;x') \Bigr\} .
\label{transverse}
\end{equation}
Its propagator equation reflects this transversality \cite{Katuwal:2021kry,
Tsamis:2006gj},
\begin{eqnarray}
\lefteqn{ \sqrt{-g} \Bigl[ \square^{\mu\nu} - R^{\mu\nu} - M^2 g^{\mu\nu}\Bigr]
i \Bigl[\mbox{}_{\nu} \Delta_{\rho}\Bigr](x;x') } \nonumber \\
& & \hspace{3.5cm} = \delta^{\mu}_{~\rho} i\delta^D(x \!-\! x') + \sqrt{-g(x)} \,
g^{\mu\nu}(x) \partial_{\nu} \partial'_{\rho} i\Delta(x;x') \; . \qquad
\label{constMprop}
\end{eqnarray}
Here $\square^{\mu\nu}$ is the vector d'Alembertian, $R^{\mu\nu}$ is the Ricci 
tensor and $i\Delta(x;x')$ is the propagator of a massless, minimally coupled 
scalar,
\begin{equation}
\partial_{\mu} \Bigl[\sqrt{-g} \, g^{\mu\nu} \partial_{\nu} i\Delta(x;x') \Bigr]
= i\delta^D(x \!-\! x') \; . \label{MMCSprop}
\end{equation}

The solution to (\ref{transverse}-\ref{constMprop}) can be expressed as a spatial
Fourier mode sum over three sorts of polarizations \cite{Katuwal:2021kry},
\begin{eqnarray}
\lefteqn{ i\Bigl[\mbox{}_{\mu} \Delta_{\rho}\Bigr](x;x') = \int \!\! 
\frac{d^{D-1}k}{(2\pi)^{D-1}} \sum_{\lambda = t,u,v} s_{\lambda} \Biggl\{ 
\theta(\Delta \eta) \mathcal{A}_{\mu}(x;\vec{k},\lambda) \mathcal{A}_{\nu}^*(x';
\vec{k},\lambda) } \nonumber \\
& & \hspace{6cm} + \theta(-\Delta \eta) \mathcal{A}_{\mu}^*(x;\vec{k},\lambda)
\mathcal{A}_{\nu}(x';\vec{k},\lambda) \Biggr\} , \qquad \label{constMmodesum}
\end{eqnarray}
where $\Delta \eta \equiv \eta - \eta'$. Longitudinal photons correspond to
$\lambda = t$ and have $s_t = -1$ with,
\begin{equation}
\mathcal{A}_{\mu}(x;\vec{k},t) = \frac{\partial_{\mu}}{M} \Bigl[ t(\eta,k)
e^{i\vec{k} \cdot \vec{x}} \Bigr] \;\; , \;\; \Bigl[ \mathcal{D} \partial_0 +
k^2\Bigr] t = 0 \;\; , \;\; t \cdot \partial_0 t^* - \partial_0 t \cdot t^* = 
\frac{i}{a^{D-2}} , \label{tmodes}
\end{equation}
where $\mathcal{D} \equiv \partial_0 + (D-2) a H$. Temporal photons correspond
to $\lambda = u$ and have $s_u = +1$ with,
\begin{eqnarray}
\mathcal{A}_{\mu}(x;\vec{k},u) = \frac{\overline{\partial}_{\mu}}{M} \Bigl[ 
u(\eta,k) e^{i\vec{k} \cdot \vec{x}} \Bigr] & , & \overline{\partial}_0 \equiv k
\;\; , \;\; \overline{\partial}_m \equiv \frac{-i k_m}{k} \mathcal{D} \; , 
\qquad \\
\Bigl[ \partial_0 \mathcal{D} + k^2 + a^2 M^2\Bigr] u = 0 & , & u 
\cdot \partial_0 u^* - \partial_0 u \cdot u^* = \frac{i}{a^{D-2}} . \qquad
\label{umodes}
\end{eqnarray}
Transverse spatial photons correspond to $\lambda = v$ and have $s_v = +1$ with,
\begin{eqnarray}
\mathcal{A}_{\mu}(x;\vec{k},v) = \epsilon_{\mu}(\vec{k},v) \, v(\eta,k) 
e^{i\vec{k} \cdot \vec{x}} & , & \epsilon_0 = 0 \;\; , \;\; k_m \epsilon_m = 0 
\; , \qquad \label{vmodesA} \\
\Bigl[ \partial^2_0 + (D \!-\! 4) a H \partial_0 + k^2 + a^2 M^2\Bigr] v = 0 & , & 
v \cdot \partial_0 v^* - \partial_0 v \cdot v^* = \frac{i}{a^{D-4}} , \qquad
\label{vmodesB}
\end{eqnarray}
where the sum over the $(D-2)$ spatial polarizations gives,
\begin{equation}
\sum_{v} \epsilon_i(\vec{k},v) \times \epsilon^*_j(\vec{k},v) = \delta_{ij} - 
\frac{k_i k_j}{k^2} \; . \label{polsum}
\end{equation}

\subsection{Time-Dependent Mass}

To understand the case of a time-dependent mass we must consider the vector
and scalar field equations,
\begin{eqnarray}
\lefteqn{ \frac{\delta S}{\delta A_{\mu}} = \partial_{\nu} \Bigl[ \sqrt{-g} \,
g^{\nu\rho} g^{\mu\sigma} F_{\rho\sigma} \Bigr] } \nonumber \\
& & \hspace{2.5cm} + iq \Bigl[ \varphi \!\cdot\! \Bigl( \partial_{\nu} \!+\! i q
A_{\nu}\Bigr) \varphi^* - \Bigl( \partial_{\nu} \!-\! i q A_{\nu}\Bigr) \varphi
\!\cdot\! \varphi^* \Bigr] g^{\mu\nu} \sqrt{-g} \; , \qquad \label{vector} \\
\lefteqn{ \frac{\delta S}{\delta \varphi^*} = \Bigl(\partial_{\mu} \!-\! i q
A_{\mu}\Bigr) \Bigl[ \sqrt{-g} \, g^{\mu\nu} \Bigl(\partial_{\nu} \!-\! i q
A_{\nu}\Bigr) \varphi \Bigr] - \varphi V'(\varphi \varphi^*) \sqrt{-g} \; .} 
\label{scalar}
\end{eqnarray}
The $0$-th order inflaton is $\varphi_0(\eta)$ which is real and obeys 
the equation,
\begin{equation}
\partial_0 \Bigl[ a^{D-2} \partial_0 \varphi_0\Bigr] + a^{D} \varphi_0
V'(\varphi_0^2) = 0 \; . \label{0order}
\end{equation}
The first order perturbations are $A_{\mu}(x)$ and the real fields 
$\alpha(x)$ and $\beta(x)$,
\begin{equation}
\varphi(x) = \varphi_0(\eta) + \alpha(x) + i \beta(x) \; .
\end{equation}
The first order contribution to the vector equation (\ref{vector}) is,
\begin{equation}
\partial_{\nu} \Bigl[ \sqrt{-g} \, g^{\nu\rho} g^{\mu\sigma} F_{\rho\sigma}
\Bigr] - 2 q^2 \varphi_0^2 \Bigl[ A_{\nu} - \partial_{\nu} \Bigl( 
\frac{\beta}{q \varphi_0}\Bigr) \Bigr] \sqrt{-g} \, g^{\nu\mu} = 0 \; . 
\label{vector1}
\end{equation}
The photon mass is $M^2 \equiv 2 q^2 \varphi_0^2$. Note from equation 
(\ref{vector1}) that antisymmetry of the field strength tensor implies,
\begin{equation} 
\partial_{\mu} \Bigl[ M^2 \sqrt{-g} \, g^{\mu\nu} \Bigl(A_{\nu} - 
\partial_{\nu} \Bigl( \frac{\beta}{q \varphi_0} \Bigr) \Bigr] = 0 \; . 
\label{betaeqn}
\end{equation}
This constraint is identical to the imaginary part of the first order
contribution to the scalar equation (\ref{scalar}). The analogous real 
part is,
\begin{equation}
\partial_{\mu} \Bigl[ \sqrt{-g} \, g^{\mu\nu} \partial_{\nu} \alpha\Bigr] 
- \sqrt{-g} \Bigl[ V'(\varphi_0^2) + 2 \varphi_0^2 V''(\varphi_0^2)\Bigr] 
\alpha = 0 \; . \label{alphaeqn}
\end{equation}

Relations (\ref{vector1}) and (\ref{betaeqn}) demonstrate that the Higgs 
mechanism continues to function when the scalar background $\varphi_0$ 
depends upon spacetime. To simplify the subsequent analysis, we will
absorb (``eat'') the imaginary part of the scalar perturbation into the 
vector field as usual,
\begin{equation}
A_{\mu} - \partial_{\mu} \Bigl( \frac{\beta}{q \varphi_0} \Bigr) 
\longrightarrow A_{\mu} \; . \label{eating}
\end{equation}
We can also use the conformal coordinate relation $g_{\mu\nu} = a^2 
\eta_{\mu\nu}$ to provide simple expressions for (\ref{vector1}) and
(\ref{betaeqn}),
\begin{equation}
\partial_{\nu} \Bigl[ a^{D-4} F^{\nu\mu} \Bigr] - M^2 a^{D-2} A^{\mu} 
= 0 \qquad \Longrightarrow \qquad \partial_{\mu} \Bigl[ M^2 a^{D-2} 
A^{\mu} \Bigr] = 0 \; , \label{vector2} 
\end{equation}
where $F^{\nu\mu} \equiv \eta^{\nu\rho} \eta^{\mu\sigma} F_{\rho\sigma}$
and $A^{\mu} \equiv \eta^{\mu\nu} A_{\nu}$. The $3+1$ decomposition of
the constraint on the right hand side of (\ref{vector2}) is,
\begin{equation}
\Bigl[ \mathcal{D} + \frac{2 \partial_0 M}{M}\Bigr] A_0 - 
\partial_m A_m = 0 \qquad , \qquad \mathcal{D} \equiv \partial_0 + 
(D\!-\!2) a H \; . \label{constraint}
\end{equation}
Relation (\ref{constraint}) permits us to $3+1$ decompose the left hand 
side of (\ref{vector2}) to,
\begin{eqnarray}
\Bigl[ \partial_0 \Bigl( \mathcal{D} + \frac{2 \partial_0 M}{M} \Bigr)
- \nabla^2 + a^2 M^2\Bigr] A_0 &\!\!\! = \!\!\!& 0 \; , 
\label{vector3A} \\
2 \Bigl( a H + \frac{\partial_0 M}{M}\Bigr) \partial_m A_0 + \Bigl[
\partial_0^2 + (D\!-\!4) a H \partial_0 - \nabla^2 + a^2 M^2\Bigr] A_m
&\!\!\! = \!\!\!& 0 \; . \label{vector3B}
\end{eqnarray}

Equations (\ref{constraint}-\ref{vector3B}) are satisfied by three 
polarizations of spatial plane waves whose associated mode functions
are $t(\eta,k)$, $u(\eta,k)$ and $v(\eta,k)$. Our notation is that a
``tilde'' over a differential operator such as $\partial_0$ or $\mathcal{D}$
indicates the addition of $\partial_0 M/M$, whereas a ``hat'' denotes
subtraction of the same quantity,
\begin{equation}
\widetilde{\mathcal{D}} \equiv \mathcal{D} + \frac{\partial_0 M}{M} 
\qquad , \qquad \widehat{\partial}_0 \equiv \partial_0 - 
\frac{\partial_0 M}{M} \; . \label{notation}
\end{equation}
What we term {\it Longitudinal photons} have the form,
\begin{equation}
\mathcal{A}_0(x;\vec{k},t) = \frac{\widehat{\partial}_0 t(\eta,k)}{M(\eta)} \,
e^{i \vec{k} \cdot \vec{x}} \qquad , \qquad \mathcal{A}_m(x;\vec{k},t) =
\frac{i k_m t(\eta,k)}{M(\eta)} \, e^{i \vec{k} \cdot \vec{x}} \; ,
\label{tAs}
\end{equation}
where the mode function $t(\eta,k)$ obeys,\footnote{Although 
$\mathcal{A}_{\mu}(x;\vec{k},t)$ satisfies (\ref{constraint}), it does not 
quite obey equations (\ref{vector3A}-\ref{vector3B}), but rather the relation
$\partial_{\nu} [a^{D-4} \mathcal{F}^{\nu\mu}(x;\vec{k},t)] = 0$.}
\begin{equation}
\Bigl[ \widetilde{\mathcal{D}} \widehat{\partial}_0 + k^2\Bigr] t = 0 \qquad , 
\qquad t \cdot \partial_0 t^* - \partial_0 t \cdot t^* = \frac{i}{a^{D-2}} \; .
\label{teqn}
\end{equation}
{\it Temporal photons} take the form,
\begin{equation}
\mathcal{A}_0(x;\vec{k},u) = \frac{k u(\eta,k)}{M(\eta)} \, e^{i \vec{k} \cdot \vec{x}} 
\qquad , \qquad \mathcal{A}_m(x;\vec{k},u) = - \frac{i k_m \widetilde{\mathcal{D}} 
u(\eta,k)}{k M(\eta)} \, e^{i \vec{k} \cdot \vec{x}} \; , \label{uAs}
\end{equation}
where the mode function $u(\eta,k)$ obeys,
\begin{equation}
\Bigl[ \widehat{\partial}_0 \widetilde{\mathcal{D}} + k^2 + a^2 M^2 \Bigr] u = 0 
\qquad , \qquad u \cdot \partial_0 u^* - \partial_0 u \cdot u^* = \frac{i}{a^{D-2}} 
\; . \label{ueqn}
\end{equation}
The tendency for longitudinal and temporal photons to diverge when the mass 
$M(\eta)$ passes through zero is obvious from expressions (\ref{tAs}) and 
(\ref{uAs}). In contrast, the time-dependent mass makes no change at all in 
relations (\ref{vmodesA}-\ref{polsum}) for the {\it Transverse spatial photons}, 
and these polarizations remain finite as the mass passes through zero.

A time-dependent mass makes no change in mode sum (\ref{constMmodesum}) for 
the propagator. However, the propagator obeys a revised version of the
constraint equation (\ref{transverse}),
\begin{equation}
\partial^{\mu} \Bigl\{a^{D-2} M^2 i \Bigl[\mbox{}_{\mu} \Delta_{\rho}\Bigr](x;x')
\Bigr\} = 0 = \partial^{\prime \rho} \Bigl\{ {a'}^{D-2} {M'}^2 i\Bigl[ \mbox{}_{\mu}
\Delta_{\rho}\Bigr](x;x') \Bigr\} \; . \label{newconstraint}
\end{equation}
The propagator equations analogous to (\ref{constMprop}-\ref{MMCSprop}) can be
given in terms of the massive photon kinetic operator,
\begin{equation}
\mathcal{D}^{\mu\nu} \equiv \partial_{\alpha} \Bigl[ a^{D-4} \Bigl( \eta^{\mu\nu}
\partial^{\alpha} - \eta^{\alpha\nu} \partial^{\mu}\Bigr) \Bigr] - a^{D-2} M^2
\eta^{\mu\nu} \; . \label{kineticop}
\end{equation}
The revised versions of (\ref{constMprop}-\ref{MMCSprop}) are,
\begin{eqnarray}
\mathcal{D}^{\mu\nu} i\Bigl[\mbox{}_{\nu} \Delta_{\rho}\Bigr](x;x') & \!\!\! =
\!\!\!& \delta^{\mu}_{~\rho} i\delta^D(x \!-\! x') + \frac{a^{D-2} M}{M'}
\widehat{\partial}^{\mu} \widehat{\partial}'_{\rho} i \Delta_{t}(x;x') \; ,
\qquad \label{newprop} \\
\frac1{M} \partial^{\mu} \Bigl[ a^{D-2} M \widehat{\partial}_{\mu} 
i\Delta_{t}(x;x') \Bigr] &\!\!\! = \!\!\!& i\delta^D(x \!-\! x') \; . \qquad
\label{newtprop}
\end{eqnarray}

\section{Approximating the Amplitudes}

The purpose of this section is to develop analytic approximations for the
crucial mode functions $t(\eta,k)$ and $u(\eta,k)$. We begin by giving a
dimensionless formulation of the problem. This formalism is then employed
to derive good analytic approximations for first, the longitudinal amplitude
and then, the temporal amplitude. At each stage these approximations are
checked against explicit numerical evolution in a simple mode of inflation.

\subsection{Dimensionless Formulation}

It is best to change the evolution variable from conformal time $\eta$ to
the number of e-foldings from the start of inflation, $n \equiv \ln[a(\eta)]$,
\begin{equation}
\partial_0 = a H \frac{\partial}{\partial n} \qquad , \qquad \partial_0^2
= a^2 H^2 \Bigl[ \frac{\partial^2}{\partial n^2} + (1 \!-\! \epsilon)
\frac{\partial}{\partial n}\Bigr] \; .
\end{equation}
We can also use factors of $8 \pi G$ to make the inflaton, the Hubble parameter
and the scalar potential dimensionless,
\begin{equation}
\psi(n) \equiv \sqrt{8\pi G} \, \varphi_0(\eta) \quad , \quad
\chi(n) \equiv \sqrt{8\pi G} \, H(\eta) \quad , \quad
U(\psi^2) \equiv (8\pi G)^2 V(\varphi^2_0) \; . \label{dimgeom}
\end{equation}
This gives dimensionless forms for the classical Friedmann equations, and for
the inflaton evolution equation, 
\begin{eqnarray}
\frac12 (D \!-\! 2) (D \!-\! 1) \chi^2 &\!\!\! = \!\!\!& \chi^2 {\psi'}^2 + 
U(\psi^2) \; , \qquad \label{Friedmann1} \\
-\frac12 (D\!-\!2) \Bigl[ (D\!-\! 1) - 2 \epsilon\Bigr] \chi^2 &\!\!\! = 
\!\!\!& \chi^2 {\psi'}^2 - U(\psi^2) \; , \qquad \label{Friedmann2} \\
0 &\!\!\! = \!\!\!& \chi^2 \Bigl[ \psi'' + (D\!-\!1\!-\!\epsilon) \psi'\Bigr] 
+ \psi U'(\psi^2) \; . \qquad \label{inflatoneqn}
\end{eqnarray}

Factors of $8\pi G$ can be extracted to give similar dimensionless forms for
the time-dependent mass $M^2(\eta) \equiv 2 q^2 \varphi_0^2(\eta)$ and the wave
number $k^2$,
\begin{equation}
\mu^2(n) \equiv 8\pi G M^2(\eta) = 2 q^2 \psi^2(n) \qquad , \qquad \kappa^2
\equiv 8 \pi G k^2 \; . \label{dimparams}
\end{equation}
We define the dimensionless Longitudinal and Temporal amplitudes as,
\begin{equation}
\mathcal{T}(n,\kappa) \equiv \ln\Bigl[ \frac{\vert t(\eta,k)\vert^2}{
\sqrt{8\pi G}}\Bigr] \qquad , \qquad \mathcal{U}(n,\kappa) \equiv
\ln\Bigl[\frac{\vert u(\eta,k)\vert^2}{\sqrt{8\pi G}}\Bigr] \; . \label{Amps}
\end{equation}
By combining the mode equations and Wronskians (\ref{teqn}) and (\ref{ueqn}) 
for each mode we can infer a single nonlinear relation for the associated
amplitudes \cite{Romania:2011ez,Romania:2012tb,Brooker:2015iya},
\begin{eqnarray}
\mathcal{T}'' + \frac12 {\mathcal{T}'}^2 + (D \!-\! 1 \!-\! \epsilon) 
\mathcal{T}' + \frac{2 \kappa^2 e^{-2n}}{\chi^2} + \frac{2 \mu_t^2}{\chi^2}
- \frac{e^{-2 [\mathcal{T} + (D-1) n]}}{2 \chi^2} &\!\!\! = \!\!\! & 0 
\; , \qquad \label{Teqn} \\
\mathcal{U}'' + \frac12 {\mathcal{U}'}^2 + (D \!-\! 1 \!-\! \epsilon) 
\mathcal{U}' + \frac{2 \kappa^2 e^{-2n}}{\chi^2} + \frac{2 \mu_u^2}{\chi^2}
- \frac{e^{-2 [\mathcal{U} + (D-1) n]}}{2 \chi^2} &\!\!\! = \!\!\! & 0 
\; , \qquad \label{Ueqn}
\end{eqnarray}
where a prime denotes differentiation with respect to $n$ and the two 
masses are,
\begin{eqnarray}
\frac{\mu^2_t}{\chi^2} &\!\!\! \equiv \!\!\!& -(D\!-\!1\!-\!\epsilon) 
\frac{\mu'}{\mu} - \frac{\mu''}{\mu} \; , \qquad \label{tmass} \\
\frac{\mu^2_u}{\chi^2} &\!\!\! \equiv \!\!\!& (D\!-\!2) (1\!-\! \epsilon)
+ \frac{\mu^2}{\chi^2} - (D\!-\!3\!+\!\epsilon) \frac{\mu'}{\mu} + 
\Bigl( \frac{\mu'}{\mu}\Bigr)' - \Bigl( \frac{\mu'}{\mu}\Bigr)^2 \; . 
\qquad \label{umass}
\end{eqnarray}
Because $\mu^2(n) = 2 q^2 \psi^2(n)$ we can use the inflaton $0$-mode
equation (\ref{inflatoneqn}) to simplify the $t$-mode mass,
\begin{equation}
\frac{\mu^2_{t}}{\chi^2} = -\frac{[\psi'' + (D\!-\!1\!-\!\epsilon) \psi']}{
\psi} = \frac{U'(\psi^2)}{\chi^2} \; . \label{tmasssimp}
\end{equation}

In order to follow the amplitudes numerically one must use a specific 
model of inflation. For simplicity we have chosen the quadratic mass
model, $U = c^2 \psi^2$, even though its prediction for the 
tensor-to-scalar ratio is disfavored by the data \cite{Aghanim:2018eyx,
Tristram:2020wbi}. The Slow Roll Approximation gives analytic expressions 
for this model which are accurate until almost the end of inflation,
\begin{equation}
\psi(n) \simeq \sqrt{\psi_0^2 \!-\! 2n} \quad , \quad \chi(n) \simeq
\frac{c}{\sqrt{3}} \sqrt{\psi_0^2 \!-\! 2n} \quad ,\quad \epsilon(n)
\simeq \frac1{\psi_0^2 \!-\! 2n} \; , \label{slowroll}
\end{equation}
where $\psi_0$ is the initial value of the dimensionless inflaton 
$0$-mode. About 56 e-foldings of inflation results from the choice 
$\psi_0 = 10.6$. To estimate the constant $c$, note that modes which 
experience 1st horizon crossing at e-folding $n_1$ (that is, $\kappa = 
\chi(n_1) e^{n_1}$) have the following approximate scalar power spectrum
and spectral index,
\begin{equation}
\Delta^2_{\mathcal{R}}(n_1) \simeq \frac1{8\pi^2} \frac{\chi^2(n_1)}{\epsilon(n_1)}
\qquad \Longrightarrow \qquad 1 - n_s \simeq 2 \epsilon +
\frac{\epsilon'}{\epsilon} \; . \label{CMB}
\end{equation}
Hence the observed scalar spectral index is consistent with $\psi_0 =
10.6$, and the observed scalar amplitude with the choice of $c = 7.1 
\times 10^{-6}$ \cite{Aghanim:2018eyx,Tristram:2020wbi}. We must also
choose a specific value for the charge $q$. Using $q^2 = 1/137$ would cause
the classical potential of $U = c^2 \psi^2$ to be completely overwhelmed
by the 1-loop Coleman-Weinberg correction of $\Delta U \simeq 3/64\pi^2 
\times \mu^4 \ln(\mu^2/s^2)$, where $s$ is the dimensionless 
renormalization scale \cite{Miao:2015oba}. Choosing the much smaller 
value of $q = 1.2 \times 10^{-6}$ reduces the 1-loop correction to a 
negligible tenth of a percent effect at the start of inflation.

\begin{figure}[H]
\centering
\includegraphics[width=4.3cm]{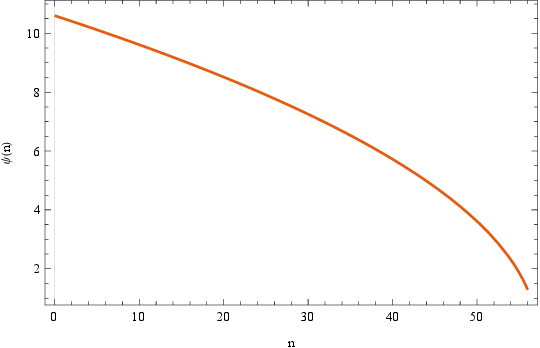}
\includegraphics[width=4.3cm]{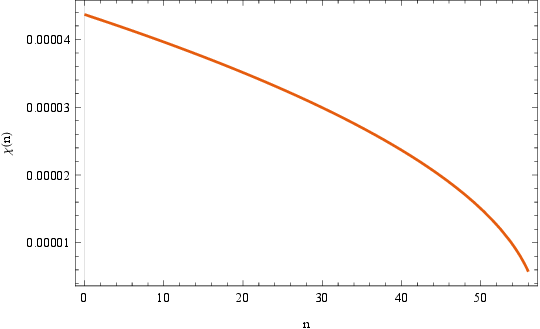}
\includegraphics[width=4.3cm]{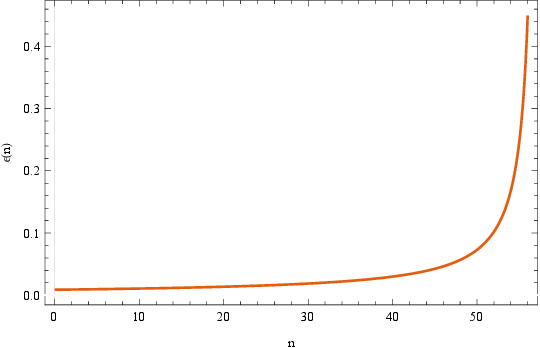}
\caption[Dimensionless inflaton field, Hubble parameter and slowroll parameter for $0\leq n\leq 56$]{\footnotesize Plots of $\psi(n)$ --- on the left --- $\chi(n)$ ---
in the center --- and $\epsilon(n)$ --- on the right --- for $0 \leq n \leq 
56$.}
\label{Earlygeom}
\end{figure}

\begin{figure}[H]
\centering
\includegraphics[width=4.3cm]{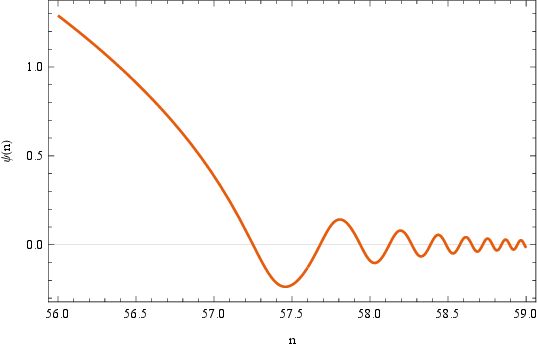}
\includegraphics[width=4.3cm]{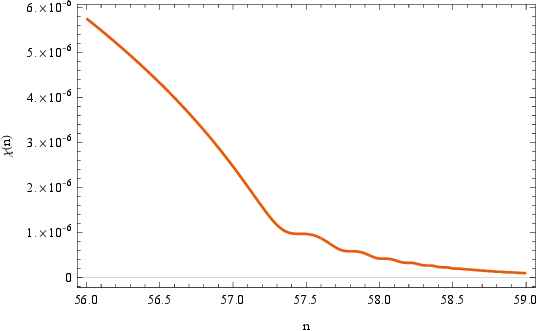}
\includegraphics[width=4.3cm]{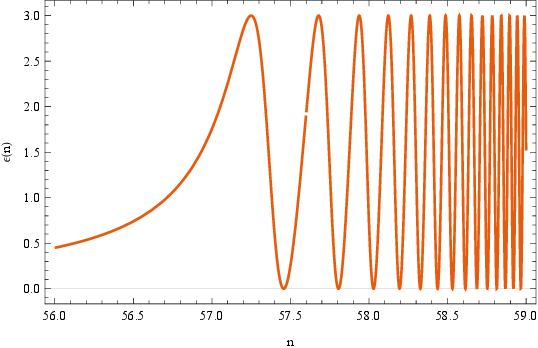}
\caption[Dimensionless inflaton field, Hubble parameter and slowroll parameter for $56\leq n\leq 59$]{\footnotesize Plots of $\psi(n)$ --- on the left --- $\chi(n)$ ---
in the center --- and $\epsilon(n)$ --- on the right --- for $56 \leq n 
\leq 59$.}
\label{Lategeom}
\end{figure}

Once we have a specific model it is possible to understand the magnitudes
of the various terms. Figure~\ref{Earlygeom} shows the dimensionless scalar, 
the dimensionless Hubble parameter and the first slow roll parameter while
inflation is occurring ($\epsilon < 1$). The slow roll approximations 
(\ref{slowroll}) are excellent during this period.
\noindent Figure~\ref{Lategeom} shows the same three quantities through the
end of inflation (which occurs at $n_e \simeq 56.7$) under the assumption
that the classical relations (\ref{Friedmann1}-\ref{inflatoneqn}) are not
corrected by the quantum effects we seek to incorporate. During this phase 
the inflaton oscillates around $\psi = 0$ with decreasing amplitude and
increasing frequency, while the first slow roll parameter oscillates in the 
range $0 \leq \epsilon \leq 3$. Because $\epsilon = {\psi'}^2$, the first 
slow roll parameter vanishes at extrema of $\psi(n)$, and it reaches its 
maximum (of $\epsilon(n) = 3$) when $\psi(n) = 0$. Of course the dimensionless 
Hubble parameter is monotonically decreasing; this decrease is rapid when 
$\epsilon \simeq 3$, and slow when $\epsilon \simeq 0$. 

\begin{figure}[H]
\centering
\includegraphics[width=4.3cm]{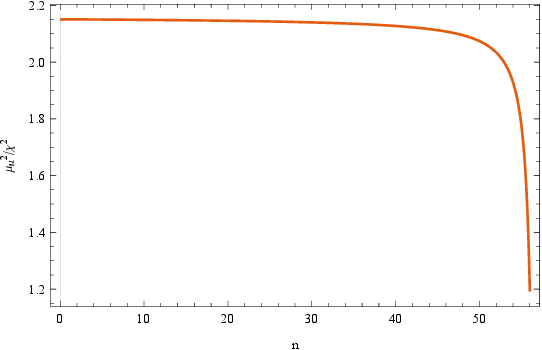}
\includegraphics[width=4.3cm]{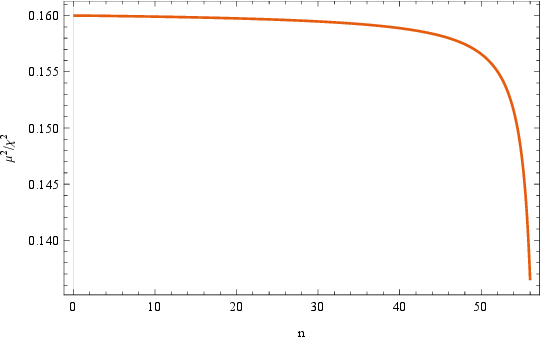}
\includegraphics[width=4.3cm]{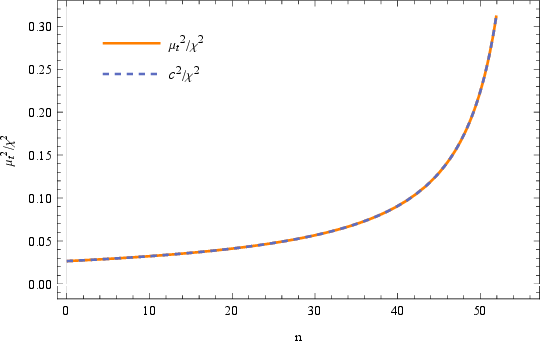}
\caption[Plots of $\mu_u^2(n)/\chi^2(n)$,  
$\mu^2(n)/\chi^2(n)$ and $\mu^2_t(n)/\chi^2(n)$ for $0 \leq n \leq 56$]{\footnotesize Plots of $\mu_u^2(n)/\chi^2(n)$ --- on the left --- 
$\mu^2(n)/\chi^2(n)$ --- in the center --- and $\mu^2_t(n)/\chi^2(n)$ --- 
on the right --- for $0 \leq n \leq 56$.}
\label{Earlymass}
\end{figure}

\begin{figure}[H]
\centering
\includegraphics[width=4.3cm]{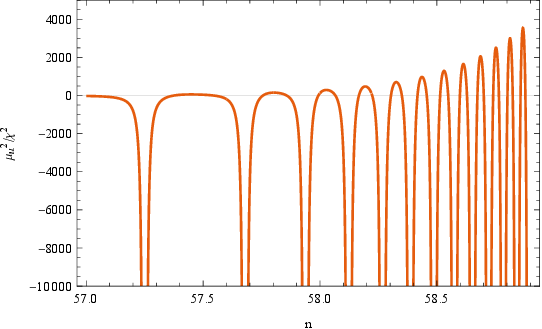}
\includegraphics[width=4.3cm]{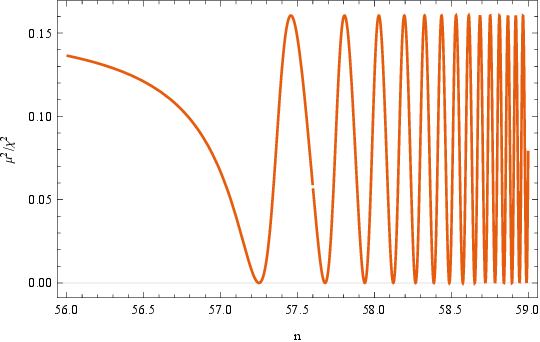}
\includegraphics[width=4.3cm]{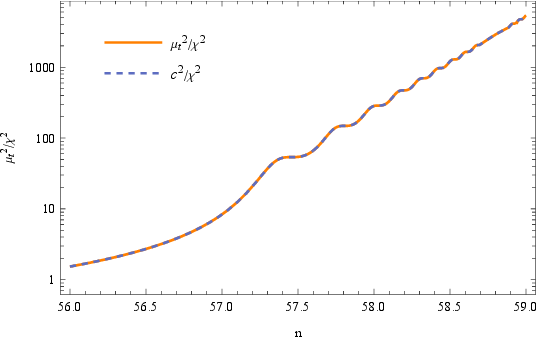}
\caption[Plots of $\mu_u^2(n)/\chi^2(n)$,  
$\mu^2(n)/\chi^2(n)$ and $\mu^2_t(n)/\chi^2(n)$ for $56 \leq n \leq 59$]{\footnotesize Plots of $\mu_u^2(n)/\chi^2(n)$ --- on the left --- 
$\mu^2(n)/\chi^2(n)$ --- in the center --- and $\mu^2_t(n)/\chi^2(n)$ --- 
on the right --- for $56 \leq n \leq 59$.}
\label{Latemass}
\end{figure}

\begin{figure}[H]
\centering
\includegraphics[width=6.5cm]{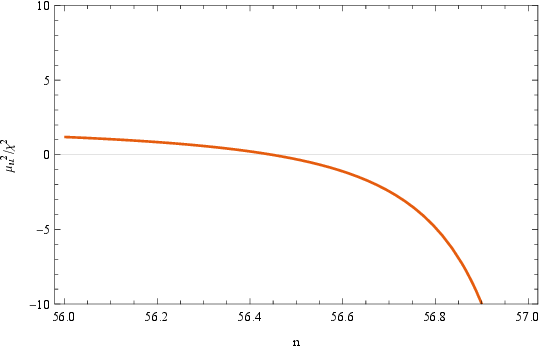}
\includegraphics[width=6.5cm]{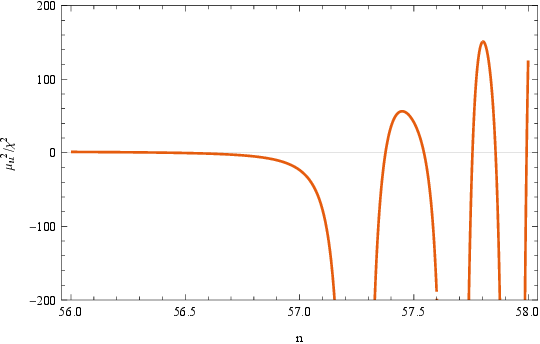}
\caption[Late time zoomed-in plots for $\mu_u^2(n)/\chi^2(n)$]{\footnotesize The left hand plot shows $\mu_u^2(n)/\chi^2(n)$ 
on the range $56 \leq n \leq 57$, just before the first zero of $\mu(n)$. The 
right hand plot depicts $\mu_u^2(n)/\chi^2(n)$ over the slightly larger range
of $56 \leq n \leq 58$, which includes the first three zeroes of $\mu(n)$.}
\label{Zoomumass}
\end{figure}

The slow roll approximation (\ref{slowroll}) tells us that $\psi' 
\simeq -1/\psi$ and $\chi(n) \simeq c/\sqrt{3} \times \psi(n)$. Setting
$D=4$, and using our values of $c = 7.1 \times 10^{-6}$ and $q = 1.2 
\times 10^{-6}$, gives the mass hierarchy,
\begin{equation}
\frac{\mu^2_u}{\chi^2} \simeq 2 + \frac{6 q^2}{c^2} \simeq 2.16 \quad >
\quad \frac{\mu^2}{\chi^2} \simeq \frac{6 q^2}{c^2} \simeq 0.16 \quad >
\quad \frac{\mu^2_t}{\chi^2} \simeq \frac{3}{\psi^2} \; . \label{masses}
\end{equation}
Figure~\ref{Earlymass} shows the various masses through inflation.
\noindent As one can just see from the larger $n$ values of 
Figure~\ref{Earlymass}, the hierarchy of equation (\ref{masses}) becomes 
inverted after the end of inflation. Figure~\ref{Latemass} shows the 
behavior after the end of inflation. During this phase $\mu^2_u/\chi^2$ 
is mostly tachyonic, and actually diverges at points where $\psi(n) = 0$. 
On the other hand, $\mu^2/\chi^2$ oscillates between $0$ and the small
value of $0.16$, while $\mu^2_t/\chi^2$ grows monotonically to large, 
positive values. The $u$-mode mass is the most important of the three, 
and its evolution is the most complex. Figure~\ref{Zoomumass} shows its 
behavior in more detail.

\subsection{Approximating the Longitudinal Amplitude}

\begin{figure}[H]
\centering
\includegraphics[width=4.3cm]{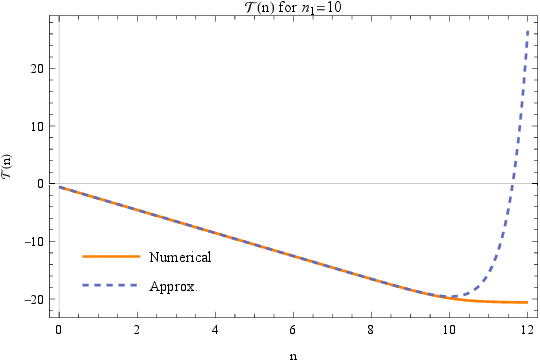}
\includegraphics[width=4.3cm]{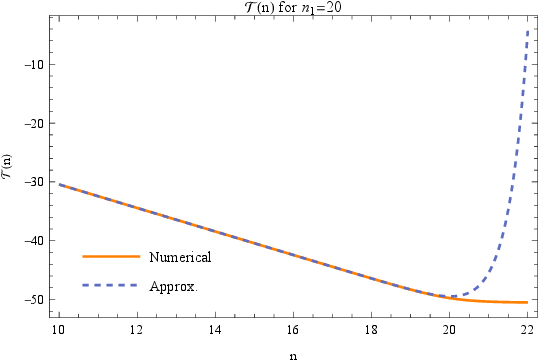}
\includegraphics[width=4.3cm]{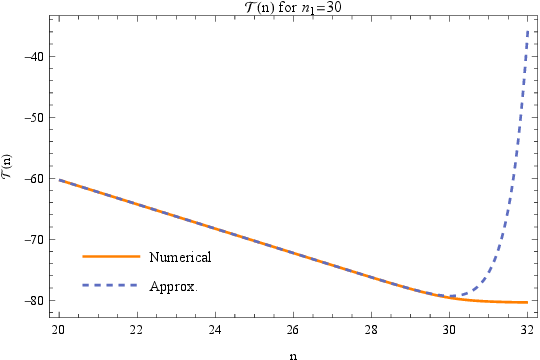}
\caption[Comparison of $\mathcal{T}(n,\kappa)$ with its UV Approximation]{\footnotesize Plot comparing $\mathcal{T}(n,\kappa)$ with the
ultraviolet form (\ref{TUVexp}) for modes which experience first horizon
crossing at $n_1 = 10$ (left), $n_1 = 20$ (center) and $n_1 = 30$ (right).}
\label{EarlyT}
\end{figure}

Equation (\ref{Teqn}) for $\mathcal{T}(n,\kappa)$ contains six terms. The 
ultraviolet regime is defined by the condition $\kappa \gg \chi(n) e^{n}$. 
In this regime equation (\ref{Teqn}) is dominated by the 4th and 6th terms, 
$2\kappa^2 e^{-2n}/\chi^2$ and $-e^{-2 [\mathcal{T} + (D-1)n]}/2 \chi^2$, 
and the amplitude takes the form,
\begin{eqnarray}
\lefteqn{\mathcal{T}(n,\kappa) = \ln\Bigl[\frac1{2\kappa}\Bigr] - (D\!-\!2) n }
\nonumber \\
& & \hspace{2.7cm} + \Bigl[ \frac12 (D\!-\!2) (D \!-\! 2\epsilon) - 
\frac{2 \mu_t^2}{\chi^2}\Bigr] \Bigl( \frac{\chi e^n}{2 \kappa}\Bigr)^2 + 
O\Biggl( \Bigl(\frac{\chi e^n}{2 \kappa}\Bigr)^4 \Biggr) \; . \qquad 
\label{TUVexp}
\end{eqnarray}
Figure~\ref{EarlyT} compares numerical evolution of the exact equation 
(\ref{Teqn}) with the ultraviolet form (\ref{TUVexp}) for wave numbers
which experience first horizon crossing at $n_1 = 10$, $n_1 = 20$, and
$n_1 = 30$. The agreement is excellent up to horizon crossing.

\begin{figure}[H]
\centering
\includegraphics[width=4.3cm]{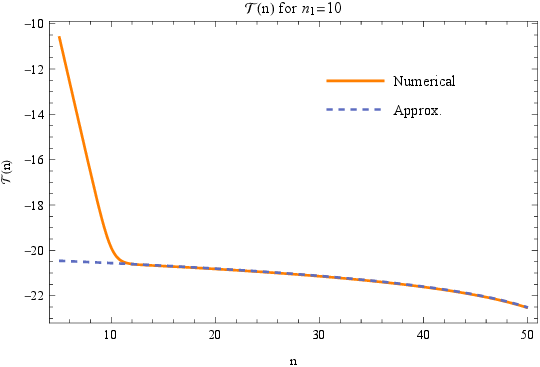}
\includegraphics[width=4.3cm]{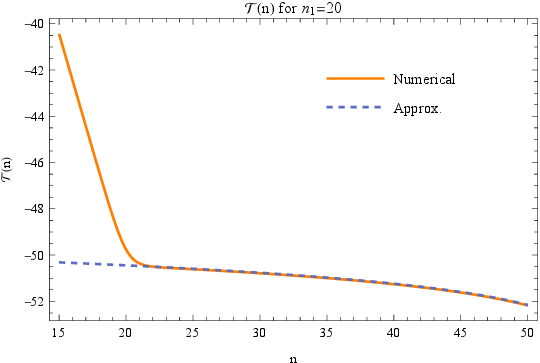}
\includegraphics[width=4.3cm]{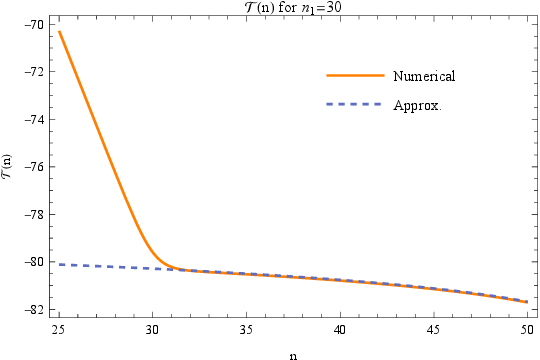}
\caption[Comparison of $\mathcal{T}(n,\kappa)$ with its late time Approximation]{\footnotesize Plot comparing $\mathcal{T}(n,\kappa)$ with the
late time form (\ref{TIRform}) for modes which experience first horizon
crossing at $n_1 = 10$ (left), $n_1 = 20$ (center) and $n_1 = 30$ (right).}
\label{LateT}
\end{figure}

After 1st horizon crossing the 4th and 6th terms of (\ref{Teqn}) effectively 
drop out and the relation simplifies to,
\begin{equation}
\mathcal{T}'' + \frac12 {\mathcal{T}'}^2 + (D \!-\! 1 \!-\! \epsilon)
\mathcal{T}' - 2 (D\!-\!1\!-\!\epsilon) \frac{\mu'}{\mu} - 2 \frac{\mu''}{\mu}
\simeq 0 \; . \label{Teqnsimp}
\end{equation}
This is an equation for $\mathcal{T}'$, and it is easy to see that a 
particular solution is,
\begin{equation}
\mathcal{T}' = 2 \frac{\mu'}{\mu} \; . \label{Tprime}
\end{equation}
Integrating (\ref{Tprime}), and using the tensor power spectrum to infer 
the integration constant to all orders in the slow roll approximation 
\cite{Brooker:2015iya}, implies,\footnote{The integration constant in relation
(\ref{TIRform}) suffices for smooth inflationary potentials. When features are
present the constant can be supplemented by known corrections which depend
nonlocally on the expansion history before first horizon crossing 
\cite{Brooker:2017kij}.}
\begin{equation}
\mathcal{T}(n,\kappa) \simeq \ln\Biggl[ \frac{\chi_1^2 C(\epsilon_1)}{2 \kappa^3}
\times \frac{\mu^2(n)}{\mu^2_1} \Biggr] \; , \label{TIRform}
\end{equation}
where the function $C(\epsilon)$ is,
\begin{equation}
C(\epsilon) = \frac1{\pi} \Gamma^2\Bigl( \frac12 \!+\! \frac1{1 \!-\! \epsilon}
\Bigr) \Bigl[2 (1 \!-\! \epsilon)\Bigr]^{\frac2{1-\epsilon}} \; . \label{Cdef}
\end{equation}
Figure~\ref{LateT} compares the exact numerical result with the infrared form
(\ref{TIRform}) for modes which experience horizon crossing at $n_1 = 10$,
$n_1 = 20$, and $n_1 = 30$. Agreement is excellent.

Although one can see from Figure~\ref{LateT} that the approximate solution 
(\ref{TIRform}) is highly accurate, it cannot be exact for two reasons:
\begin{enumerate}
\item{We neglected the 4th and 6th terms in simplifying equation (\ref{Teqn})
to reach (\ref{Teqnsimp}); and}
\item{Just because (\ref{Tprime}) is {\it a} solution to (\ref{Teqnsimp}) does 
not mean it is {\it the} solution.}
\end{enumerate}
To find the {\it general} solution to (\ref{Teqnsimp}) we substitute 
$\mathcal{T}' = 2 \mu'/\mu + f(n)$,
\begin{equation}
f' + 2 \frac{\mu'}{\mu} f + \frac12 f^2 + (D \!-\! 1 \!-\! \epsilon) f \simeq 
0 \; . \label{feqn1}
\end{equation}
Now divide by $\mu^2 e^{(D-1) n} \chi f^2$ to reach the form,
\begin{equation}
\frac{\partial}{\partial n} \Biggl[ \frac1{\mu^2 e^{(D-1) n} \chi f}\Biggr] =
\frac1{2 \mu^2 e^{(D-1)n} \chi} \; . \label{feqn2}
\end{equation}
Integrating equation (\ref{feqn2}) from some point $n_2$ gives the general
solution,
\begin{eqnarray}
\lefteqn{f(n) = f_2 \Biggl[ e^{(D-1)(n-n_2)} \Bigl[ \frac{\chi(n)}{\chi_2}\Bigr] 
\Bigl[ \frac{\mu(n)}{\mu_2}\Bigr]^2 } \nonumber \\
& & \hspace{4cm} + \frac12 f_2 \int_{n_2}^{n} \!\!\!\!\! dn' \, e^{(D-1)(n-n')} 
\Bigl[ \frac{\chi(n)}{\chi(n')}\Bigr] \Bigl[ \frac{\mu(n)}{\mu(n')} \Bigr]^2 
\Biggr]^{-1} \; . \qquad \label{fsol}
\end{eqnarray}

Careful consideration of (\ref{fsol}) reveals that $\mathcal{T}(n,\kappa)$ actually
has a finite limit as the mass vanishes. To see this, assume $n$ is such that 
$\mu(n) \rightarrow 0$, and expand the integral of (\ref{fsol}) for small $\mu(n)$,
\begin{eqnarray}
\lefteqn{\int_{n_2}^{n} \!\!\!\!\! dn' \, e^{(D-1)(n-n')} \Bigl[ 
\frac{\chi(n)}{\chi(n')}\Bigr] \Bigl[ \frac{\mu(n)}{\mu(n')} \Bigr]^2 \!\!= 
-\frac{\mu(n)}{\mu'(n)} } \nonumber \\
& & \hspace{2cm} - \frac12 \Bigl[D\!-\!1 \!-\! \epsilon(n) \!+\! 
\frac{\mu''(n)}{\mu'(n)} \Bigr] \Bigl[ \frac{\mu(n)}{\mu'(n)}\Bigr]^2 \ln[\mu^2(n)] 
+ O(1) \; . \qquad \label{singexp}
\end{eqnarray}
Near the point where $\mu(n) \rightarrow 0$ we therefore have,
\begin{equation}
f(n) \longrightarrow -\frac{2 \mu'(n)}{\mu(n)} + \Bigl[D \!-\! 1 \!-\! \epsilon(n)
\!+\! \frac{\mu''(n)}{\mu'(n)}\Bigr] \ln[\mu^2(n)] + O(1) \; . \label{fexp}
\end{equation}
Hence we have,
\begin{equation}
\mathcal{T}'(n,\kappa) \longrightarrow \Bigl[D\!-\!1 \!-\! \epsilon(n) \!+\!
\frac{\mu''(n)}{\mu'(n)} \Bigr] \ln[\mu^2(n)] + O(1) \; . \label{Tpfull}
\end{equation}
Although expression (\ref{Tpfull}) diverges as $\mu(n)$ goes to zero, the 
singularity is integrable, which means that $\mathcal{T}(n,\kappa)$ remains
finite.
\begin{figure}[H]
\centering
\includegraphics[width=6cm]{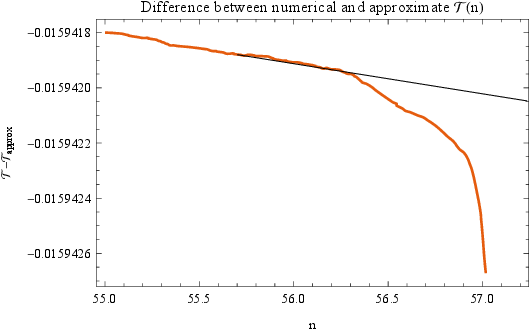}
\caption[Numerical determination of the constant $f_2$]{\footnotesize Numerical determination of the constant $f_2$ from
the difference of $\mathcal{T}(n,\kappa)$ and expression (\ref{TIRform}).
In this case $\kappa$ was chosen to experience first horizon crossing at
$n_1 = 10$.}
\label{f2plot}
\end{figure} 
As we see from Figure~\ref{f2plot}, the constant $f_2$ in expression 
(\ref{fsol}) represents the difference between the actual value of 
$\mathcal{T}'(n_2,\kappa)$ and its approximate form (\ref{Tprime}) $2 
\mu'(n_2)/\mu(n_2)$. Because the approximate form is quite accurate, $f_2$ 
is a very small number, about $f_2 \sim -10^{-7}$. The fact that $f_2$ drops 
out of the asymptotic form (\ref{fexp}) means that the ultimate finiteness 
of $\mathcal{T}(n,\kappa)$ is a robust conclusion. However, $\mu(n)$ must 
be {\it very} close to zero before the integral (\ref{singexp}) begins to 
dominate over the first term in the denominator of (\ref{fsol}), which has 
a relative enhancement of $e^{(D-1)(n-n_2)}$. If we take $n_2 = 56$ and 
define $n_* \simeq 57.25$ as the first zero of $\mu(n)$, the point $n_f$ at 
which expressions (\ref{fexp}-\ref{Tpfull}) become valid approximately obeys,
\begin{equation}
n_* - n_f \simeq \frac12 f_2 \, e^{-3 (n_* - n_2)} \Bigl( \frac{\chi_2}{\chi_*} 
\Bigr) \Bigl(\frac{\mu_2}{\mu_*'} \Bigr)^2 \simeq 10^{-9} \; . \label{fdomn}
\end{equation}
We can therefore estimate the minimum value of $\mathcal{T}(n,\kappa)$ as
\begin{equation}
\mathcal{T}_{\rm min} \simeq \ln\Biggl[ \frac{\chi_1^2 C(\epsilon_1)}{2 \kappa^3}
\!\times\! \frac14 f_2^2 \, e^{-6 (n_* - n_2)} \Bigl( \frac{\chi_2}{\chi_*}\Bigr)^2
\Bigl( \frac{\mu_2}{\mu_1}\Bigr)^2 \Bigl( \frac{\mu_2}{\mu_*'}\Big)^2 \Biggr] \; . 
\label{Tmin}
\end{equation}

\subsection{Approximating the Temporal Amplitude}

\begin{figure}[H]
\centering
\includegraphics[width=4.3cm]{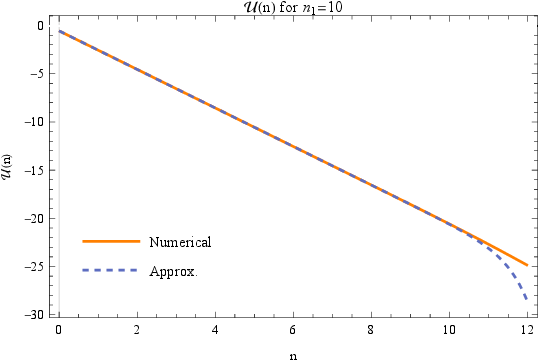}
\includegraphics[width=4.3cm]{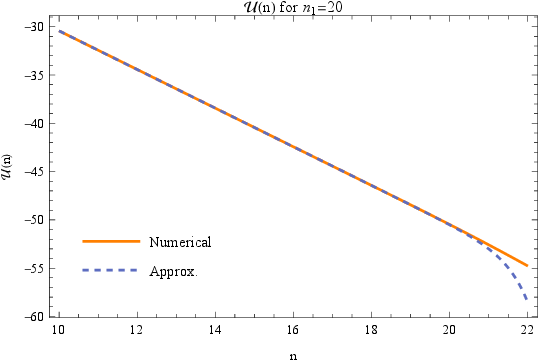}
\includegraphics[width=4.3cm]{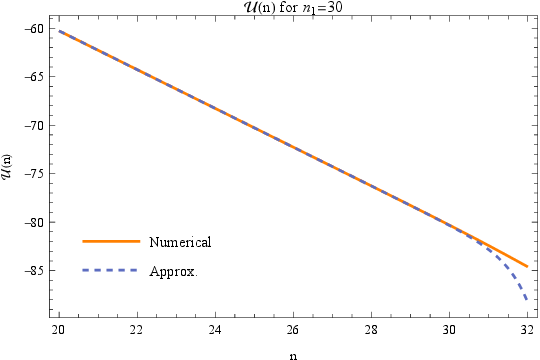}
\caption[Comparison of $\mathcal{U}(n,\kappa)$ with its UV Approximation]{\footnotesize Plot comparing $\mathcal{U}(n,\kappa)$ with the
ultraviolet form (\ref{UUVexp}) for modes which experience first horizon
crossing at $n_1 = 10$ (left), $n_1 = 20$ (center) and $n_1 = 30$ (right).}
\label{EarlyU}
\end{figure}

Equation (\ref{Ueqn}) for $\mathcal{U}(n,\kappa)$ contains the same six 
terms as (\ref{Teqn}). In the ultraviolet it is also dominated by the 4th 
and 6th terms, $2\kappa^2 e^{-2n}/\chi^2$ and $-e^{-2 [\mathcal{U} + 
(D-1)n]}/2 \chi^2$. Hence the ultraviolet expansion of $\mathcal{U}(n,\kappa)$
takes the same form as (\ref{TUVexp}),
\begin{eqnarray}
\lefteqn{\mathcal{U}(n,\kappa) = \ln\Bigl[\frac1{2\kappa}\Bigr] - (D\!-\!2) n }
\nonumber \\
& & \hspace{2.7cm} + \Bigl[ \frac12 (D\!-\!2) (D \!-\! 2\epsilon) - 
\frac{2 \mu_u^2}{\chi^2}\Bigr] \Bigl( \frac{\chi e^n}{2 \kappa}\Bigr)^2 + 
O\Biggl( \Bigl(\frac{\chi e^n}{2 \kappa}\Bigr)^4 \Biggr) \; . \qquad 
\label{UUVexp}
\end{eqnarray}
Figure~\ref{EarlyU} compares the exact numerical solution with the ultraviolet
form (\ref{UUVexp}) for modes which experience first horizon crossing at $n_1 
= 10$, $n_1 = 20$ and $n_1 = 30$. Agreement is excellent up to first horizon 
crossing, just as it was in the analogous comparison of Figure~\ref{EarlyT} for 
$\mathcal{T}(n,\kappa)$.

\begin{figure}[H]
\centering
\includegraphics[width=4.3cm]{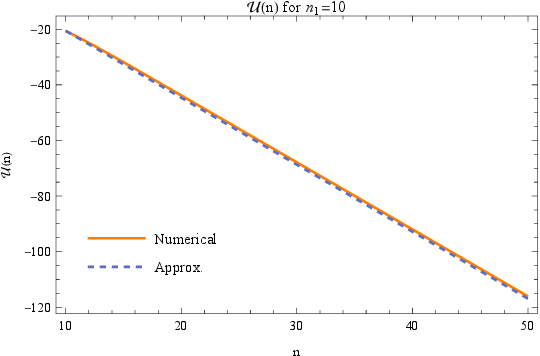}
\includegraphics[width=4.3cm]{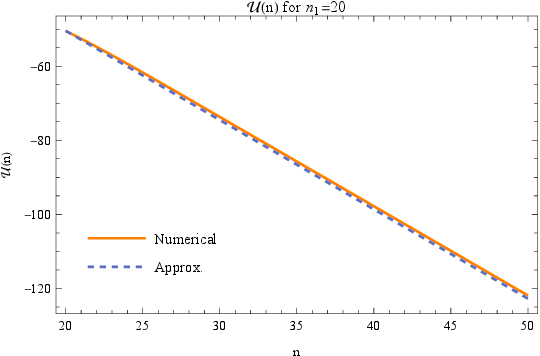}
\includegraphics[width=4.3cm]{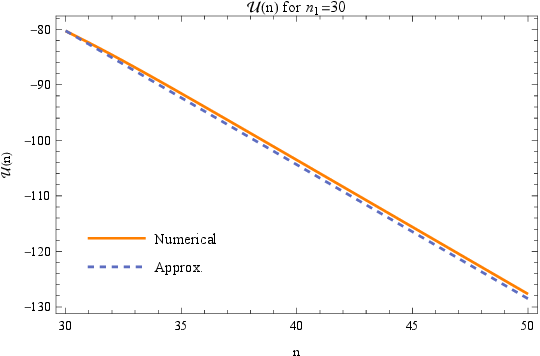}
\caption[Comparison of $\mathcal{U}(n,\kappa)$ with its late time Approximation]{\footnotesize Plot comparing $\mathcal{U}(n,\kappa)$ with the
late time form (\ref{UIRform}) for modes which experience first horizon
crossing at $n_1 = 10$ (left), $n_1 = 20$ (center) and $n_1 = 30$ (right).}
\label{LateU}
\end{figure}

The 4th and 6th terms of (\ref{Ueqn}) drop out after first horizon crossing,
and the relation simplifies to,
\begin{equation}
\mathcal{U}'' + \frac12 {\mathcal{U}'}^2 + (D\!-\!1\!-\! \epsilon) \mathcal{U}'
+ \frac{2 \mu^2_u}{\chi^2} \simeq 0 \; . \label{Ueqnsimp}
\end{equation}
Recall from Figure~\ref{Earlymass} that $\mu_u^2(n)/\chi^2(n)$ is approximately
constant during inflation. This means that equation (\ref{Ueqnsimp}) can be 
roughly solved as,
\begin{equation}
\mathcal{U}'(n,\kappa) \simeq -(D\!-\!1\!-\!\epsilon) + \sqrt{(D \!-\! 1 \!-\!
\epsilon)^2 - \frac{4 \mu_u^2}{\chi^2}} \; . \label{approxUprime}
\end{equation}
With the appropriate integration constant we therefore have,
\begin{eqnarray}
\lefteqn{\mathcal{U}(n,\kappa) \simeq \ln\Biggl[ \frac{\chi_1^2 C(\epsilon_1)}{
2 \kappa^3} \!\times\! \frac{\chi_1}{\chi(n)} \Biggl] - (D \!-\! 1) (n \!-\! n_1)
} \nonumber \\
& & \hspace{5cm} + \int_{n_1}^{n} \!\!\!\! dn' \sqrt{[D\!-\! 1 \!-\! \epsilon(n')]^2
- \frac{4 \mu_u^2(n')}{\chi^2(n')} } \; . \qquad \label{UIRform}
\end{eqnarray}
Figure~\ref{LateU} compares this approximation with the numerical evolution
for modes which experience horizon crossing at $n_1 = 10$, $n_1 = 20$ and
$n_1 = 30$. 

After the end of inflation $\mu^2(n)/\chi^2(n)$ falls off whereas the 
derivative terms in $\mu^2_u/\chi^2$ become large and tachyonic. This means
we can neglect $\mu^2(n)/\chi^2(n)$,
\begin{equation}
\frac{\mu_u^2}{\chi^2} \simeq (D\!-\!2) (1\!-\! \epsilon) - (D\!-\!3\!+\!
\epsilon) \frac{\mu'}{\mu} + \Bigl( \frac{\mu'}{\mu}\Bigr)' - \Bigl(
\frac{\mu'}{\mu}\Bigr)^2 \; . \label{umasssimp2}
\end{equation}
We now make the substitution,
\begin{equation}
\mathcal{U}'(n,\kappa) = -\frac{2 \mu'(n)}{\mu(n)} - 2 (D\!-\!2) + g(n) \; ,
\label{Uansatz}
\end{equation}
in equation (\ref{Ueqnsimp}) to find,
\begin{equation}
g' - \Bigl(D \!-\! 3 \!+\! \epsilon \!+\! 2 \frac{\mu'}{\mu}\Bigr) g 
+ \frac12 g^2 = 0 \; . \label{geqn}
\end{equation}
Multiplying by $e^{(D-3)n} \mu^2(n)/[\chi(n) g^2(n)]$ makes the $g$-dependent
terms a total derivative, and permits us to write the general solution as,
\begin{eqnarray}
\lefteqn{ g(n) = g_2 \Biggl[ e^{-(D-3) (n-n_2)} \Bigl[ \frac{\chi(n)}{\chi_2}
\Bigr] \Bigl[ \frac{\mu_2}{\mu(n)}\Bigr]^2 } \nonumber \\
& & \hspace{4cm} + \frac12 g_2 \!\! \int_{n_2}^{n} \!\!\!\!\! dn' \, 
e^{-(D-3) (n-n')} \Bigl[ \frac{\chi(n)}{\chi(n')}\Bigr] \Bigl[ \frac{\mu(n')}{
\mu(n)}\Bigr]^2 \Biggr]^{-1} \; , \qquad \label{gsol}
\end{eqnarray} 
where the constant $g_2$ is determined to interpolate between (\ref{approxUprime})
and (\ref{Uansatz}),
\begin{equation}
g_2 = \frac{2 \mu_2'}{\mu_2} + (D\!-\!3\!+\!\epsilon_2) +
\sqrt{(D\!-\! 1 \!-\! \epsilon_2)^2 - \frac{4 \mu_2^2}{\chi_2^2}} \; . 
\label{g2def}
\end{equation}
Note that, whereas $f(n)$ diverges as $\mu(n)$ approaches zero, $g(n)$ goes to
zero like $\mu^2(n)$.

\begin{figure}[H]
\centering
\includegraphics[width=4.3cm]{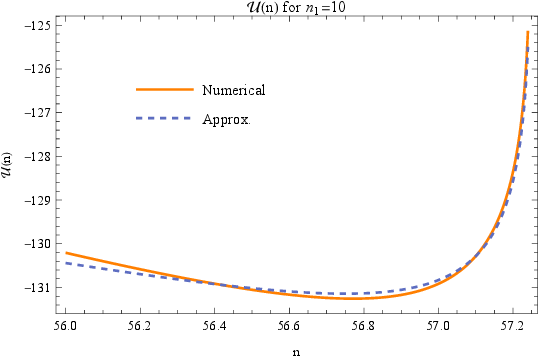}
\includegraphics[width=4.3cm]{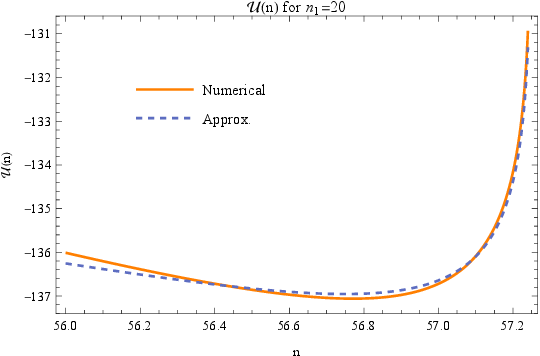}
\includegraphics[width=4.3cm]{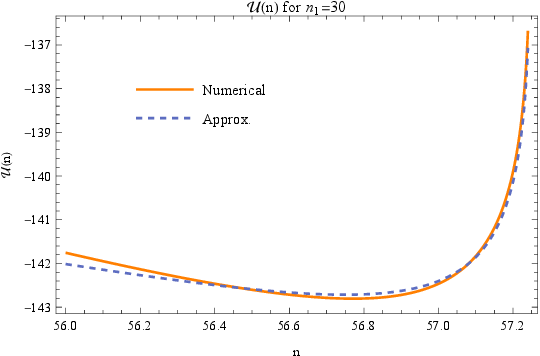}
\caption[Comparison of $\mathcal{U}(n,\kappa)$ with its post-inflationary Approximation]{\footnotesize Plot comparing $\mathcal{U}(n,\kappa)$ with the
post-inflationary form (\ref{UPostform}) for modes which experience first horizon
crossing at $n_1 = 10$ (left), $n_1 = 20$ (center) and $n_1 = 30$ (right).}
\label{UPost}
\end{figure}

Integrating equation (\ref{Uansatz}), and using (\ref{UIRform}) to supply the
integration constant, gives,
\begin{eqnarray}
\lefteqn{ \mathcal{U}(n,\kappa) \simeq \ln\Biggl[ \frac{\chi_1^2 C(\epsilon_1)}{
2 \kappa^3} \!\times\! \frac{\chi_1}{\chi_2} \!\times\! \frac{\mu_2^2}{\mu^2(n)}
\Biggr] - (D\!-\!1) (n_2 \!-\! n_1) - 2 (D\!-\!2) (n \!-\! n_2) } \nonumber \\
& & \hspace{3.2cm} + \int_{n_1}^{n_2} \!\!\!\!\! dn' \sqrt{ [D\!-\! 1 \!-\! 
\epsilon(n')]^2 - \frac{4 \mu_u^2(n')}{\chi^2(n')} } + \int_{n_2}^{n} \!\!\!\!\!
dn' \, g(n') \; . \qquad \label{UPostform}
\end{eqnarray}
Because $g(n)$ vanishes as $\mu(n) \rightarrow 0$, the $-\ln[\mu^2(n)]$ 
divergence of $\mathcal{U}(n,\kappa)$ is robust. Note that this is not even 
affected by neglecting $\mu^2(n)/\chi^2(n)$ in (\ref{umasssimp2}). 
Figure~\ref{UPost} compares the numerical solution with our analytic 
approximation (\ref{UPostform}).

\section{Quantum-Correcting the Inflaton 0-Mode}

The purpose of this section is to use the photon propagator to quantum-correct 
the classical equation for the inflaton $0$-mode from (\ref{0order}) to,
\begin{equation}
\partial_0 \Bigl[a^{D-2} \partial_0 \varphi_0\Bigr] + a^{D} \varphi_0
V'(\varphi_0^2) + q^2 \varphi_0 a^{D-2} \eta^{\mu\nu} i \Bigl[\mbox{}_{\mu}
\Delta_{\nu}\Bigr](x;x) = 0 \; . \label{new0modeeqn}
\end{equation}
We begin by deriving exact expressions for the $t$-mode and $u$-mode contributions
to trace of the photon propagator. We then use a variant of the work-energy theorem
to show how reheating occurs. 

\subsection{The Effective Force}

The $t$-mode contribution to the coincidence limit of the trace of the photon
propagator in equation (\ref{new0modeeqn}) is,
\begin{equation}
\sqrt{-g} \, g^{\mu\nu} i\Bigl[ \mbox{}_{\mu} \Delta_{\nu}\Bigr]_{t}(x;x) =
a^{D-2} \!\! \int \!\! \frac{d^{D-1} k}{(2\pi)^{D-1}} \Biggl\{ \frac1{M^2} \,
\widehat{\partial}_0 t \!\cdot\! \widehat{\partial_0} t^* - \frac{k^2}{M^2}
\, t \!\cdot \! t^* \Biggr\} \; . \label{ttrace1}
\end{equation}
The $t$-mode equation (\ref{teqn}) can be exploited to write the product of 
time derivatives in terms of the norm-squared,
\begin{equation}
\partial_0 t \!\cdot\! \partial_0 t^* = \Bigl[ \frac12 \partial_0^2 + \frac12
(D\!-\!2) a H \partial_0 + k^2 - (D\!-\!2) a H \frac{\partial_0 M}{M} -
\frac{\partial_0^2 M}{M} \Bigr] (t t^*) \; . \label{tID1}
\end{equation}
Using this identity we can re-express the $t$-mode contribution (\ref{ttrace1})
as,
\begin{eqnarray}
\lefteqn{\sqrt{-g} \, g^{\mu\nu} i\Bigl[ \mbox{}_{\mu} \Delta_{\nu}\Bigr]_{t} = 
\frac{a^{D-2}}{M^2} \!\! \int \!\! \frac{d^{D-1}k}{(2\pi)^{D-1}} \Biggl\{ 
\frac12 \partial_0^2 + \frac12 (D\!-\!2) a H \partial_0 - \frac{\partial_0 M}{M}
\partial_0 } \nonumber \\
& & \hspace{4cm} - (D\!-\!2) a H \frac{\partial_0 M}{M} - \frac{\partial_0^2 M}{M}
+ \Bigl( \frac{\partial_0 M}{M}\Bigr)^2 \Biggr\} (t t^*) \; . \qquad 
\label{ttrace2}
\end{eqnarray}
Converting to dimensionless form, and employing equation (\ref{Teqn}) to eliminate
second derivatives, gives the compact form,
\begin{eqnarray}
\lefteqn{\sqrt{-g} \, g^{\mu\nu} i\Bigl[ \mbox{}_{\mu} \Delta_{\nu}\Bigr]_{t}(x;x) 
= \frac{e^{D n} \chi^2(n)}{(8\pi G)^{\frac{D}2 -1} \mu^2(n)} \!\! \int \!\! 
\frac{d^{D-1} \kappa}{(2\pi)^{D-1}} \, e^{\mathcal{T}(n,\kappa)} } \nonumber \\
& & \hspace{1.3cm} \times \Biggl\{ \Bigl[ \frac12 \mathcal{T}'(n,\kappa) - 
\frac{\mu'(n)}{\mu(n)} \Bigr]^2 - \frac{\kappa^2 e^{-2n}}{\chi^2(n)} + 
\frac1{4 \chi^2} e^{-2 [\mathcal{T}(n,\kappa) + (D-1)n]} \Biggr\} . \qquad 
\label{ttrace3}
\end{eqnarray}

The $u$-mode contribution to the photon trace in (\ref{new0modeeqn}) is,
\begin{equation}
\sqrt{-g} \, g^{\mu\nu} i\Bigl[ \mbox{}_{\mu} \Delta_{\nu}\Bigr]_{u}(x;x) =
a^{D-2} \!\! \int \!\! \frac{d^{D-1} k}{(2\pi)^{D-1}} \Biggl\{ -\frac{k^2}{M^2} 
\, u \!\cdot\! u^* + \frac{1}{M^2} \widetilde{D} u \!\cdot\! \widetilde{D} u^*
\Biggr\} \; . \label{utrace1}
\end{equation}
We can eliminate the norm-square of $\partial_0 u$ using the $u$-mode equation
(\ref{ueqn}),
\begin{eqnarray}
\lefteqn{\partial_0 u \!\cdot\! \partial_0 u^* = \Biggl[ \frac12 \partial_0^2 + 
\frac12 (D\!-\!2) a H \partial_0 + k^2 + a^2 M^2 + (D\!-\!2) a^2 H^2 (1 \!-\! 
\epsilon) } \nonumber \\
& & \hspace{3.2cm} - (D\!-\!2) a H \frac{\partial_0 M}{M} + \partial_0 \Bigl( 
\frac{\partial_0 M}{M}\Bigr) - \Bigl(\frac{\partial_0 M}{M} \Bigr)^2 \Biggr] 
(u u^*) \; . \qquad \label{uID1}
\end{eqnarray}
Substituting (\ref{uID1}) in (\ref{utrace1}), and taking apart the factors of
$\widetilde{D} = \partial_0 + (D-2) a H + \partial_0 M/M$ gives,
\begin{eqnarray}
\lefteqn{\sqrt{-g} \, g^{\mu\nu} i\Bigl[ \mbox{}_{\mu} \Delta_{\nu}\Bigr]_{u} = 
\frac{a^{D-2}}{M^2} \!\! \int \!\! \frac{d^{D-1}k}{(2\pi)^{D-1}} \Biggl\{ 
\frac12 \partial_0^2 + \frac32 (D\!-\!2) a H \partial_0 + \frac{\partial_0 M}{M}
\partial_0 + a^2 M^2 } \nonumber \\
& & \hspace{.7cm} + (D\!-\!2) (D\!-\!1\!-\!\epsilon) a^2 H^2 + (D\!-\!2) a H 
\frac{\partial_0 M}{M} + \partial_0 \Bigl(\frac{\partial_0 M}{M} \Bigr) 
\Biggr\} (u u^*) \; . \qquad \label{utrace2}
\end{eqnarray}
The final, dimensionless form is very similar to (\ref{ttrace3}),
\begin{eqnarray}
\lefteqn{\sqrt{-g} \, g^{\mu\nu} i\Bigl[ \mbox{}_{\mu} \Delta_{\nu}\Bigr]_{u}(x;x) 
= \frac{e^{D n} \chi^2(n)}{(8\pi G)^{\frac{D}2 -1} \mu^2(n)} \!\! \int \!\! 
\frac{d^{D-1} \kappa}{(2\pi)^{D-1}} \, e^{\mathcal{U}(n,\kappa)} } \nonumber \\
& & \hspace{-0.1cm} \times \Biggl\{ \Bigl[ \frac12 \mathcal{U}'(n,\kappa) + 
\frac{\mu'(n)}{\mu(n)} + D\!-\!2 \Bigr]^2 - \frac{\kappa^2 e^{-2n}}{\chi^2(n)} 
+ \frac1{4 \chi^2} e^{-2 [\mathcal{U}(n,\kappa) + (D-1)n]} \Biggr\} . 
\qquad \label{utrace3}
\end{eqnarray}

\subsection{Reheating}

The dimensionless form of the inflaton $0$-mode equation (\ref{new0modeeqn}) 
takes the form,
\begin{equation}
e^{n} \chi \frac{\partial}{\partial n} \Bigl[ e^{(D-1) n} \chi \psi'\Bigr] =
- e^{D n} \psi \Biggl[ U'(\psi^2) + \frac{Q^2 \chi^2}{\mu^2} \!\! \int \!\!
\frac{d^{D-1}\kappa}{(2\pi)^{D-1}} \Biggl\{ \qquad \Biggr\} \Biggr] \equiv
\mathcal{F} \; . \label{force}
\end{equation}
where the term inside the curly brackets is the sum of the $t$ and $u$ contributions
from expressions (\ref{ttrace3}) and (\ref{utrace3}), and $Q^2 \equiv q^2/(8\pi G)^{
\frac{D}2 -2}$ is the dimensionless charge. Multiplying both sides of (\ref{force})
by $e^{(D-2)n} \chi \psi'$ and integrating gives a curious generalization of the
famous work-energy theorem of introductory physics,
\begin{equation}
e^{(D-1)n} \chi \psi' \frac{\partial}{\partial n} \Bigl[ e^{(D-1) n} \chi \psi'
\Bigr] = \frac12 \frac{\partial}{\partial n} \Bigl[ e^{(D-1) n} \chi \psi'\Bigr]^2
= e^{(D-2) n} \chi \psi' \!\times\! \mathcal{F} \; . \label{WEthm1}
\end{equation}
We now integrate (\ref{WEthm1}) from the beginning of reheating (at $n = n_i$) to
the end (at $n = n_f$), and use the fact that $\psi'(n_f) = 0$ at the end of 
reheating,
\begin{equation}
0 - \frac12 \Bigl[ e^{(D-1) n_i} \chi_i \psi'_i\Bigr]^2 = \int_{n_i}^{n_f} \!\!\!\!
dn \, e^{(D-2) n} \chi(n) \psi'(n) \mathcal{F}(n) \; . \label{WEthm2}
\end{equation}

Equation (\ref{WEthm2}) clearly implies that the product of $\psi'(n) \times
\mathcal{F}(n)$ must be {\it negative} in order to suck the energy out of the 
inflaton $0$-mode. The classical contribution from $\psi' \times -\psi U'(\psi^2)$
is positive, so reheating must be driven by the quantum corrections from the
$t$-modes and the $u$-modes, each of which has two positive and one negative
contribution. From expressions (\ref{ttrace3}) and (\ref{utrace3}) we see that
the desired negative contribution can only come from the $-\kappa^2 
e^{-2n}/\chi^2(n)$ terms, however, it is not clear if the dominant effect
comes from $t$-modes or $u$-modes. It is also not clear whether the largest
contributions come from super-horizon modes (with $\kappa < \chi(n_e) e^{n_e}$,
where $n_e$ denotes the end of inflation) or sub-horizon modes (with $\kappa >
\chi(n_e) e^{n_e}$). Note that discretization can only recover the longest
wavelength sub-horizon modes.

Let us first examine sub-horizon modes, for which $\chi e^{n}/\kappa$ is 
small. In this case the ultraviolet expansions (\ref{TUVexp}) and (\ref{UUVexp})
imply that the multiplicative exponentials agree to leading order,
\begin{equation}
e^{\mathcal{T}(n,\kappa)} = \frac{e^{-(D-2) n}}{2 \kappa} \Biggl\{ 1 + 
O\Bigl( \frac{\chi^2 e^{2n}}{\kappa^2}\Bigr) \Biggr\} \;\; , \;\;
e^{\mathcal{U}(n,\kappa)} = \frac{e^{-(D-2) n}}{2 \kappa} \Biggl\{ 1 + 
O\Bigl( \frac{\chi^2 e^{2n}}{\kappa^2}\Bigr) \Biggr\} \; .
\end{equation}
Substituting the same ultraviolet expansions into the curly bracketed parts
of (\ref{ttrace3}) and (\ref{utrace3}) gives,
\begin{eqnarray}
\lefteqn{ \Bigl[ \frac12 \mathcal{T}' - \frac{\mu'}{\mu}\Bigr]^2 - 
\frac{\kappa^2 e^{-2n}}{\chi^2} + \frac{e^{-2 [\mathcal{T} + (D-1) n]}}{4 \chi^2} 
} \nonumber \\
& & \hspace{2cm} = -\frac12 (D\!-\!2) (1 \!-\! \epsilon) - (1\!-\!\epsilon) 
\frac{\mu'}{\mu} - \Bigl( \frac{\mu'}{\mu}\Bigr)' + O \Bigl( \frac{\chi^2 e^{2n}}{
\kappa^2}\Bigr) \; , \qquad \\
\lefteqn{ \Bigl[ \frac12 \mathcal{U}' + \frac{\mu'}{\mu}+ D\!-\!2 \Bigr]^2 - 
\frac{\kappa^2 e^{-2n}}{\chi^2} + \frac{e^{-2 [\mathcal{U} + (D-1) n]}}{4 \chi^2} 
} \nonumber \\
& & \hspace{2cm} = \frac12 (D\!-\!2) (1 \!-\! \epsilon) + (1\!-\!\epsilon) 
\frac{\mu'}{\mu} + \Bigl( \frac{\mu'}{\mu}\Bigr)' + O \Bigl( \frac{\chi^2 e^{2n}}{
\kappa^2}\Bigr) \; . \qquad
\end{eqnarray}
Hence there is perfect cancellation between the sub-horizon $t$-mode and $u$-mode
contributions at leading order.

Super-horizon modes cannot show the same cancellation because 
$\mathcal{T}(n,\kappa)$ approaches a large, negative constant (\ref{Tmin}) as 
$\mu(n)$ goes to zero, whereas $\mathcal{U}(n,\kappa)$ diverges like 
$\mathcal{U}_* + \ln[\mu_2^2/\mu^2(n)]$.\footnote{The constant $\mathcal{U}_*$ 
can be found from expression (\ref{UPostform}) by extracting the factor of
$\ln[\mu_2^2/\mu^2(n)]$ and then setting $n = n_*$ in the remainder.} This means 
that the multiplicative exponentials take the form,
\begin{equation}
e^{\mathcal{T}(n,\kappa)} \longrightarrow e^{\mathcal{T}_{\rm min}} \qquad , \qquad
e^{\mathcal{U}(n,\kappa)} \longrightarrow e^{\mathcal{U}_*} \Bigl[ \frac{\mu_2}{\mu(n)}
\Bigr]^2 \; . 
\end{equation}
The curly bracketed terms which involve explicit factors of $\mu'/\mu$ wind up
depending on the functions $f(n)$ and $g(n)$, given in expressions (\ref{fsol}) 
and (\ref{gsol}), respectively,
\begin{eqnarray}
\Bigl[ \frac12 \mathcal{T}' - \frac{\mu'}{\mu}\Bigr]^2 &\!\!\! = \!\!\!& \frac14
f^2(n) \longrightarrow \Bigl[\frac{\mu'(n)}{\mu(n)}\Bigr]^2 \; , \\
\Bigl[ \frac12 \mathcal{U}' + \frac{\mu'}{\mu}+ D\!-\!2 \Bigr]^2 &\!\!\! = \!\!\!&
\frac14 g^2(n) \longrightarrow 0 \; .
\end{eqnarray}
This means that the $t$-modes contribute positively, while the $u$-modes make
a negative contribution,
\begin{eqnarray}
e^{\mathcal{T}} \!\times\! \Biggl\{ \qquad \Biggr\} &\!\!\! \longrightarrow \!\!\!&
e^{\mathcal{T}_{\rm min}} \Bigl[ \frac{\mu'(n)}{\mu(n)}\Bigr]^2 \; , \label{tlead} \\
e^{\mathcal{U}} \!\times\! \Biggl\{ \qquad \Biggr\} &\!\!\! \longrightarrow \!\!\!&
e^{\mathcal{U}_*} \Bigl[ \frac{\mu_2}{\mu(n)} \Bigr]^2 \!\times\! -\frac{\kappa^2
e^{-2n_*}}{\chi_*^2} \; . \qquad \label{ulead}
\end{eqnarray}
How large the relative coefficients are depends on the integration constant $f_2$,
for which we do not yet have an analytic form.

Whether (\ref{tlead}) or (\ref{ulead}) dominates, it is significant that both terms
diverge like $1/\mu^2(n)$. Because the effective force contains another factor of 
$1/\mu^2(n)$, this means that the quantum correction diverges like $1/\mu^4(n)$ near
the point $n_*$ at which $\mu(n)$ vanishes. The measure factor in (\ref{WEthm2})
softens this somewhat, but not enough,
\begin{equation}
Q^2 \psi'(n) \psi(n) dn = \frac14 d\mu^2 \; . \label{measure}
\end{equation} 
The integral (\ref{WEthm2}) therefore diverges before $\mu(n) = 0$, which
presumably brings reheating to an end. 

\section{Conclusions}

Ema et al. have shown that coupling a charged inflaton to electromagnetism 
provides the most efficient reheating \cite{Ema:2016dny}. The mechanism is
that the inflaton's evolution induces a time-dependent photon mass through
the Higgs mechanism. Nothing special changes about the transverse spatial
polarizations, but inverse powers of the mass appear in the 
longitudinal-temporal polarizations (\ref{tAs}) and (\ref{uAs}), which result 
from the photon having ``eaten'' the phase of the inflaton field. These 
factors diverge when the inflaton passes through zero. The effect is 
strengthened by factors of $\mu'(n)/\mu(n)$ which appear in the mass terms 
(\ref{tmass}-\ref{umass}) of the two modes.

This chapter represents an effort to improve on previous excellent numerical 
studies of this process based on discretizing space \cite{Bezrukov:2020txg}. 
Although that method can accommodate arbitrarily strong photon fields, it is of 
course limited to a finite range of sub-horizon modes. In contrast, we use the 
trace of the coincident photon propagator to study the inflaton $0$-mode 
equation (\ref{new0modeeqn}). Our expressions (\ref{ttrace2}-\ref{ttrace3}) and 
(\ref{utrace2}-\ref{utrace3}) for the longitudinal and temporal contributions 
to this trace are exact. They can be used to include the effects of super-horizon
modes, and of arbitrarily short wave length modes. In fact, our use of dimensional
regularization means that the far ultraviolet can be included as well, through the
use of expansions (\ref{TUVexp}) and (\ref{UUVexp}).
 
We have also derived good analytic approximations for the amplitudes. Before
first horizon crossing these are (\ref{TUVexp}) and (\ref{UUVexp}), respectively.
After first crossing the $t$-mode amplitude is well approximated by expression
(\ref{TIRform}) until close to the point at which $\mu(n)=0$. However, expression 
(\ref{Tpfull}) shows that the $t$-mode amplitude remains finite when $\mu(n) = 0$. 

Two forms are required to approximate the $u$-mode amplitude after first
horizon crossing, owing to its dependence on the complicated behavior of the 
$u$-mode mass term (\ref{umass}), which is evident from Figures~\ref{Latemass} 
and \ref{Zoomumass}. During inflation, the near constancy of $\epsilon(n)$
and $\mu^2_{u}(n)/\chi^2(n)$, result in expression (\ref{UIRform}) giving a
good approximation. After the end of inflation the better approximation is
provided by expression (\ref{UPostform}). Because this last form becomes exact
as $\mu(n) \rightarrow 0$, we know that the $u$-mode amplitude diverges like
$-\ln[\mu^2(n)]$, which provides an extra factor of $1/\mu^2(n)$ in the trace
of the photon propagator (\ref{utrace3}).

The obvious next step is to exploit the powerful analytic expressions we have 
derived to make a detailed numerical study of reheating in a realistic model,
such as Starobinsky inflation \cite{Starobinsky:1980te}, Higgs inflation 
\cite{Bezrukov:2007ep}, or a hybrid model \cite{Bezrukov:2020txg}. Such an
analysis would begin by renormalizing equation (\ref{new0modeeqn}), and then
focus on determining whether the dominant effect for $\mu(n) \rightarrow 0$ 
comes from sub-horizon or super-horizon modes, and whether it is the $t$-modes
or the $u$-modes which contribute more strongly. Another key issue is whether
or not the effect is so strong that the inflaton is precluded from making 
even a single oscillation. Right now, it seems as if the strongest effect 
comes from super-horizon $u$-modes, and this contribution is so strong that 
the inflaton $0$-mode is prevented from passing through zero.

Finally, our extension of the vector propagator to include time-dependent 
masses in cosmological backgrounds has two obvious applications in addition 
to reheating. The first of these is the study of quantum corrections to the
expansion history of classical inflation \cite{Miao:2015oba,Liao:2018sci,
Kyriazis:2019xgj,Miao:2019bnq,Miao:2020zeh,Sivasankaran:2020dzp,Katuwal:2021kry}.
Another obvious application is for the study of phase transitions in the early 
universe \cite{Ema:2016dny}.
\chapter{SUMMARY AND CONCLUSIONS} \label{conclusion}
In this dissertation, we considered various aspects of perturbative quantum gravity and its application to early universe cosmology.

After an introduction, in Chapter \ref{gaugePaper}, we considered quantum gravitational corrections to Maxwell’s equations on flat space background. Although the vacuum polarization is highly gauge dependent, we explicitly showed
that this gauge dependence is canceled by contributions from the source which disturbs the
effective field and the observer who measures it. Our final result is a gauge independent, real
and causal effective field equation that can be used in the same way as the classical Maxwell
equation.

In Chapter \ref{perturbative-scalar-coupling}, we discussed how physicists working on atom interferometers are interested in scalar couplings to electromagnetism of dimensions 5 and 6 which might be induced by quantum gravity. There is a widespread belief that such couplings can only be induced by conjectured non-perturbative effects, resulting in unknown coupling strengths. However, we exhibited a completely perturbative mechanism through which quantum gravity induces dimension six couplings with precisely calculable coefficients.

Next, in Chapter \ref{coleman-weinberg}, we accurately approximated the contribution that photons make to the effective potential of a charged inflaton for inflationary geometries with an arbitrary first slow roll parameter $\epsilon$. We found a small, nonlocal contribution and a numerically larger, local part. The local part involves first and second derivatives of $\epsilon$, coming exclusively from the constrained part of the electromagnetic field which carries the long range interaction. This causes the effective potential induced by electromagnetism to respond more strongly to geometrical evolution than for either scalars, which have no derivatives, or spin one half particles, which have only one derivative. For $\epsilon=0$ our final result agreed with that of Allen on de Sitter background\cite{Allen:1983dg}, while the flat space limit agrees with the classic result of Coleman and Weinberg\cite{Coleman:1973jx}.

Finally, in Chapter \ref{reheatingChapter} we considered reheating for a charged inflaton which is minimally coupled to electromagnetism. The evolution of such an inflaton induces a time-dependent mass for the photon. We showed how the massive photon propagator can be expressed as a spatial Fourier mode sum involving three different sorts of mode functions, just like the constant mass case in Chapter \ref{coleman-weinberg}. We developed accurate analytic approximations for these mode functions, and used them to approximate the effective force exerted on the inflaton 0-mode. This effective force allows one to simply compute the evolution of the inflaton 0-mode and to follow the progress of reheating.



\end{document}